\documentclass[10pt, a4paper]{article}
\usepackage{latexsym, amsfonts, euscript,amssymb,amsmath}
\usepackage{listings}
\usepackage{ragged2e}
\usepackage{hyperref}
\usepackage[a4paper,bindingoffset=0.2in,%
            left=1in,right=1.6in,top=1in,bottom=1in,%
            footskip=.25in]{geometry}

\def\FF{{\mathbb F}}
\def\Z{{\mathbb Z}}

\def\Fq{{\mathbb F}_q}

\def\nkq{{[n,k]_q}}

\def\a{{\alpha}}

\newtheorem{theorem}{Theorem}

\renewcommand{\rightmark}

\begin{document}
\begin{center}
{\LARGE Gr\"obner Bases for Schubert Codes}\\[.5cm]
{\large Arunkumar R. Patil\footnote{\scriptsize Department of Mathematics, Shri Guru Gobind Singhji Institute of Engineering and Technology, Nanded, India, Email: arun.iitb@gmail.com}, 
Nitin S. Darkunde\footnote{\scriptsize School of Mathematical Sciences, Swami Ramanand Teerth Marathwada University, Nanded, India, Email: darkundenitin@gmail.com}}
\end{center}
\begin{abstract}

We consider the problem of determining Gr\"obner bases of binomial ideals associated with linear error correcting codes. Computation of Gr\"obner bases of linear codes have become a topic of interest to many researchers in coding theory because of its several applications in decoding and error corrections. In  this paper, Gr\"obner bases of linear codes associated to Grassmann varieties and Schubert varieties over a binary field have been obtained. We also use them to study the decoding of binary Schubert codes.
\end{abstract}
{\bf Keywords:} Grassmann Variety, Schubert Variety, linear code,  Gr\"obner basis, Grassmann code, Schubert code, binomial ideal.

\section{Introduction}
\label{sec:intro}

\label{sect.intro}
The main goal of this paper is to compute Gr\"obner bases of linear codes associated to Grassmann and Schubert varieties over binary fields and to study the decoding of Schubert codes arising from some special Schubert varieties over binary field. 
In recent years, Gr\"{o}bner bases \cite{CLO} have become a topic of interests for many researchers in mathematics, 
computer science and allied areas. The techniques of finding Gr\"{o}bner bases  is widely used in computational
algebraic geometry for solving a system of polynomial equations in several variables, ideal membership problem,
ideal description problem and implicitization problem etc. Gr\"obner bases have also been used in algebraic coding theory
to study structure of linear codes. 
	In \cite{Edgar}, M. Borges-Quintana et al proved that each linear code can be described by a binomial ideal which is the sum of toric ideal and a non prime ideal. In the same paper, authors have determined Gr\"{o}bner bases of these ideals over $\mathbb{F}_2$. 
	 The linear codes constructed from higher dimensional algebraic varieties have seen to be of great interest to the people working in algebraic geometry and coding theory. In \cite{TV3},  Tsfasman, Vl\u{a}dut and Zink determined the existence of a sequence of algebraic-geometric (or Goppa) codes that exceeded the Gilbert-Varshamov bound on the minimum distance. Linear codes from Hermitian varieties, Grassmann varieties, Schubert varieties, Determinantal varieties have been studied quite extensively. Grasmann codes have been introduced by C.T.Ryan \cite{CTR1,CTR2} and studied extensively by D.Yu.Nogin\cite{DYN},	Ghorpade, Lachaud\cite{SRGL}, Tsfasman, Vl\u{a}dut\cite{TV}, Patil, Pillai\cite{P}, Kaipa\cite{SRGKAIPA}, Hansen, 	Johnsen, Ranestad\cite{HJR}.
    
		In this paper we find  the Gr\"obner basis of the binary linear code $C_{\alpha}{(\ell,m)}$ associated with
		Schubert variety $\Omega_{\alpha}$\cite{SRGL,KL}. Further, we discuss the decoding of these codes based on some classical decoding techniques like Maximum likelihood decoding\cite{Xing} or Nearest neighbour decoding\cite{Xing} and another decoding technique which has been employed and ellaborated in \cite{Edgar}.

\section{Preliminaries}
\label{sec:Prelim}

\subsection{Coding Theory}
  Let $\Fq$ be a finite field with $q$ elements, where $q$ is some power of prime. For $0\neq x \in \Fq^n$, we denote its {\em norm} by $\Vert x \Vert$ and define it as the number of non zero entries in an n-tuple $x=(x_1,x_2,\cdots,x_n)$. A {\em  $\nkq$-linear code} $C$ is a $k$-dimensional subspace of $\Fq^n$ over $\Fq$. For a linear code $C$, its (minimum) distance is denoted by $d=d(C)$ and defined as $\min\left\{\Vert x \Vert : x\neq 0, x \in C\right\}$. 
  We call a linear code of length $n$, dimension $k$, and (minimum) distance $d$ as $[n,k,d]_q$-linear code. For more details one can  refer \cite{MS, Xing}. An $[n,k]_q$-linear code is called as {\em non-degenerate linear code}, if it is not contained in a co-ordinate hyperplane of $\Fq^{n}$\cite{TV}. There are several bounds in coding theory, but the well-known bound is given by Singleton\cite{Singleton}, known as {\em Singleton bound}, which states that: for a linear $[n,k,d]$ code over $\Fq$, we have $k+d \leq n+1$. And the code which attains the Singleton bound, i.e. for which $k+d = n+1$ is called as {\em Maximum Distance Separable}(MDS) code\cite{Xing}. A {\em generator matrix}\cite{Xing} for a linear code
  $C$ is a matrix $G$ whose rows form a basis for $C$, whereas a {\em parity check matrix}\cite{Xing} $H$ for a linear code $C$ is a matrix
  whose rows form a basis for the dual code of $C$, usually denoted by $C^\perp$. If $C$ is $[n,k]_q$-linear code, then $C^\perp$
  is $[n,n-k]_q$-linear code. Consequently $G$ is a $k\times n$ matrix whereas $H$ is $(n-k)\times n$ matrix. Also,  $v\in C \Leftrightarrow vH^T=0$ and $v\in C^\perp\Leftrightarrow vG^T=0$\cite{Xing}. A linear $[n,k]_q$ code $C$ has distance $d$ if and only if any $(d-1)$ columns of $H$ are linearly independent and $H$ possesses $d$ columns which are linearly dependent\cite{Xing}.  A linear code of distance $d$ is $a$-{\em error-detecting}
  if and only if $d\geq a+1$, whereas a code $C$ is $b$-{\em error-correcting} if and only if $d\geq 2b+1$. Hence, the error correcting capability(\cite{Edgar, MS}) of a code is $t=[(d-1)/2]$, where $[.]$ denote the greatest integer function. Therefore, one always strive for   a linear code to have a distance as large as possible. According to nearest neighbour decoding technique, if a codeword   $v$ is transmitted and the word $w$ is received, then one defines the error pattern as   $e=w-v\in w+C$, hence the error pattern $e$ and the received word $w$ are in the same coset of $C$. However, as soon as the lenth $n$ becomes large, the nearest decoding technique takes a lot of time for decoding. Therefore, to minimize the time for decoding, syndrome decoding technique is used. If $H$ is a  parity check matrix of an $\nkq$ linear code, then the vector  $S_H(w):=wH^T \in \Fq^{n-k}$ is called the {\em syndrome }\cite{Xing} of  $w \in \Fq^n$.
   For  $u, v \in \Fq^n$, one can easily prove that $S(u+v)=S(u)+S(v)$,  and $S(u)=0\Leftrightarrow u\in C$,
  $S(u)=S(v)\Leftrightarrow$ $u$ and $v$ are in the same coset of $C$. Generally, in syndrome decoding\cite{MS,Xing}, one used to find $S(w)$, where $w$ is the received word. Next we find the coset leader $w'$ next to the syndrome $S(w) = S(w')$ in the syndrome look-up table and finally decode $w$ as $v= w-w'$.

\subsection{Gr\"obner Bases}
 Let $k$ be a field  and $k[X]:= k[x_1,x_2,\cdots,x_n]$ be a polynomial ring in $n$ variables $x_1,x_2,\cdots, \\x_n$. If $\alpha := (\alpha_1, \alpha_2,\cdots,\alpha_n)\in \mathbb{Z}^{n}_{\geq{0}}$, where $\Z_{\geq 0}$ denote the set of non negative integers, then the monomial in $k[X]$ is denoted by $X^\alpha:=x_1^{\alpha_1}x_2^{\alpha_2}\cdots x_n^{\alpha_n}$,  and the sum $\alpha_1+\alpha_2+\cdots+\alpha_n$ is called the {\em total degree} of monomial. 
    A {\em monomial order}\cite{CLO}  $\prec$  on $k[X]$ is a relation $\prec$ on the set of monomials 
    ${X}^\alpha$, such that $\prec$ is a total ordering on $\mathbb{Z}^{n}_{\geq{0}}$, for any $\gamma\in\mathbb{Z}^{n}_{\geq{0}}$, $\alpha\prec\beta$ implies that $\alpha+\gamma\prec\beta+\gamma$, 
   and  every non empty subset of $\mathbb{Z}^{n}_{\geq{0}}$ has a smallest element under $\prec$. One can define infinitely many orderings on $k[X]$. The commonly used monomial orders are {\em lex} order, {\em revlex} order, {\em graded reverse lex} order. For more details one can refer \cite{CLO}. The ordering of immense importance to us is degree reverse lexicographic ordering, which is degree compatible ordering(\cite{Edgar, Irene}).
     Given a monomial order, and a non zero polynomial $f \in k[X]$, let $multideg(f)$ denote the multi degree of $f$ which is defined as $max\{\alpha\in\mathbb{Z}^{n}_{\geq{0}}\colon a_{\alpha}\neq {0}\}$. Also, one can define(cf.\cite{CLO, Edgar}), the leading coefficient of $f$, leading monomial of $f$, leading term of $f$, and total degree of $f$ denoted by $LC(f)$, $LM(f)$, $LT(f)$ and $TD(f)$ respectively .
           For a non-zero ideal $I\subset k[X]$, we have $LT(I):=\{cx^\alpha$ $\colon$ $\exists  f\in I $ with $LT(f):=cx^{\alpha}\}$. And $\langle LT(I) \rangle$ := ideal generated by the elements of $LT(I)$ .

    For a fixed monomial order\cite{CLO}, a finite subset $\mathcal{G}=\{g_1,g_2,\ldots,g_s\}$ of an ideal $I$ is said to be a {\em Gr\"obner basis}\cite{Xing}
     if $\langle LT(I)\rangle = \langle LT(g_1),LT(g_2),\ldots,LT(g_s)\rangle $ . Moreover every ideal in a 
     ring $k[X]$ possesses a Gr\"obner basis. If one uses Gr\"obner basis for dividing a 
     polynomial $f\in k[X]$, then one can get rid of non-uniqueness of remainder in multivariate 
     polynomial ring.
    A Gr\"obner basis $\mathcal{G}$ for an ideal $I$ is said to be {\em reduced} if, $LC(g)=1,\forall g\in \mathcal{G} $ and no monomial in $\mathcal{G}$
    can divide the leading term of any other monomial in $\mathcal{G}$ . For detailed literature on Gr\"obner bases, one may refer
    \cite{CLO}. 
   
\subsection{Algebraic Geometry}
    Let $k^n$ denotes an $n$-dimensional {\em Affine Space}\cite{CLO}, for an integer $n\geq 1$. Further for polynomials $g_1, g_2,\cdots,g_t \in k[X]$ , we denote the {\em Affine Variety}\cite{CLO} defined by $g_1, g_2,\cdots,g_t$ by $V(g_1, g_2,\cdots,g_t)$ and define it as $V(g_1, g_2,\cdots,g_t):= \{(b_1,b_2,\cdots,b_n)\in k^n : g_i(b_1,b_2,\cdots,b_n)\\=0, \text{for all}\hspace{0.09in}  1\leq i \leq t\}$. Affine variety is nothing but a set of common solutions of a bunch of polynomials thereof. There is one more space of quite importance called as {\em Projective Space}\cite{CLO}, which we denote by $\mathbb{P}^n(k)$ and define it as $\mathbb{P}^n(k):= k^{(n+1)}-0/ \sim$, where $\sim$ is an equivalence relation on the non-zero points of $k^{n+1}$, which is given by $(y_0, y_1,\cdots, y_n)\sim (y'_0, y'_1,\cdots, y'_n)$ if and only if there is $\eta \in k^*$ such that $(y_0, y_1,\cdots, y_n)= \eta  (y'_0, y'_1,\cdots, y'
_n)$.\\ Now we will take a look at projective systems\cite{TV} and Grassmannians: An $[n,k]_q$-{\em projective system} is a collection of $n$ not necessarily distinct points in the $(k-1)$-
	  dimensional projective space $\mathbb{P}_{\Fq}^{k-1}$ . For more details about this and
	  for non-degeneracy of such systems, one may refer \cite{TV}. Also, there is a one-to-one correspondence
	  between the equivalence classes of nondegenerate $[n,k]_q$ projective systems and the equivalence classes
	  of linear $[n,k]_q$ codes(\cite{TV}). The case in study undertaken here is of {\em Grassmannians}. For a fixed $\ell$ , $1\leq \ell \leq m $, we denote the set of all $\ell$-
	  dimensional subspaces of an $m$-dimensional vector space
	  $\Fq^m$ over $\Fq$ by $G_{\ell,m}$. This {\em Grassmannian} $G_{\ell,m}$ over $\Fq$ defines an 
	  $[n,k]_q$ projective system, with following parameters\cite{SRGL}:\\
$$
 n=|G_{\ell,m}(\Fq)|= {{m}\brack{\ell}}_q :=\frac{(q^m-1)(q^m-q)\cdots(q^m-q^{\ell-1})}{(q^{\ell}-1)(q^{\ell}-q)\cdots
(q^{\ell}-q^{\ell-1})}.
$$ and
$\quad k := 
{{m}\choose{\ell}}$ .
Let $
I(\ell,m):=\{ \a = (\a_1, \dots , \a_{\ell} )\in \Z^{\ell}  : 
1\le \a_1 < \dots < \a_\ell  \le m \} 
$. Let $W \in G_{\ell,m}(\Fq)$ and $A_W$ be an $\ell \times m$ matrix whose rows gives a basis of $W$. Then through {\em Pl\"ucker embedding} set up which is described in\cite{SRGL}, we obtain the corresponding Pl\"ucker coordinate 
$
(p_{\alpha}(A_W))_{\alpha\in I(\ell,m)} \in \mathbb{P}_{\Fq}^{k-1}
$, where $p_{\alpha}(A):=det(a_{ij})_{1\leq i,j\leq \ell}$. These Pl\"ucker coordinates satisfies certain quadratic polynomial\cite{KL}. One may refer\cite{SRGL,KL,hp} for its details.

 The non-degenerate linear $[n,k]_q$ code corresponding to the projective system defined by $G_{\ell,m}(\Fq)$ with Pl\"ucker embedding
is denoted by $C(\ell,m)$ which is called as {\em Grassmann code} , where the parameters of this code are known to be
$n={{m}\brack{\ell}}_q$,
$k={{m}\choose{\ell}}$ and $d=q^\delta$, where $\delta=\ell(m-\ell)$(cf.\cite{SRGL}).
    Now we will take a look at subvarieties of Grassmannians viz. Schubert varieties \cite{KL}. Fix any 
    $\alpha=(\alpha_{1},\ldots,\alpha_{\ell})\in I(\ell,m)$. Let $C_1\subset\ldots \subset C_{\ell}$ be the sequence of subspaces of
    $\Fq^m$ associated to $\alpha$. Let 
$ C_i = span\{e_1,\ldots,e_{\alpha_{i}}\}$ , $1\leq i \leq\ell $,
where $\{e_1,\ldots,e_{\alpha_{i}}\}$ denotes the canonical basis of $\Fq^m$. Then $\Omega_{\alpha}(\Fq)=\{W\in G_{\ell,m}(\Fq):dim(W\cap C_i) \geq i$ for $ i=1,\ldots,\ell\} $ is the Schubert subvariety of  $G_{\ell,m}(\Fq)$
associated to $\alpha$. As a projective variety, $\Omega_{\alpha}(\Fq):=\{p\in G_{\ell,m}(\Fq) : p_{\beta} = 0$   for all  $ \beta \nleq \alpha \} $(\cite{SRGL,KL}) which gives a non-degenerate $[n_{\alpha}, k_{\alpha}]_q$ projective system, via induced Pl\"ucker embedding,
and hence, a linear $[n_{\alpha}, k_{\alpha}]_q$ code. This code is called as a Schubert code and denoted by 
$C_{\alpha}{(\ell,m)}$. The parameters of this code are known to be $n_{\alpha}=\#\Omega_{\alpha}(\Fq)$, $ k_{\alpha}= \#\{\beta \in I(\ell,m): \beta \leq \alpha\}$ and $d(C_{\alpha}{(\ell,m)})
=q^{\delta_{\alpha}}$, where, $ \delta_{\alpha}=\displaystyle{\sum_{i=1}^{\ell}}(\alpha_{i}-i)$ (\cite{SRGL}). As an example, we describe $C_{(1,4)}(2,5)(\FF_{2})$ in 
$G_{2,5}(\FF_{2})$ as follows: $\Omega_{(1,4)}=\{p\in G_{2,5}:p_{\beta}=0$, \text{for all}    $\beta \nleq (1,4)\}$. Therefore, $p \in \Omega_{(1,4)} \Longrightarrow  p= (p_{12},p_{13},p_{14},0,0,0,0,0,0,0)$, hence this
 code is $[7,3,4]_2$-linear code . Its generator matrix is:
 $$A_{(1,4)}=\begin{bmatrix}
   1 & 0 & 0 & 0 & 1 & 1 & 1 \\
   0 & 1 & 0 & 1 & 1 & 1 & 0 \\
   0 & 0 & 1 & 1 & 1 & 0 & 1 \\
  \end{bmatrix}_{3\times 7}.
$$\\
On the similar lines, we can find generator matrices for reamaining Schubert codes such as: $C_{(1,5)}(2,5)(\FF_2)$, $C_{(2,3)}(2,5)(\FF_2)$, $C_{(2,4)}(2,5)(\FF_2)$. 


\section{Binomial Ideals of Linear Codes}
\label{sec:cd}
In this section, we are collecting some known results from \cite{Edgar} regarding Gr\"obner bases for binary codes and thereby we will apply these results to find out Gr\"obner bases for Schubert codes listed in section 2. Moreover, we are going to compute the minimum distance of Schubert codes in accordance with \cite{Edgar} and will discuss about their decoding.

M.Borges-Quintana et al \cite{Edgar} defined an ideal associated with the binary linear code, denoted by
    $I(C)$ and defined it as:
    $$
    I(C):=\langle\{\small{X}^a - X^b :(\psi(X^a),(\psi(X^b))\in R\}\rangle\subset k[X]
    $$
    where define, the map $\psi$ as:
    \begin{equation*}
     \psi:[X] \rightarrow  \FF_2^n  , 
     \end{equation*}
     where,
     \begin{center}
      $  \psi(X^a)=\psi\left(\displaystyle\prod_{i=1}^n{x_i^{a_i}}\right)=(a_1 \text{mod 2},\ldots,a_n \text{mod 2})$
     \end{center}
and 

\begin{equation}
\left((\psi(X^a),(\psi(X^b)\right)\in R \Leftrightarrow \psi(X^a)-(\psi(X^b)\in C.
\end{equation}

\begin{theorem}
{\em \cite{Edgar}} Let $C$ be a linear code and $R$ be an equivalence relation(defined in equation (1) above). The ideal associated with $C$ is
 $I(C)$ which is given by :
\begin{equation}
I(C):=\langle\{\small{X}^a - X^b :(\psi(X^a),(\psi(X^b))\in R\}\rangle\subset k[X]
\end{equation}  
which is a binomial ideal.
\end{theorem}
Further M.Borges-Quintana et. al. \cite{Edgar} proved that above ideal $I(C)$ equals another ideal in a simpler form, which
is denoted by $I$ and it is defined as:
\begin{equation}
 I=\langle\{X^{w_1}-1,\ldots,X^{w_k}-1\}\cup\{x_i^2-1:i=1,\ldots,n\}\rangle.
\end{equation}
where $w_1,\ldots,w_k$ are rows of generator matrix of given $[n,k]_2$ linear code.
\begin{theorem}
{\em \cite{Edgar}} With the notations as above:  I = I(C) .
\end{theorem}
\begin{theorem}
{\em \cite{Edgar}} Let $C$ be a linear code and let $\mathcal{G}$ be its associated reduced Gr\"obner basis with respect to $\prec$ . If $X^a \in [X]$ satisfies the condition  $wt(\psi(\text{Can}(X^a,\mathcal{G})))\leq t$, then
$\psi(\text{Can}(X^a, \mathcal{G}))$ is the error vector corresponding to $\psi(X^a)$, that is
$$\psi(X^a)-\psi(\text{Can}(X^a, \mathcal{G}))$$
is the closest codeword to $\psi({X^a})$. On the other hand, if 
$wt(\psi(\text{Can}(X^a,\mathcal{G})))> t$, then $\psi(X^a)$ contains more than $t$ errors.
\end{theorem}


\begin{theorem}
{\em \cite{Edgar}} (Error Correcting Capability of a given linear code $C$ over $\FF_2$)\\
$$
t= \text{min}\{TD(f): f\in \mathcal{G}-\{x_i^2-1:i=1,\ldots,n\}\}-1
$$
\end{theorem}
Now, we will employ above theorems in order to find out ideal associated to corresponding Schubert Codes and further compute their reduced Gr\"obner bases.

\section{Gr\"obner Bases of Schubert Codes}
Let us calculate Gr\"obner bases for Schubert codes, which we had discussed in section $2$.

\subsection{Gr\"obner Basis for the Schubert Code $\bf C_{(1,4)}(2,5)(\FF_2)$:}
Firstly, for the Schubert code $C_{(1,4)}(2,5)(\FF_2)$, we have a generator matrix, which we calculated in above section 2, which is given by,\\
$$A_{(1,4)}=\begin{bmatrix}
   1 & 0 & 0 & 0 & 1 & 1 & 1 \\
   0 & 1 & 0 & 1 & 1 & 1 & 0 \\
   0 & 0 & 1 & 1 & 1 & 0 & 1 \\
  \end{bmatrix}_{3\times 7}
$$ \\
By theorem 1 and 2 above(\cite{Edgar}), we can write an ideal associated with this Schubert Code, which is given by:
\begin{align*}
I=I(C_{(1,4)}(2,5)(\FF_2)) =&\langle x_{1}x_{5}x_{6}x_{7}-1, x_{2}x_{4}x_{5}x_{6}-1, x_{3}x_{4}x_{5}x_{7}-1, x_{1}^2-1, x_{2}^2-1,\\
 &  x_{3}^2-1,x_{4}^2-1, x_{5}^2-1,x_{6}^2-1,x_{7}^2-1\rangle
\end{align*}
We denote the reduced Gr\"obner basis of this ideal by $\mathcal{G}_{(1,4)}(2,5)$, which is given by:
\begin{align*}
\mathcal{G}_{(1,4)}(2,5)= &\langle x_1^2-1, x_1x_2-x_4x_7, x_2^2-1,x_1x_3-x_4x_6, x_2x_3-x_6x_7, x_3^2-1,\\ 
 &x_1x_4-x_2x_7, x_2x_4-x_1x_7, x_3x_4-x_5x_7,x_4^2-1,x_1x_5-x_6x_7, x_2x_5-x_4x_6,\\
 &x_3x_5-x_4x_7, x_4x_5-x_3x_7,
x_5^2-1,x_1x_6-x_5x_7, x_2x_6-x_3x_7, x_3x_6-x_2x_7,\\ 
&x_5x_6-x_1x_7,
x_6^2-1,x_7^2-1\rangle
\end{align*}

\subsection{Error Correcting capability of the Schubert Code $\bf C_{(1,4)}(2,5)(\FF_2)$:}
By theorem $4$ above, we can conclude that, the error correcting capability of this Schubert Code is 1, which is in accordance with our known results(\cite{Edgar,MS}), because for this Schubert Code, we know that $d=4$, and hence $t=1 (\text{since}, t=[(d-1)/2] )$.

\subsection{Decoding with $[7,3,4]$-linear Schubert Code $\bf C_{(1,4)}(2,5)(\FF_2)$:}
By using theorem 3 above from \cite{Edgar}, we can tackle the decoding problem upto certain extent by using Gr\"obner bases as follows:
\begin{center}
\begin{tabular}{|c|c|c|}
\hline
Received word & Canonical form of  &Received word is decoded to\\ 
 & this word w.r.t. $\mathcal{G}_{(1,4)}(2,5)$ & \\
\hline
$x_1x_2x_3x_4x_5$	&	$x_4$	&	$x_1x_2x_3x_5$\\
\hline
$x_1x_2x_3$			&	$x_5$	&	$x_1x_2x_3x_5$\\
\hline
$x_2x_3x_4x_5x_6$	&	$x_3$	&	$x_2x_4x_5x_6$\\
\hline
$x_2x_3x_4x_5x_7$	&	$x_2$	&	$x_3x_4x_5x_7$\\
\hline
$x_3x_4x_5x_6x_7$	&	$x_6$	&	$x_3x_4x_5x_7$\\
\hline

\end{tabular}
\end{center}
As far as decoding of remaining tyes of received words is concerned, one can employ classical decoding techniques like Nearest neighbour decoding technique\cite{Xing}.

\subsection{Gr\"obner Basis for the Schubert Code $\bf C_{(1,5)}(2,5)(\FF_2)$:}
For the case of Schubert code $C_{(1,5)}(2,5)(\FF_2)$, we have a generator matrix given below:\\
$$A_{(1,5)}=
\left[{\begin{array}{ccccccccccccccc}
1 & 0 & 0 & 0 & 0 & 0 & 0 & 0 & 1 & 1 & 1 & 1 & 1 & 1 & 1 \\
0 & 1 & 0 & 0 & 0 & 1 & 1 & 1 & 1 & 1 & 1 & 1 & 0 & 0 & 0 \\
0 & 0 & 1 & 0 & 1 & 0 & 1 & 1 & 1 & 1 & 0 & 0 & 1 & 0 & 1 \\
0 & 0 & 0 & 1 & 1 & 1 & 0 & 1 & 0 & 1 & 1 & 0 & 1 & 1 & 0  
\end{array}}\right]_{4\times15}$$\\
By using techniques in\cite{Edgar}, we can write an ideal $I(C_{(1,5)}(2,5)(\FF_2))$ associated with code $C_{(1,5)}(2,5)(\FF_2)$. We denote the reduced Gr\"obner basis of this ideal by $\mathcal{G}_{(1,5)}(2,5)$, which is given by:\\
$\langle x_1x_2x_3x_4 - x_8x_9x_{11}x_{13}, x_1x_2x_3x_5 - x_6x_9x_{10}x_{14}, x_1x_2x_4x_5 -
x_7x_{10}x_{11}x_{15}, x_1x_2x_3x_6 - x_5x_9x_{10}x_{14}, x_1x_3x_4x_6 -
x_7x_{10}x_{12}x_{13}, x_1x_2x_5x_6 - x_3x_9x_{10}x_{14}, x_1x_3x_5x_6 -
x_2x_9x_{10}x_{14}, x_2x_3x_5x_6 - x_{11}x_{12}x_{13}x_{15}, x_1x_4x_5x_6 -
x_8x_9x_{12}x_{15}, x_1x_2x_4x_7 - x_5x_{10}x_{11}x_{15}, x_1x_3x_4x_7 -
x_6x_{10}x_{12}x_{13}, x_1x_2x_5x_7 - x_4x_{10}x_{11}x_{15}, x_1x_3x_5x_7 -
x_8x_{11}x_{12}x_{14}, x_1x_4x_5x_7 - x_2x_{10}x_{11}x_{15}, x_2x_4x_5x_7 -
x_1x_{10}x_{11}x_{15}, x_1x_2x_6x_7 - x_8x_{13}x_{14}x_{15}, x_1x_3x_6x_7 -
x_4x_{10}x_{12}x_{13}, x_1x_4x_6x_7 - x_3x_{10}x_{12}x_{13}, x_3x_4x_6x_7 -
x_9x_{11}x_{14}x_{15}, x_1x_2x_3x_8 - x_4x_9x_{11}x_{13}, x_1x_2x_4x_8 -
x_3x_9x_{11}x_{13}, x_1x_3x_4x_8 - x_2x_9x_{11}x_{13}, x_2x_3x_4x_8 -
x_{10}x_{12}x_{14}x_{15}, x_1x_3x_5x_8 - x_7x_{11}x_{12}x_{14}, x_1x_4x_5x_8 -x_6x_9x_{12}x_{15}, x_1x_2x_6x_8 - x_7x_{13}x_{14}x_{15}, x_1x_4x_6x_8 -x_5x_9x_{12}x_{15}, x_1x_5x_6x_8 - x_4x_9x_{12}x_{15}, x_4x_5x_6x_8 - x_1x_9x_{12}x_{15},x_1x_2x_7x_8 - x_6x_{13}x_{14}x_{15}, x_1x_3x_7x_8 - x_5x_{11}x_{12}x_{14}, x_1x_5x_7x_8- x_3x_{11}x_{12}x_{14}, x_3x_5x_7x_8 - x_9x_{10}x_{13}x_{15}, x_1x_6x_7x_8 -x_2x_{13}x_{14}x_{15}, x_2x_6x_7x_8 - x_1x_{13}x_{14}x_{15}, x_1x_2x_3x_9 -x_5x_6x_{10}x_{14}, x_1x_2x_4x_9 - x_3x_8x_{11}x_{13}, x_1x_3x_4x_9 - x_2x_8x_{11}x_{13},x_2x_3x_4x_9 - x_1x_8x_{11}x_{13}, x_1x_2x_5x_9 - x_3x_6x_{10}x_{14}, x_1x_3x_5x_9 -x_2x_6x_{10}x_{14}, x_2x_3x_5x_9 - x_1x_6x_{10}x_{14}, x_1x_4x_5x_9 - x_6x_8x_{12}x_{15},
x_2x_4x_5x_9 - x_7x_{12}x_{13}x_{14}, x_1x_2x_6x_9 - x_3x_5x_{10}x_{14}, x_1x_3x_6x_9 -x_2x_5x_{10}x_{14}, x_2x_3x_6x_9 - x_1x_5x_{10}x_{14}, x_1x_4x_6x_9 - x_5x_8x_{12}x_{15},x_3x_4x_6x_9 - x_7x_{11}x_{14}x_{15}, x_1x_5x_6x_9 - x_4x_8x_{12}x_{15}, x_2x_5x_6x_9 -x_1x_3x_{10}x_{14}, x_3x_5x_6x_9 - x_1x_2x_{10}x_{14}, x_4x_5x_6x_9 - x_1x_8x_{12}x_{15},
x_2x_4x_7x_9 - x_5x_{12}x_{13}x_{14}, x_3x_4x_7x_9 - x_6x_{11}x_{14}x_{15}, x_2x_5x_7x_9- x_4x_{12}x_{13}x_{14}, x_3x_5x_7x_9 - x_8x_{10}x_{13}x_{15}, x_4x_5x_7x_9 -x_2x_{12}x_{13}x_{14}, x_2x_6x_7x_9 - x_8x_{10}x_{11}x_{12}, x_3x_6x_7x_9 -x_4x_{11}x_{14}x_{15}, x_4x_6x_7x_9 - x_3x_{11}x_{14}x_{15}, x_1x_2x_8x_9 -x_3x_4x_{11}x_{13}, x_1x_3x_8x_9 - x_2x_4x_{11}x_{13}, x_2x_3x_8x_9 - x_1x_4x_{11}x_{13},x_1x_4x_8x_9 - x_5x_6x_{12}x_{15}, x_2x_4x_8x_9 - x_1x_3x_{11}x_{13}, x_3x_4x_8x_9 -x_1x_2x_{11}x_{13}, x_1x_5x_8x_9 - x_4x_6x_{12}x_{15}, x_3x_5x_8x_9 -x_7x_{10}x_{13}x_{15}, x_4x_5x_8x_9 - x_1x_6x_{12}x_{15}, x_1x_6x_8x_9 -x_4x_5x_{12}x_{15}, x_2x_6x_8x_9 - x_7x_{10}x_{11}x_{12}, x_4x_6x_8x_9 -x_1x_5x_{12}x_{15}, x_5x_6x_8x_9 - x_1x_4x_{12}x_{15}, x_2x_7x_8x_9 -x_6x_{10}x_{11}x_{12}, x_3x_7x_8x_9 - x_5x_{10}x_{13}x_{15}, x_5x_7x_8x_9 -x_3x_{10}x_{13}x_{15}, x_6x_7x_8x_9 - x_2x_{10}x_{11}x_{12}, x_1x_2x_3x_{10} -x_5x_6x_9x_{14}, x_1x_2x_4x_{10} - x_5x_7x_{11}x_{15}, x_1x_3x_4x_{10} -x_6x_7x_{12}x_{13}, x_2x_3x_
4x_{10} - x_8x_{12}x_{14}x_{15}, x_1x_2x_5x_{10} -x_4x_7x_{11}x_{15}, x_1x_3x_5x_{10} - x_2x_6x_9x_{14}, x_2x_3x_5x_{10} - x_1x_6x_9x_{14},
x_1x_4x_5x_{10} - x_2x_7x_{11}x_{15}, x_2x_4x_5x_{10} - x_1x_7x_{11}x_{15}, x_1x_2x_6x_{10}- x_3x_5x_9x_{14}, x_1x_3x_6x_{10} - x_2x_5x_9x_{14}, x_2x_3x_6x_{10} -x_1x_5x_9x_{14}, x_1x_4x_6x_{10} - x_3x_7x_{12}x_{13}, x_3x_4x_6x_{10} -x_1x_7x_{12}x_{13}, x_1x_5x_6x_{10} - x_2x_3x_9x_{14}, x_2x_5x_6x_{10} - x_1x_3x_9x_{14},x_3x_5x_6x_{10} - x_1x_2x_9x_{14}, x_4x_5x_6x_{10} - x_8x_{11}x_{13}x_{14}, x_1x_2x_7x_{10}- x_4x_5x_{11}x_{15}, x_1x_3x_7x_{10} - x_4x_6x_{12}x_{13}, x_1x_4x_7x_{10} -x_2x_5x_{11}x_{15}, x_2x_4x_7x_{10} - x_1x_5x_{11}x_{15}, x_3x_4x_7x_{10} -x_1x_6x_{12}x_{13}, x_1x_5x_7x_{10} - x_2x_4x_{11}x_{15}, x_2x_5x_7x_{10} -x_1x_4x_{11}x_{15}, x_3x_5x_7x_{10} - x_8x_9x_{13}x_{15}, x_4x_5x_7x_{10} -x_1x_2x_{11}x_{15}, x_1x_6x_7x_{10} - x_3x_4x_{12}x_{13}, x_2x_6x_7x_{10} -x_8x_9x_{11}x_{12}, x_3x_6x_7x_{10} - x_1x_4x_{12}x_{13}, x_4x_6x_7x_{10} -x_1x_3x_{12}x_{13}, x_2x_3x_8x_{10} - x_4x_{12}x_{14}x_{15}, x_2x_4x_8x_{10} -x_3x_{12}x_{14}x_{15}, x_3x_4x_8x_{10} - x_2x_{12}x_{14}x_{15}, x_3x_5x_8x_{10} 
-x_7x_9x_{13}x_{15}, x_4x_5x_8x_{10} - x_6x_{11}x_{13}x_{14}, x_2x_6x_8x_{10} -x_7x_9x_{11}x_{12}, x_4x_6x_8x_{10} - x_5x_{11}x_{13}x_{14}, x_5x_6x_8x_{10} -x_4x_{11}x_{13}x_{14}, x_2x_7x_8x_{10} - x_6x_9x_{11}x_{12}, x_3x_7x_8x_{10} -x_5x_9x_{13}x_{15}, x_5x_7x_8x_{10} - x_3x_9x_{13}x_{15}, x_6x_7x_8x_{10} -x_2x_9x_{11}x_{12}, x_1x_2x_9x_{10} - x_3x_5x_6x_{14}, x_1x_3x_9x_{10} - x_2x_5x_6x_{14},x_2x_3x_9x_{10} - x_1x_5x_6x_{14}, x_1x_5x_9x_{10} - x_2x_3x_6x_{14}, x_2x_5x_9x_{10} -x_1x_3x_6x_{14}, x_3x_5x_9x_{10} - x_7x_8x_{13}x_{15}, x_1x_6x_9x_{10} - x_2x_3x_5x_{14},x_2x_6x_9x_{10} - x_1x_3x_5x_{14}, x_3x_6x_9x_{10} - x_1x_2x_5x_{14}, x_5x_6x_9x_{10} -x_1x_2x_3x_{14}, x_2x_7x_9x_{10} - x_6x_8x_{11}x_{12}, x_3x_7x_9x_{10} -x_5x_8x_{13}x_{15}, x_5x_7x_9x_{10} - x_3x_8x_{13}x_{15}, x_6x_7x_9x_{10} -x_2x_8x_{11}x_{12}, x_2x_8x_9x_{10} - x_6x_7x_{11}x_{12}, x_3x_8x_9x_{10} -x_5x_7x_{13}x_{15}, x_5x_8x_9x_{10} - x_3x_7x_{13}x_{15}, x_6x_8x_9x_{10} -x_2x_7x_{11}x_{12}, x_7x_8x_9x_{10} - x_3x_5x_{13}x_{15}, x_1x_
2x_3x_{11} -x_4x_8x_9x_{13}, x_1x_2x_4x_{11} - x_5x_7x_{10}x_{15}, x_1x_3x_4x_{11} - x_2x_8x_9x_{13},
x_2x_3x_4x_{11} - x_1x_8x_9x_{13}, x_1x_2x_5x_{11} - x_4x_7x_{10}x_{15}, x_1x_3x_5x_{11}- x_7x_8x_{12}x_{14}, x_2x_3x_5x_{11} - x_6x_{12}x_{13}x_{15}, x_1x_4x_5x_{11} -x_2x_7x_{10}x_{15}, x_2x_4x_5x_{11} - x_1x_7x_{10}x_{15}, x_2x_3x_6x_{11} -x_5x_{12}x_{13}x_{15}, x_3x_4x_6x_{11} - x_7x_9x_{14}x_{15}, x_2x_5x_6x_{11} -x_3x_{12}x_{13}x_{15}, x_3x_5x_6x_{11} - x_2x_{12}x_{13}x_{15}, x_4x_5x_6x_{11} -x_8x_{10}x_{13}x_{14}, x_1x_2x_7x_{11} - x_4x_5x_{10}x_{15}, x_1x_3x_7x_{11} -x_5x_8x_{12}x_{14}, x_1x_4x_7x_{11} - x_2x_5x_{10}x_{15}, x_2x_4x_7x_{11} -x_1x_5x_{10}x_{15}, x_3x_4x_7x_{11} - x_6x_9x_{14}x_{15}, x_1x_5x_7x_{11} -x_2x_4x_{10}x_{15}, x_2x_5x_7x_{11} - x_1x_4x_{10}x_{15}, x_3x_5x_7x_{11} -x_1x_8x_{12}x_{14}, x_4x_5x_7x_{11} - x_1x_2x_{10}x_{15}, x_2x_6x_7x_{11} -x_8x_9x_{10}x_{12}, x_3x_6x_7x_{11} - x_4x_9x_{14}x_{15}, x_4x_6x_7x_{11} -x_3x_9x_{14}x_{15}, x_1x_2x_8x_{11} - x_3x_4x_9x_{13}, x_1x_3x_8x_{11} -x_5x_7x_{12}x_{14}, x_2x_3x_8x_{11} - x_1x_4x_9x_{13}, x_1x_4x_8x_{11} - x_2x_3x_9x_{13},x_2x_4x_
8x_{11} - x_1x_3x_9x_{13}, x_3x_4x_8x_{11} - x_1x_2x_9x_{13}, x_1x_5x_8x_{11} -x_3x_7x_{12}x_{14}, x_3x_5x_8x_{11} - x_1x_7x_{12}x_{14}, x_4x_5x_8x_{11} -x_6x_{10}x_{13}x_{14}, x_2x_6x_8x_{11} - x_7x_9x_{10}x_{12}, x_4x_6x_8x_{11} -x_5x_{10}x_{13}x_{14}, x_5x_6x_8x_{11} - x_4x_{10}x_{13}x_{14}, x_1x_7x_8x_{11} -x_3x_5x_{12}x_{14}, x_2x_7x_8x_{11} - x_6x_9x_{10}x_{12}, x_3x_7x_8x_{11} -x_1x_5x_{12}x_{14}, x_5x_7x_8x_{11} - x_1x_3x_{12}x_{14}, x_6x_7x_8x_{11} -x_2x_9x_{10}x_{12}, x_1x_2x_9x_{11} - x_3x_4x_8x_{13}, x_1x_3x_9x_{11} - x_2x_4x_8x_{13},x_2x_3x_9x_{11} - x_1x_4x_8x_{13}, x_1x_4x_9x_{11} - x_2x_3x_8x_{13}, x_2x_4x_9x_{11} -x_1x_3x_8x_{13}, x_3x_4x_9x_{11} - x_6x_7x_{14}x_{15}, x_2x_6x_9x_{11} -x_7x_8x_{10}x_{12}, x_3x_6x_9x_{11} - x_4x_7x_{14}x_{15}, x_4x_6x_9x_{11} -x_3x_7x_{14}x_{15}, x_2x_7x_9x_{11} - x_6x_8x_{10}x_{12}, x_3x_7x_9x_{11} -x_4x_6x_{14}x_{15}, x_4x_7x_9x_{11} - x_3x_6x_{14}x_{15}, x_6x_7x_9x_{11} -x_3x_4x_{14}x_{15}, x_1x_8x_9x_{11} - x_2x_3x_4x_{13}, x_2x_8x_9x_{11} - x_1x_3x_4x_{13}
,x_3x_8x_9x_{11} - x_1x_2x_4x_{13}, x_4x_8x_9x_{11} - x_1x_2x_3x_{13}, x_6x_8x_9x_{11} -x_2x_7x_{10}x_{12}, x_7x_8x_9x_{11} - x_2x_6x_{10}x_{12}, x_1x_2x_{10}x_{11} -x_4x_5x_7x_{15}, x_1x_4x_{10}x_{11} - x_2x_5x_7x_{15}, x_2x_4x_{10}x_{11} -x_1x_5x_7x_{15}, x_1x_5x_{10}x_{11} - x_2x_4x_7x_{15}, x_2x_5x_{10}x_{11} -x_1x_4x_7x_{15}, x_4x_5x_{10}x_{11} - x_1x_2x_7x_{15}, x_2x_6x_{10}x_{11} -x_7x_8x_9x_{12}, x_4x_6x_{10}x_{11} - x_5x_8x_{13}x_{14}, x_5x_6x_{10}x_{11} -x_4x_8x_{13}x_{14}, x_1x_7x_{10}x_{11} - x_2x_4x_5x_{15}, x_2x_7x_{10}x_{11} -x_1x_4x_5x_{15}, x_4x_7x_{10}x_{11} - x_1x_2x_5x_{15}, x_5x_7x_{10}x_{11} -x_1x_2x_4x_{15}, x_6x_7x_{10}x_{11} - x_2x_8x_9x_{12}, x_2x_8x_{10}x_{11} -x_6x_7x_9x_{12}, x_4x_8x_{10}x_{11} - x_5x_6x_{13}x_{14}, x_5x_8x_{10}x_{11} -x_4x_6x_{13}x_{14}, x_6x_8x_{10}x_{11} - x_4x_5x_{13}x_{14}, x_7x_8x_{10}x_{11} -
x_2x_6x_9x_{12}, x_1x_9x_{10}x_{11} - x_{12}x_{13}x_{14}x_{15}, x_2x_9x_{10}x_{11} -x_6x_7x_8x_{12}, x_6x_9x_{10}x_{11} - x_2x_7x_8x_{12}, x_7x_9x_{10}x_{11} -x_2x_6x_8x_{12}, x_8x_9x_{10}x_{11} - x_2x_6x_7x_{12}, x_1x_3x_4x_{12} -
x_6x_7x_{10}x_{13}, x_2x_3x_4x_{12} - x_8x_{10}x_{14}x_{15}, x_1x_3x_5x_{12} -
x_7x_8x_{11}x_{14}, x_2x_3x_5x_{12} - x_6x_{11}x_{13}x_{15}, x_1x_4x_5x_{12} -
x_6x_8x_9x_{15}, x_2x_4x_5x_{12} - x_7x_9x_{13}x_{14}, x_1x_3x_6x_{12} -
x_4x_7x_{10}x_{13}, x_2x_3x_6x_{12} - x_5x_{11}x_{13}x_{15}, x_1x_4x_6x_{12} -
x_5x_8x_9x_{15}, x_3x_4x_6x_{12} - x_1x_7x_{10}x_{13}, x_1x_5x_6x_{12} - x_4x_8x_9x_{15},x_2x_5x_6x_{12} - x_3x_{11}x_{13}x_{15}, x_3x_5x_6x_{12} - x_2x_{11}x_{13}x_{15},x_4x_5x_6x_{12} - x_1x_8x_9x_{15}, x_1x_3x_7x_{12} - x_5x_8x_{11}x_{14}, x_1x_4x_7x_{12}- x_3x_6x_{10}x_{13}, x_2x_4x_7x_{12} - x_5x_9x_{13}x_{14}, x_3x_4x_7x_{12} -x_1x_6x_{10}x_{13}, x_1x_5x_7x_{12} - x_3x_8x_{11}x_{14}, x_2x_5x_7x_{12} -x_4x_9x_{13}x_{14}, x_3x_5x_7x_{12} - x_1x_8x_{11}x_{14}, x_4x_5x_7x_{12} -x_2x_9x_{13}x_{14}, x_1x_6x_7x_{12} - x_3x_4x_{10}x_{13}, x_3x_6x_7x_{12} -x_1x_4x_{10}x_{13}, x_4x_6x_7x_{12} - x_1x_3x_{10}x_{13}, x_1x_3x_8x_{12} -x_5x_7x_{11}x_{14}, x_2x_3x_8x_{12} - x_4x_{10}x_{14}x_{15}, x_1x_4x_8x_{12} -x_5x_6x_9x_{15}, x_2x_4x_8x_{12} - x_3x_{10}x_{14}x_{15}, x_3x_4x_8x_{12} -x_2x_{10}x_{14}x_{15}, x_1x_5x_8x_{12} - x_4x_6x_9x_{15}, x_3x_5x_8x_{12} -x_1x_7x_{11}x_{14}, x_4x_5x_8x_{12} - x_1x_6x_9x_{15}, x_1x_6x_8x_{12} - x_4x_5x_9x_{15},x_4x_6x_8x_{12} - x_1x_5x_9x_{15}, x_5x_6x_8x_{12} - x_1x_4x_9x_{15}, x_
1x_7x_8x_{12} -
x_3x_5x_{11}x_{14}, x_3x_7x_8x_{12} - x_1x_5x_{11}x_{14}, x_5x_7x_8x_{12} -
x_1x_3x_{11}x_{14}, x_1x_4x_9x_{12} - x_5x_6x_8x_{15}, x_2x_4x_9x_{12} -
x_5x_7x_{13}x_{14}, x_1x_5x_9x_{12} - x_4x_6x_8x_{15}, x_2x_5x_9x_{12} -
x_4x_7x_{13}x_{14}, x_4x_5x_9x_{12} - x_1x_6x_8x_{15}, x_1x_6x_9x_{12} - x_4x_5x_8x_{15},x_4x_6x_9x_{12} - x_1x_5x_8x_{15}, x_5x_6x_9x_{12} - x_1x_4x_8x_{15}, x_2x_7x_9x_{12} -x_4x_5x_{13}x_{14}, x_4x_7x_9x_{12} - x_2x_5x_{13}x_{14}, x_5x_7x_9x_{12} -x_2x_4x_{13}x_{14}, x_1x_8x_9x_{12} - x_4x_5x_6x_{15}, x_4x_8x_9x_{12} - x_1x_5x_6x_{15},x_5x_8x_9x_{12} - x_1x_4x_6x_{15}, x_6x_8x_9x_{12} - x_1x_4x_5x_{15}, x_1x_3x_{10}x_{12}
- x_4x_6x_7x_{13}, x_2x_3x_{10}x_{12} - x_4x_8x_{14}x_{15}, x_1x_4x_{10}x_{12} -x_3x_6x_7x_{13}, x_2x_4x_{10}x_{12} - x_3x_8x_{14}x_{15}, x_3x_4x_{10}x_{12} -x_2x_8x_{14}x_{15}, x_1x_6x_{10}x_{12} - x_3x_4x_7x_{13}, x_3x_6x_{10}x_{12} -x_1x_4x_7x_{13}, x_4x_6x_{10}x_{12} - x_1x_3x_7x_{13}, x_1x_7x_{10}x_{12} -x_3x_4x_6x_{13}, x_3x_7x_{10}x_{12} - x_1x_4x_6x_{13}, x_4x_7x_{10}x_{12} -x_1x_3x_6x_{13}, x_6x_7x_{10}x_{12} - x_1x_3x_4x_{13}, x_2x_8x_{10}x_{12} -x_3x_4x_{14}x_{15}, x_3x_8x_{10}x_{12} - x_2x_4x_{14}x_{15}, x_4x_8x_{10}x_{12} -x_2x_3x_{14}x_{15}, x_1x_9x_{10}x_{12} - x_{11}x_{13}x_{14}x_{15}, x_1x_3x_{11}x_{12} -x_5x_7x_8x_{14}, x_2x_3x_{11}x_{12} - x_5x_6x_{13}x_{15}, x_1x_5x_{11}x_{12} -x_3x_7x_8x_{14}, x_2x_5x_{11}x_{12} - x_3x_6x_{13}x_{15}, x_3x_5x_{11}x_{12} -x_2x_6x_{13}x_{15}, x_2x_6x_{11}x_{12} - x_3x_5x_{13}x_{15}, x_3x_6x_{11}x_{12} -x_2x_5x_{13}x_{15}, x_5x_6x_{11}x_{12} - x_2x_3x_{13}x_{15}, x_1x_7x_{11}x_{12} -x_3x_5x_8x_{14}, x_3x_7x_{11}x_{12} - x_1x_5x_8x_{14}, x_5x_7x_{11}x_{12} -x_
1x_3x_8x_{14}, x_1x_8x_{11}x_{12} - x_3x_5x_7x_{14}, x_3x_8x_{11}x_{12} -x_1x_5x_7x_{14}, x_5x_8x_{11}x_{12} - x_1x_3x_7x_{14}, x_7x_8x_{11}x_{12} -x_1x_3x_5x_{14}, x_1x_9x_{11}x_{12} - x_{10}x_{13}x_{14}x_{15}, x_1x_{10}x_{11}x_{12} -x_9x_{13}x_{14}x_{15}, x_9x_{10}x_{11}x_{12} - x_1x_{13}x_{14}x_{15}, x_2x_3x_5x_{13} -x_6x_{11}x_{12}x_{15}, x_2x_4x_5x_{13} - x_7x_9x_{12}x_{14}, x_1x_2x_6x_{13} -x_7x_8x_{14}x_{15}, x_2x_3x_6x_{13} - x_5x_{11}x_{12}x_{15}, x_2x_5x_6x_{13} -x_3x_{11}x_{12}x_{15}, x_3x_5x_6x_{13} - x_2x_{11}x_{12}x_{15}, x_4x_5x_6x_{13} -x_8x_{10}x_{11}x_{14}, x_1x_2x_7x_{13} - x_6x_8x_{14}x_{15}, x_2x_4x_7x_{13} -x_5x_9x_{12}x_{14}, x_2x_5x_7x_{13} - x_4x_9x_{12}x_{14}, x_3x_5x_7x_{13} -
x_8x_9x_{10}x_{15}, x_4x_5x_7x_{13} - x_2x_9x_{12}x_{14}, x_1x_6x_7x_{13} -
x_2x_8x_{14}x_{15}, x_2x_6x_7x_{13} - x_1x_8x_{14}x_{15}, x_1x_2x_8x_{13} -
x_6x_7x_{14}x_{15}, x_3x_5x_8x_{13} - x_7x_9x_{10}x_{15}, x_4x_5x_8x_{13} -
x_6x_{10}x_{11}x_{14}, x_1x_6x_8x_{13} - x_2x_7x_{14}x_{15}, x_2x_6x_8x_{13} -
x_1x_7x_{14}x_{15}, x_4x_6x_8x_{13} - x_5x_{10}x_{11}x_{14}, x_5x_6x_8x_{13} -
x_4x_{10}x_{11}x_{14}, x_1x_7x_8x_{13} - x_2x_6x_{14}x_{15}, x_2x_7x_8x_{13} -
x_1x_6x_{14}x_{15}, x_3x_7x_8x_{13} - x_5x_9x_{10}x_{15}, x_5x_7x_8x_{13} -
x_3x_9x_{10}x_{15}, x_6x_7x_8x_{13} - x_1x_2x_{14}x_{15}, x_2x_4x_9x_{13} -
x_5x_7x_{12}x_{14}, x_2x_5x_9x_{13} - x_4x_7x_{12}x_{14}, x_3x_5x_9x_{13} -
x_7x_8x_{10}x_{15}, x_4x_5x_9x_{13} - x_2x_7x_{12}x_{14}, x_2x_7x_9x_{13} -
x_4x_5x_{12}x_{14}, x_3x_7x_9x_{13} - x_5x_8x_{10}x_{15}, x_4x_7x_9x_{13} -
x_2x_5x_{12}x_{14}, x_5x_7x_9x_{13} - x_3x_8x_{10}x_{15}, x_3x_8x_9x_{13} -
x_5x_7x_{10}x_{15}, x_5x_8x_9x_{13} - x_3x_7x_{10}x_{15}, x_7x_8x_9x_{13} -
x_3x_5x_{10}x_{15}, x_3x_5x_{10}x_{13} - x_7x_8x_9x_{15}, x_4x_5x_{10}x_{13} -
x_6x_8x_{11}x_{14}, x_4x_6x_{10}x_{13} - x_5x_8x_{11}x_{14}, x_5x_6x_{10}x_{13} -x_4x_8x_{11}x_{14}, x_3x_7x_{10}x_{13} - x_5x_8x_9x_{15}, x_5x_7x_{10}x_{13} -x_3x_8x_9x_{15}, x_3x_8x_{10}x_{13} - x_5x_7x_9x_{15}, x_4x_8x_{10}x_{13} -x_5x_6x_{11}x_{14}, x_5x_8x_{10}x_{13} - x_3x_7x_9x_{15}, x_6x_8x_{10}x_{13} -x_4x_5x_{11}x_{14}, x_7x_8x_{10}x_{13} - x_3x_5x_9x_{15}, x_1x_9x_{10}x_{13} -x_{11}x_{12}x_{14}x_{15}, x_3x_9x_{10}x_{13} - x_5x_7x_8x_{15}, x_5x_9x_{10}x_{13} -x_3x_7x_8x_{15}, x_7x_9x_{10}x_{13} - x_3x_5x_8x_{15}, x_8x_9x_{10}x_{13} -x_3x_5x_7x_{15}, x_2x_3x_{11}x_{13} - x_5x_6x_{12}x_{15}, x_2x_5x_{11}x_{13} -x_3x_6x_{12}x_{15}, x_3x_5x_{11}x_{13} - x_2x_6x_{12}x_{15}, x_4x_5x_{11}x_{13} -x_6x_8x_{10}x_{14}, x_2x_6x_{11}x_{13} - x_3x_5x_{12}x_{15}, x_3x_6x_{11}x_{13} -x_2x_5x_{12}x_{15}, x_4x_6x_{11}x_{13} - x_5x_8x_{10}x_{14}, x_5x_6x_{11}x_{13} -x_2x_3x_{12}x_{15}, x_4x_8x_{11}x_{13} - x_5x_6x_{10}x_{14}, x_5x_8x_{11}x_{13} -x_4x_6x_{10}x_{14}, x_6x_8x_{11}x_{13} - x_4x_5x_{10}x_{14}, x_1x_9x_{11}
x_{13} -x_{10}x_{12}x_{14}x_{15}, x_1x_{10}x_{11}x_{13} - x_9x_{12}x_{14}x_{15}, x_4x_{10}x_{11}x_{13} -x_5x_6x_8x_{14}, x_5x_{10}x_{11}x_{13} - x_4x_6x_8x_{14}, x_6x_{10}x_{11}x_{13} -
x_4x_5x_8x_{14}, x_8x_{10}x_{11}x_{13} - x_4x_5x_6x_{14}, x_9x_{10}x_{11}x_{13} -x_1x_{12}x_{14}x_{15}, x_2x_3x_{12}x_{13} - x_5x_6x_{11}x_{15}, x_2x_4x_{12}x_{13} -x_5x_7x_9x_{14}, x_2x_5x_{12}x_{13} - x_3x_6x_{11}x_{15}, x_3x_5x_{12}x_{13} -x_2x_6x_{11}x_{15}, x_4x_5x_{12}x_{13} - x_2x_7x_9x_{14}, x_2x_6x_{12}x_{13} -x_3x_5x_{11}x_{15}, x_3x_6x_{12}x_{13} - x_2x_5x_{11}x_{15}, x_5x_6x_{12}x_{13} -x_2x_3x_{11}x_{15}, x_2x_7x_{12}x_{13} - x_4x_5x_9x_{14}, x_4x_7x_{12}x_{13} -x_2x_5x_9x_{14}, x_5x_7x_{12}x_{13} - x_2x_4x_9x_{14}, x_1x_9x_{12}x_{13} -x_{10}x_{11}x_{14}x_{15}, x_2x_9x_{12}x_{13} - x_4x_5x_7x_{14}, x_4x_9x_{12}x_{13} -x_2x_5x_7x_{14}, x_5x_9x_{12}x_{13} - x_2x_4x_7x_{14}, x_7x_9x_{12}x_{13} -x_2x_4x_5x_{14}, x_1x_{10}x_{12}x_{13} - x_9x_{11}x_{14}x_{15}, x_9x_{10}x_{12}x_{13} -x_1x_{11}x_{14}x_{15}, x_1x_{11}x_{12}x_{13} - x_9x_{10}x_{14}x_{15}, x_2x_{11}x_{12}x_{13} -x_3x_5x_6x_{15}, x_3x_{11}x_{12}x_{13} - x_2x_5x_6x_{15}, x_5x_{11}x_{12}x_{13} -x_2x_3x_6x_{15}, x_6x_{11}x_{12}x_{13} - x_2x_3x_
5x_{15}, x_9x_{11}x_{12}x_{13} -x_1x_{10}x_{14}x_{15}, x_{10}x_{11}x_{12}x_{13} - x_1x_9x_{14}x_{15}, x_2x_3x_4x_{14} -
x_8x_{10}x_{12}x_{15}, x_1x_2x_6x_{14} - x_7x_8x_{13}x_{15}, x_3x_4x_6x_{14} -
x_7x_9x_{11}x_{15}, x_1x_2x_7x_{14} - x_6x_8x_{13}x_{15}, x_3x_4x_7x_{14} -
x_6x_9x_{11}x_{15}, x_1x_6x_7x_{14} - x_2x_8x_{13}x_{15}, x_2x_6x_7x_{14} -
x_1x_8x_{13}x_{15}, x_3x_6x_7x_{14} - x_4x_9x_{11}x_{15}, x_4x_6x_7x_{14} -
x_3x_9x_{11}x_{15}, x_1x_2x_8x_{14} - x_6x_7x_{13}x_{15}, x_2x_3x_8x_{14} -
x_4x_{10}x_{12}x_{15}, x_2x_4x_8x_{14} - x_3x_{10}x_{12}x_{15}, x_3x_4x_8x_{14} -x_2x_{10}x_{12}x_{15}, x_1x_6x_8x_{14} - x_2x_7x_{13}x_{15}, x_2x_6x_8x_{14} -x_1x_7x_{13}x_{15}, x_1x_7x_8x_{14} - x_2x_6x_{13}x_{15}, x_2x_7x_8x_{14} -x_1x_6x_{13}x_{15}, x_6x_7x_8x_{14} - x_1x_2x_{13}x_{15}, x_3x_4x_9x_{14} -x_6x_7x_{11}x_{15}, x_3x_6x_9x_{14} - x_4x_7x_{11}x_{15}, x_4x_6x_9x_{14} -x_3x_7x_{11}x_{15}, x_3x_7x_9x_{14} - x_4x_6x_{11}x_{15}, x_4x_7x_9x_{14} -x_3x_6x_{11}x_{15}, x_6x_7x_9x_{14} - x_3x_4x_{11}x_{15}, x_2x_3x_{10}x_{14} -x_4x_8x_{12}x_{15}, x_2x_4x_{10}x_{14} - x_3x_8x_{12}x_{15}, x_3x_4x_{10}x_{14} -x_2x_8x_{12}x_{15}, x_2x_8x_{10}x_{14} - x_3x_4x_{12}x_{15}, x_3x_8x_{10}x_{14} -x_2x_4x_{12}x_{15}, x_4x_8x_{10}x_{14} - x_2x_3x_{12}x_{15}, x_1x_9x_{10}x_{14} -x_{11}x_{12}x_{13}x_{15}, x_3x_4x_{11}x_{14} - x_6x_7x_9x_{15}, x_3x_6x_{11}x_{14} -x_4x_7x_9x_{15}, x_4x_6x_{11}x_{14} - x_3x_7x_9x_{15}, x_3x_7x_{11}x_{14} -x_4x_6x_9x_{15}, x_4x_7x_{11}x_{14} - x_3x_6x_9x_{15}, x_6x_7x_{11}x_{14} -x_3x_4x_
9x_{15}, x_1x_9x_{11}x_{14} - x_{10}x_{12}x_{13}x_{15}, x_3x_9x_{11}x_{14} -x_4x_6x_7x_{15}, x_4x_9x_{11}x_{14} - x_3x_6x_7x_{15}, x_6x_9x_{11}x_{14} -x_3x_4x_7x_{15}, x_7x_9x_{11}x_{14} - x_3x_4x_6x_{15}, x_1x_{10}x_{11}x_{14} -x_9x_{12}x_{13}x_{15}, x_9x_{10}x_{11}x_{14} - x_1x_{12}x_{13}x_{15}, x_2x_3x_{12}x_{14} -x_4x_8x_{10}x_{15}, x_2x_4x_{12}x_{14} - x_3x_8x_{10}x_{15}, x_3x_4x_{12}x_{14} -x_2x_8x_{10}x_{15}, x_2x_8x_{12}x_{14} - x_3x_4x_{10}x_{15}, x_3x_8x_{12}x_{14} -
x_2x_4x_{10}x_{15}, x_4x_8x_{12}x_{14} - x_2x_3x_{10}x_{15}, x_1x_9x_{12}x_{14} -x_{10}x_{11}x_{13}x_{15}, x_1x_{10}x_{12}x_{14} - x_9x_{11}x_{13}x_{15}, x_2x_{10}x_{12}x_{14} -x_3x_4x_8x_{15}, x_3x_{10}x_{12}x_{14} - x_2x_4x_8x_{15}, x_4x_{10}x_{12}x_{14} -x_2x_3x_8x_{15}, x_8x_{10}x_{12}x_{14} - x_2x_3x_4x_{15}, x_9x_{10}x_{12}x_{14} -x_1x_{11}x_{13}x_{15}, x_1x_{11}x_{12}x_{14} - x_9x_{10}x_{13}x_{15}, x_9x_{11}x_{12}x_{14} -x_1x_{10}x_{13}x_{15}, x_{10}x_{11}x_{12}x_{14} - x_1x_9x_{13}x_{15}, x_1x_2x_{13}x_{14} -x_6x_7x_8x_{15}, x_1x_6x_{13}x_{14} - x_2x_7x_8x_{15}, x_2x_6x_{13}x_{14} -
x_1x_7x_8x_{15}, x_1x_7x_{13}x_{14} - x_2x_6x_8x_{15}, x_2x_7x_{13}x_{14} -
x_1x_6x_8x_{15}, x_6x_7x_{13}x_{14} - x_1x_2x_8x_{15}, x_1x_8x_{13}x_{14} -
x_2x_6x_7x_{15}, x_2x_8x_{13}x_{14} - x_1x_6x_7x_{15}, x_6x_8x_{13}x_{14} -
x_1x_2x_7x_{15}, x_7x_8x_{13}x_{14} - x_1x_2x_6x_{15}, x_1x_9x_{13}x_{14} -
x_{10}x_{11}x_{12}x_{15}, x_1x_{10}x_{13}x_{14} - x_9x_{11}x_{12}x_{15}, x_9x_{10}x_{13}x_{14} -x_1x_{11}x_{12}x_{15}, x_1x_{11}x_{13}x_{14} - x_9x_{10}x_{12}x_{15}, x_9x_{11}x_{13}x_{14} -x_1x_{10}x_{12}x_{15}, x_{10}x_{11}x_{13}x_{14} - x_1x_9x_{12}x_{15}, x_1x_{12}x_{13}x_{14} -x_9x_{10}x_{11}x_{15}, x_9x_{12}x_{13}x_{14} - x_1x_{10}x_{11}x_{15}, x_{10}x_{12}x_{13}x_{14} -x_1x_9x_{11}x_{15}, x_{11}x_{12}x_{13}x_{14} - x_1x_9x_{10}x_{15}, x_1^2 - 1, x_2^2 - 1, x_3^2- 1, x_4^2 - 1, x_5^2 - 1, x_6^2 - 1, x_7^2 - 1, x_8^2 - 1, x_9^2 - 1, x_{10}^2 -1,x_{11}^2 - 1, x_{12}^2 - 1, x_{13}^2 - 1, x_{14}^2 - 1, x_{15}^2 - 1\rangle$.

\subsection{Error Correcting capability of the Schubert Code $\bf C_{(1,5)}(2,5)(\FF_2)$: }
The minimum of the total degree of each of the above binomial except binomials of the form $x_i^2-1$, for $ i=1,2,\ldots,15$ is $4$ and hence by, theorem $4$ above\cite{Edgar}, we get $t=3$, which is in accordance with our known results(\cite{Edgar, MS}), because for this Schubert code, we know that $d=8$, and hence $t=3$.

\subsection{Decoding with $[15,4,8]$-linear Schubert Code $\bf C_{(1,5)}(2,5)(\FF_2)$:}
By using theorem 3 above from \cite{Edgar}, we can tackle the decoding problem as follows:
\begin{center}
\begin{tabular}{|c|c|c|c|}
\hline
Received word & Canonical form of & Received word is decoded to\\
& this word w.r.t. $\mathcal{G}_{(1,5)}(2,5)$ &   \\
\hline
$x_7x_8x_9x_{10}x_{11}x_{12}x_{13}x_{14}x_{15}$			&	$x_1x_7x_8$		&	$x_1x_9x_{10}x_{11}x_{12}x_{13}x_{14}x_{15}$\\
\hline
$x_9x_{10}x_{11}x_{12}x_{13}x_{14}x_{15}$				&	$x_1$			&	$x_1x_9x_{10}x_{11}x_{12}x_{13}x_{14}x_{15}$\\
\hline
$x_3x_9x_{10}x_{11}x_{12}x_{13}x_{14}x_{15}$			&	$x_1x_3$	&	$x_1x_9x_{10}x_{11}x_{12}x_{13}x_{14}x_{15}$\\
\hline
$x_2x_9x_{10}x_{11}x_{12}x_{13}x_{14}x_{15}$			&	$x_1x_2$	&	$x_1x_9x_{10}x_{11}x_{12}x_{13}x_{14}x_{15}$\\
\hline
$x_4x_9x_{10}x_{11}x_{12}x_{13}x_{14}x_{15}$			&	$x_1x_4$	&	$x_1x_9x_{10}x_{11}x_{12}x_{13}x_{14}x_{15}$\\
\hline
$x_5x_9x_{10}x_{11}x_{12}x_{13}x_{14}x_{15}$			&	$x_1x_5$	&	$x_1x_9x_{10}x_{11}x_{12}x_{13}x_{14}x_{15}$\\
\hline
$x_{10}x_{11}x_{12}x_{13}x_{14}x_{15}$					&	$x_1x_9$	&	$x_1x_9x_{10}x_{11}x_{12}x_{13}x_{14}x_{15}$\\
\hline
\end{tabular}
\end{center}
For decoding of remaining tyes of received words is concerned, one can use minimum distance deccoding technique\cite{Xing}.

\subsection{Gr\"obner Basis for the Schubert Code  $\bf C_{(2,3)}(2,5)(\FF_2)$:}

Generator matrix for the Schubert code  $C_{(2,3)}(2,5)(\FF_2)$ is denoted by $A_{(2,3)}$ which is given as follows:\\
$$A_{(2,3)}=\begin{bmatrix}
   1 & 0 & 0 & 0 & 1 & 1 & 1 \\
   0 & 1 & 0 & 1 & 1 & 0 & 1 \\
   0 & 0 & 1 & 1 & 0 & 1 & 1 \\
  \end{bmatrix}_{3\times 7}
$$ \\
By using techniques in\cite{Edgar}, we can write an ideal $I(C_{(2,3)}(2,5)(\FF_2))$ associated with code $C_{(2,3)}(2,5)(\FF_2)$,
therefore Gr\"obner basis of an ideal associated with this code, which is denoted by $\mathcal{G}_{(2,3)}(2,5)$  is given by:\\
$
\langle x_1^2 - 1, x_1x_2 - x_3x_7, x_2^2 - 1, x_1x_3 - x_2x_7, x_2x_3 - x_1x_7, x_3^2 -
1, x_1x_4 - x_2x_6, x_2x_4 - x_5x_7, x_3x_4 - x_6x_7, x_4^2 - 1, x_1x_5 - x_6x_7,
x_2x_5 - x_4x_7, x_3x_5 - x_2x_6, x_4x_5 - x_2x_7, x_5^2 - 1, x_1x_6 - x_5x_7,
x_3x_6-x_4x_7,x_4x_6-x_3x_7,x_5x_6-x_1x_7,x_6^2 - 1,x_7^2 - 1 \rangle$.

\subsection{Error Correcting capability of the Schubert Code $\bf C_{(2,3)}(2,5)(\FF_2)$:}
The minimum of the total degree of each of the above binomial except binomials of the form $x_i^2-1$, for  $ i=1,2,\ldots,7$ is $2$ and hence by, theorem $4$ above, we get $t=1$, which is in accordance with our known results(\cite{Edgar, MS}), because for this Schubert code, we know that $d=4$, and hence $t=1$.

\subsection{Decoding with $[7,3,4]$-linear Schubert Code $\bf C_{(2,3)}(2,5)(\FF_2)$:}
By using theorem $3$ above from \cite{Edgar}, we can tackle the decoding problem as follows:
\begin{center}
\begin{tabular}{|c|c|c|}
\hline
Received word & Canonical form of & Received word is decoded to
\\ & this word w.r.t. $\mathcal{G}_{(2,3)}(2,5)$ &  \\
\hline
$x_1x_2x_3x_4x_5$	&	$x_2$		&		$x_1x_3x_4x_5$\\
\hline
$x_1x_2x_3$			&	$x_7$		&		$x_1x_2x_3x_7$\\
\hline
$x_1x_2x_7$			&	$x_3$		&		$x_1x_2x_3x_7$\\
\hline
$x_2x_5x_7$			&	$x_4$		&		$x_2x_4x_5x_7$\\
\hline
$x_2x_5x_6$			&	$x_3$		&		$x_2x_3x_5x_6$\\
\hline
$x_3x_6x_7$			&	$x_4$		&		$x_3x_4x_6x_7$\\
\hline
$x_5x_6x_7$			&	$x_1$		&		$x_1x_5x_6x_7$\\
\hline
\end{tabular}
\end{center}
Of course, for decoding of remaining types of received words, one can use Minimum distance decoding rule(Completer/Incomplete)\cite{Edgar} or probabilistic approach i.e. maximum likelihood decoding rule\cite{Edgar}.

\subsection{Gr\"obner Basis for the Schubert Code  $\bf C_{(2,4)}(2,5)(\FF_2)$:}
For this code generator matrix is denoted by $A_{(2,4)}$ which is given below:$$A_{(2,4)}=
\left[{\begin{array}{ccccccccccccccccccc}
1 & 0 & 0 & 0 & 0 & 0 & 0 & 0 & 0 & 0 & 1 & 1 & 1 & 1 & 1 & 1 & 1 & 1 & 1 \\
0 & 1 & 0 & 0 & 0 & 0 & 0 & 1 & 1 & 1 & 1 & 1 & 1 & 1 & 0 & 0 & 0 & 0 & 0 \\
0 & 0 & 1 & 0 & 0 & 0 & 1 & 0 & 1 & 1 & 1 & 1 & 0 & 0 & 1 & 0 & 0 & 0 & 1 \\
0 & 0 & 0 & 1 & 0 & 1 & 0 & 1 & 0 & 1 & 0 & 1 & 1 & 0 & 0 & 1 & 0 & 1 & 0 \\
0 & 0 & 0 & 0 & 1 & 1 & 1 & 0 & 0 & 1 & 0 & 1 & 0 & 0 & 1 & 1 & 1 & 0 & 0 
\end{array}}\right]_{5\times19}$$
Hence, Gr\"obner basis of an ideal associated with this code, which is denoted by $\mathcal{G}_{(2,4)}(2,5)$ and is given by:\\
$
\langle x_{2} x_{3} x_{6} x_{10} x_{12} x_{14} -  x_{4} x_{7} x_{15} x_{16} x_{18} x_{19}, x_{2} x_{4} x_{7} x_{10} x_{12} x_{14} -  x_{3} x_{6} x_{15} x_{16} x_{18} x_{19}, x_{2} x_{4} x_{10} x_{12} x_{14} x_{15} -  x_{3} x_{6} x_{7} x_{16} x_{18} x_{19}, x_{2} x_{5} x_{8} x_{10} x_{14} x_{16} -  x_{3} x_{6} x_{12} x_{13} x_{17} x_{19}, x_{2} x_{3} x_{10} x_{12} x_{14} x_{16} -  x_{4} x_{6} x_{7} x_{15} x_{18} x_{19}, x_{5} x_{8} x_{9} x_{13} x_{14} x_{16} -  x_{2} x_{6} x_{7} x_{11} x_{15} x_{17}, x_{2} x_{5} x_{12} x_{13} x_{14} x_{16} -  x_{3} x_{6} x_{8} x_{10} x_{17} x_{19}, x_{2} x_{3} x_{4} x_{10} x_{15} x_{16} -  x_{6} x_{7} x_{12} x_{14} x_{18} x_{19}, x_{3} x_{4} x_{8} x_{11} x_{15} x_{16} -  x_{6} x_{7} x_{9} x_{13} x_{18} x_{19}, x_{3} x_{4} x_{9} x_{13} x_{15} x_{16} -  x_{6} x_{7} x_{8} x_{11} x_{18} x_{19}, x_{3} x_{4} x_{12} x_{14} x_{15} x_{16} -  x_{2} x_{6} x_{7} x_{10} x_{18} x_{19}, x_{2} x_{5} x_{7} x_{10} x_{11} x_{17} -  x_{4} x_{9} x_{12} x_{14} x_{15} x_{18}, x_{2} x_{3} x_{6} x_{
10} x_{12} x_{17} -  x_{5} x_{8} x_{13} x_{14} x_{16} x_{19}, x_{2} x_{4} x_{7} x_{10} x_{12} x_{17} -  x_{5} x_{9} x_{11} x_{14} x_{15} x_{18}, x_{3} x_{5} x_{8} x_{10} x_{12} x_{17} -  x_{4} x_{9} x_{11} x_{13} x_{18} x_{19}, x_{4} x_{5} x_{9} x_{10} x_{12} x_{17} -  x_{3} x_{8} x_{11} x_{13} x_{18} x_{19}, x_{4} x_{5} x_{10} x_{11} x_{12} x_{17} -  x_{3} x_{8} x_{9} x_{13} x_{18} x_{19}, x_{2} x_{5} x_{6} x_{10} x_{13} x_{17} -  x_{3} x_{8} x_{12} x_{14} x_{16} x_{19}, x_{3} x_{5} x_{10} x_{12} x_{13} x_{17} -  x_{4} x_{8} x_{9} x_{11} x_{18} x_{19}, x_{2} x_{7} x_{10} x_{11} x_{14} x_{17} -  x_{4} x_{5} x_{9} x_{12} x_{15} x_{18}, x_{2} x_{6} x_{8} x_{12} x_{14} x_{17} -  x_{3} x_{5} x_{10} x_{13} x_{16} x_{19}, x_{5} x_{6} x_{8} x_{12} x_{14} x_{17} -  x_{2} x_{3} x_{10} x_{13} x_{16} x_{19}, x_{2} x_{7} x_{9} x_{12} x_{14} x_{17} -  x_{4} x_{5} x_{10} x_{11} x_{15} x_{18}, x_{5} x_{7} x_{9} x_{12} x_{14} x_{17} -  x_{2} x_{4} x_{10} x_{11} x_{15} x_{18}, x_{2} x_{6} x_{10} x_{12} x_{14} x_{17} -  x_{3} 
x_{5} x_{8} x_{13} x_{16} x_{19}, x_{3} x_{6} x_{10} x_{12} x_{14} x_{17} -  x_{2} x_{5} x_{8} x_{13} x_{16} x_{19}, x_{2} x_{7} x_{10} x_{12} x_{14} x_{17} -  x_{4} x_{5} x_{9} x_{11} x_{15} x_{18}, x_{4} x_{7} x_{10} x_{12} x_{14} x_{17} -  x_{2} x_{5} x_{9} x_{11} x_{15} x_{18}, x_{3} x_{8} x_{10} x_{12} x_{14} x_{17} -  x_{2} x_{5} x_{6} x_{13} x_{16} x_{19}, x_{5} x_{8} x_{10} x_{12} x_{14} x_{17} -  x_{2} x_{3} x_{6} x_{13} x_{16} x_{19}, x_{4} x_{9} x_{10} x_{12} x_{14} x_{17} -  x_{2} x_{5} x_{7} x_{11} x_{15} x_{18}, x_{5} x_{9} x_{10} x_{12} x_{14} x_{17} -  x_{2} x_{4} x_{7} x_{11} x_{15} x_{18}, x_{5} x_{10} x_{11} x_{12} x_{14} x_{17} -  x_{2} x_{4} x_{7} x_{9} x_{15} x_{18}, x_{2} x_{6} x_{10} x_{13} x_{14} x_{17} -  x_{3} x_{5} x_{8} x_{12} x_{16} x_{19}, x_{5} x_{10} x_{12} x_{13} x_{14} x_{17} -  x_{2} x_{3} x_{6} x_{8} x_{16} x_{19}, x_{2} x_{10} x_{12} x_{14} x_{15} x_{17} -  x_{4} x_{5} x_{7} x_{9} x_{11} x_{18}, x_{2} x_{5} x_{12} x_{13} x_{16} x_{17} -  x_{3} x_{6} x_{8} x_{10} x_{14} x_
{19}, x_{5} x_{8} x_{10} x_{14} x_{16} x_{17} -  x_{2} x_{3} x_{6} x_{12} x_{13} x_{19}, x_{2} x_{10} x_{12} x_{14} x_{16} x_{17} -  x_{3} x_{5} x_{6} x_{8} x_{13} x_{19}, x_{3} x_{4} x_{6} x_{8} x_{11} x_{18} -  x_{7} x_{9} x_{13} x_{15} x_{16} x_{19}, x_{3} x_{4} x_{7} x_{8} x_{11} x_{18} -  x_{6} x_{9} x_{13} x_{15} x_{16} x_{19}, x_{3} x_{6} x_{7} x_{8} x_{11} x_{18} -  x_{4} x_{9} x_{13} x_{15} x_{16} x_{19}, x_{4} x_{6} x_{7} x_{8} x_{11} x_{18} -  x_{3} x_{9} x_{13} x_{15} x_{16} x_{19}, x_{3} x_{4} x_{8} x_{9} x_{11} x_{18} -  x_{5} x_{10} x_{12} x_{13} x_{17} x_{19}, x_{4} x_{5} x_{8} x_{9} x_{11} x_{18} -  x_{3} x_{10} x_{12} x_{13} x_{17} x_{19}, x_{4} x_{7} x_{8} x_{9} x_{11} x_{18} -  x_{3} x_{6} x_{13} x_{15} x_{16} x_{19}, x_{3} x_{4} x_{8} x_{10} x_{11} x_{18} -  x_{5} x_{9} x_{12} x_{13} x_{17} x_{19}, x_{4} x_{5} x_{8} x_{10} x_{11} x_{18} -  x_{3} x_{9} x_{12} x_{13} x_{17} x_{19}, x_{4} x_{5} x_{9} x_{10} x_{11} x_{18} -  x_{3} x_{8} x_{12} x_{13} x_{17} x_{19}, x_{2} x_{3} x_{6} x_{10} x_
{12} x_{18} -  x_{4} x_{7} x_{14} x_{15} x_{16} x_{19}, x_{2} x_{4} x_{7} x_{10} x_{12} x_{18} -  x_{3} x_{6} x_{14} x_{15} x_{16} x_{19}, x_{3} x_{5} x_{8} x_{10} x_{12} x_{18} -  x_{4} x_{9} x_{11} x_{13} x_{17} x_{19}, x_{4} x_{5} x_{9} x_{10} x_{12} x_{18} -  x_{3} x_{8} x_{11} x_{13} x_{17} x_{19}, x_{3} x_{4} x_{8} x_{11} x_{12} x_{18} -  x_{5} x_{9} x_{10} x_{13} x_{17} x_{19}, x_{3} x_{5} x_{8} x_{11} x_{12} x_{18} -  x_{4} x_{9} x_{10} x_{13} x_{17} x_{19}, x_{4} x_{5} x_{9} x_{11} x_{12} x_{18} -  x_{3} x_{8} x_{10} x_{13} x_{17} x_{19}, x_{4} x_{5} x_{10} x_{11} x_{12} x_{18} -  x_{3} x_{8} x_{9} x_{13} x_{17} x_{19}, x_{3} x_{4} x_{6} x_{9} x_{13} x_{18} -  x_{7} x_{8} x_{11} x_{15} x_{16} x_{19}, x_{3} x_{4} x_{7} x_{9} x_{13} x_{18} -  x_{6} x_{8} x_{11} x_{15} x_{16} x_{19}, x_{3} x_{6} x_{7} x_{9} x_{13} x_{18} -  x_{4} x_{8} x_{11} x_{15} x_{16} x_{19}, x_{4} x_{6} x_{7} x_{9} x_{13} x_{18} -  x_{3} x_{8} x_{11} x_{15} x_{16} x_{19}, x_{3} x_{4} x_{8} x_{9} x_{13} x_{18} -  x_{5} x_{10} x_{
11} x_{12} x_{17} x_{19}, x_{3} x_{5} x_{8} x_{9} x_{13} x_{18} -  x_{4} x_{10} x_{11} x_{12} x_{17} x_{19}, x_{3} x_{6} x_{8} x_{9} x_{13} x_{18} -  x_{4} x_{7} x_{11} x_{15} x_{16} x_{19}, x_{3} x_{5} x_{8} x_{10} x_{13} x_{18} -  x_{4} x_{9} x_{11} x_{12} x_{17} x_{19}, x_{3} x_{4} x_{9} x_{10} x_{13} x_{18} -  x_{5} x_{8} x_{11} x_{12} x_{17} x_{19}, x_{3} x_{5} x_{9} x_{10} x_{13} x_{18} -  x_{4} x_{8} x_{11} x_{12} x_{17} x_{19}, x_{3} x_{4} x_{8} x_{11} x_{13} x_{18} -  x_{5} x_{9} x_{10} x_{12} x_{17} x_{19}, x_{3} x_{5} x_{8} x_{11} x_{13} x_{18} -  x_{4} x_{9} x_{10} x_{12} x_{17} x_{19}, x_{3} x_{6} x_{8} x_{11} x_{13} x_{18} -  x_{4} x_{7} x_{9} x_{15} x_{16} x_{19}, x_{3} x_{4} x_{9} x_{11} x_{13} x_{18} -  x_{5} x_{8} x_{10} x_{12} x_{17} x_{19}, x_{4} x_{5} x_{9} x_{11} x_{13} x_{18} -  x_{3} x_{8} x_{10} x_{12} x_{17} x_{19}, x_{4} x_{7} x_{9} x_{11} x_{13} x_{18} -  x_{3} x_{6} x_{8} x_{15} x_{16} x_{19}, x_{3} x_{5} x_{8} x_{12} x_{13} x_{18} -  x_{4} x_{9} x_{10} x_{11} x_{17} x_{19}, x_{3}
 x_{4} x_{9} x_{12} x_{13} x_{18} -  x_{5} x_{8} x_{10} x_{11} x_{17} x_{19}, x_{4} x_{5} x_{9} x_{12} x_{13} x_{18} -  x_{3} x_{8} x_{10} x_{11} x_{17} x_{19}, x_{3} x_{5} x_{10} x_{12} x_{13} x_{18} -  x_{4} x_{8} x_{9} x_{11} x_{17} x_{19}, x_{2} x_{6} x_{10} x_{12} x_{14} x_{18} -  x_{3} x_{4} x_{7} x_{15} x_{16} x_{19}, x_{3} x_{6} x_{10} x_{12} x_{14} x_{18} -  x_{2} x_{4} x_{7} x_{15} x_{16} x_{19}, x_{2} x_{7} x_{10} x_{12} x_{14} x_{18} -  x_{3} x_{4} x_{6} x_{15} x_{16} x_{19}, x_{4} x_{7} x_{10} x_{12} x_{14} x_{18} -  x_{2} x_{3} x_{6} x_{15} x_{16} x_{19}, x_{2} x_{4} x_{6} x_{7} x_{15} x_{18} -  x_{3} x_{10} x_{12} x_{14} x_{16} x_{19}, x_{3} x_{4} x_{6} x_{7} x_{15} x_{18} -  x_{2} x_{10} x_{12} x_{14} x_{16} x_{19}, x_{3} x_{4} x_{6} x_{8} x_{15} x_{18} -  x_{7} x_{9} x_{11} x_{13} x_{16} x_{19}, x_{3} x_{4} x_{6} x_{9} x_{15} x_{18} -  x_{7} x_{8} x_{11} x_{13} x_{16} x_{19}, x_{4} x_{6} x_{7} x_{9} x_{15} x_{18} -  x_{3} x_{8} x_{11} x_{13} x_{16} x_{19}, x_{3} x_{6} x_{8} x_{9} x_{15} x_{
18} -  x_{4} x_{7} x_{11} x_{13} x_{16} x_{19}, x_{4} x_{6} x_{8} x_{9} x_{15} x_{18} -  x_{3} x_{7} x_{11} x_{13} x_{16} x_{19}, x_{2} x_{4} x_{6} x_{10} x_{15} x_{18} -  x_{3} x_{7} x_{12} x_{14} x_{16} x_{19}, x_{3} x_{4} x_{6} x_{10} x_{15} x_{18} -  x_{2} x_{7} x_{12} x_{14} x_{16} x_{19}, x_{2} x_{4} x_{7} x_{10} x_{15} x_{18} -  x_{3} x_{6} x_{12} x_{14} x_{16} x_{19}, x_{3} x_{4} x_{6} x_{11} x_{15} x_{18} -  x_{7} x_{8} x_{9} x_{13} x_{16} x_{19}, x_{4} x_{6} x_{7} x_{11} x_{15} x_{18} -  x_{3} x_{8} x_{9} x_{13} x_{16} x_{19}, x_{3} x_{4} x_{8} x_{11} x_{15} x_{18} -  x_{6} x_{7} x_{9} x_{13} x_{16} x_{19}, x_{6} x_{7} x_{8} x_{11} x_{15} x_{18} -  x_{3} x_{4} x_{9} x_{13} x_{16} x_{19}, x_{4} x_{8} x_{9} x_{11} x_{15} x_{18} -  x_{3} x_{6} x_{7} x_{13} x_{16} x_{19}, x_{6} x_{8} x_{9} x_{11} x_{15} x_{18} -  x_{3} x_{4} x_{7} x_{13} x_{16} x_{19}, x_{2} x_{3} x_{6} x_{12} x_{15} x_{18} -  x_{4} x_{7} x_{10} x_{14} x_{16} x_{19}, x_{3} x_{4} x_{6} x_{12} x_{15} x_{18} -  x_{2} x_{7} x_{10} x_{14} x_
{16} x_{19}, x_{2} x_{4} x_{7} x_{12} x_{15} x_{18} -  x_{3} x_{6} x_{10} x_{14} x_{16} x_{19}, x_{2} x_{4} x_{10} x_{12} x_{15} x_{18} -  x_{3} x_{6} x_{7} x_{14} x_{16} x_{19}, x_{3} x_{4} x_{6} x_{13} x_{15} x_{18} -  x_{7} x_{8} x_{9} x_{11} x_{16} x_{19}, x_{3} x_{4} x_{9} x_{13} x_{15} x_{18} -  x_{6} x_{7} x_{8} x_{11} x_{16} x_{19}, x_{6} x_{7} x_{9} x_{13} x_{15} x_{18} -  x_{3} x_{4} x_{8} x_{11} x_{16} x_{19}, x_{6} x_{8} x_{9} x_{13} x_{15} x_{18} -  x_{3} x_{4} x_{7} x_{11} x_{16} x_{19}, x_{3} x_{6} x_{11} x_{13} x_{15} x_{18} -  x_{4} x_{7} x_{8} x_{9} x_{16} x_{19}, x_{4} x_{6} x_{11} x_{13} x_{15} x_{18} -  x_{3} x_{7} x_{8} x_{9} x_{16} x_{19}, x_{6} x_{8} x_{11} x_{13} x_{15} x_{18} -  x_{3} x_{4} x_{7} x_{9} x_{16} x_{19}, x_{4} x_{9} x_{11} x_{13} x_{15} x_{18} -  x_{3} x_{6} x_{7} x_{8} x_{16} x_{19}, x_{6} x_{9} x_{11} x_{13} x_{15} x_{18} -  x_{3} x_{4} x_{7} x_{8} x_{16} x_{19}, x_{4} x_{6} x_{7} x_{14} x_{15} x_{18} -  x_{2} x_{3} x_{10} x_{12} x_{16} x_{19}, x_{2} x_{6} x_{10} x_{
14} x_{15} x_{18} -  x_{3} x_{4} x_{7} x_{12} x_{16} x_{19}, x_{3} x_{6} x_{10} x_{14} x_{15} x_{18} -  x_{2} x_{4} x_{7} x_{12} x_{16} x_{19}, x_{4} x_{7} x_{10} x_{14} x_{15} x_{18} -  x_{2} x_{3} x_{6} x_{12} x_{16} x_{19}, x_{2} x_{6} x_{12} x_{14} x_{15} x_{18} -  x_{3} x_{4} x_{7} x_{10} x_{16} x_{19}, x_{4} x_{6} x_{12} x_{14} x_{15} x_{18} -  x_{2} x_{3} x_{7} x_{10} x_{16} x_{19}, x_{4} x_{7} x_{12} x_{14} x_{15} x_{18} -  x_{2} x_{3} x_{6} x_{10} x_{16} x_{19}, x_{2} x_{10} x_{12} x_{14} x_{15} x_{18} -  x_{3} x_{4} x_{6} x_{7} x_{16} x_{19}, x_{4} x_{10} x_{12} x_{14} x_{15} x_{18} -  x_{2} x_{3} x_{6} x_{7} x_{16} x_{19}, x_{2} x_{3} x_{6} x_{7} x_{16} x_{18} -  x_{4} x_{10} x_{12} x_{14} x_{15} x_{19}, x_{3} x_{4} x_{6} x_{7} x_{16} x_{18} -  x_{2} x_{10} x_{12} x_{14} x_{15} x_{19}, x_{3} x_{4} x_{7} x_{8} x_{16} x_{18} -  x_{6} x_{9} x_{11} x_{13} x_{15} x_{19}, x_{3} x_{6} x_{7} x_{8} x_{16} x_{18} -  x_{4} x_{9} x_{11} x_{13} x_{15} x_{19}, x_{3} x_{4} x_{7} x_{9} x_{16} x_{18} -  x_{6} x_{8}
 x_{11} x_{13} x_{15} x_{19}, x_{3} x_{4} x_{8} x_{9} x_{16} x_{18} -  x_{6} x_{7} x_{11} x_{13} x_{15} x_{19}, x_{3} x_{7} x_{8} x_{9} x_{16} x_{18} -  x_{4} x_{6} x_{11} x_{13} x_{15} x_{19}, x_{4} x_{7} x_{8} x_{9} x_{16} x_{18} -  x_{3} x_{6} x_{11} x_{13} x_{15} x_{19}, x_{2} x_{3} x_{6} x_{10} x_{16} x_{18} -  x_{4} x_{7} x_{12} x_{14} x_{15} x_{19}, x_{2} x_{3} x_{7} x_{10} x_{16} x_{18} -  x_{4} x_{6} x_{12} x_{14} x_{15} x_{19}, x_{3} x_{4} x_{7} x_{10} x_{16} x_{18} -  x_{2} x_{6} x_{12} x_{14} x_{15} x_{19}, x_{3} x_{4} x_{7} x_{11} x_{16} x_{18} -  x_{6} x_{8} x_{9} x_{13} x_{15} x_{19}, x_{3} x_{4} x_{8} x_{11} x_{16} x_{18} -  x_{6} x_{7} x_{9} x_{13} x_{15} x_{19}, x_{6} x_{7} x_{8} x_{11} x_{16} x_{18} -  x_{3} x_{4} x_{9} x_{13} x_{15} x_{19}, x_{7} x_{8} x_{9} x_{11} x_{16} x_{18} -  x_{3} x_{4} x_{6} x_{13} x_{15} x_{19}, x_{2} x_{3} x_{6} x_{12} x_{16} x_{18} -  x_{4} x_{7} x_{10} x_{14} x_{15} x_{19}, x_{2} x_{4} x_{7} x_{12} x_{16} x_{18} -  x_{3} x_{6} x_{10} x_{14} x_{15} x_{19}, x_{3}
 x_{4} x_{7}x_{12} x_{16} x_{18} -  x_{2} x_{6} x_{10} x_{14} x_{15} x_{19}, x_{2} x_{3} x_{10} x_{12} x_{16} x_{18} -  x_{4} x_{6} x_{7} x_{14} x_{15} x_{19}, x_{3} x_{4} x_{7} x_{13} x_{16} x_{18} -  x_{6} x_{8} x_{9} x_{11} x_{15} x_{19}, x_{3} x_{6} x_{7} x_{13} x_{16} x_{18} -  x_{4} x_{8} x_{9} x_{11} x_{15} x_{19}, x_{3} x_{4} x_{9} x_{13} x_{16} x_{18} -  x_{6} x_{7} x_{8} x_{11} x_{15} x_{19}, x_{6} x_{7} x_{9} x_{13} x_{16} x_{18} -  x_{3} x_{4} x_{8} x_{11} x_{15} x_{19}, x_{3} x_{8} x_{9} x_{13} x_{16} x_{18} -  x_{4} x_{6} x_{7} x_{11} x_{15} x_{19}, x_{7} x_{8} x_{9} x_{13} x_{16} x_{18} -  x_{3} x_{4} x_{6} x_{11} x_{15} x_{19}, x_{3} x_{4} x_{11} x_{13} x_{16} x_{18} -  x_{6} x_{7} x_{8} x_{9} x_{15} x_{19}, x_{3} x_{7} x_{11} x_{13} x_{16} x_{18} -  x_{4} x_{6} x_{8} x_{9} x_{15} x_{19}, x_{4} x_{7} x_{11} x_{13} x_{16} x_{18} -  x_{3} x_{6} x_{8} x_{9} x_{15} x_{19}, x_{3} x_{8} x_{11} x_{13} x_{16} x_{18} -  x_{4} x_{6} x_{7} x_{9} x_{15} x_{19}, x_{7} x_{8} x_{11} x_{13} x_{16} x_{18} -  
x_{3} x_{4}x_{6} x_{9} x_{15} x_{19}, x_{7} x_{9} x_{11} x_{13} x_{16} x_{18} -  x_{3} x_{4} x_{6} x_{8} x_{15} x_{19}, x_{3} x_{6} x_{7} x_{14} x_{16} x_{18} -  x_{2} x_{4} x_{10} x_{12} x_{15} x_{19}, x_{3} x_{6} x_{10} x_{14} x_{16} x_{18} -  x_{2} x_{4} x_{7} x_{12} x_{15} x_{19}, x_{2} x_{7} x_{10} x_{14} x_{16} x_{18} -  x_{3} x_{4} x_{6} x_{12} x_{15} x_{19}, x_{4} x_{7} x_{10} x_{14} x_{16} x_{18} -  x_{2} x_{3} x_{6} x_{12} x_{15} x_{19}, x_{3} x_{6} x_{12} x_{14} x_{16} x_{18} -  x_{2} x_{4} x_{7} x_{10} x_{15} x_{19}, x_{2} x_{7} x_{12} x_{14} x_{16} x_{18} -  x_{3} x_{4} x_{6} x_{10} x_{15} x_{19}, x_{3} x_{7} x_{12} x_{14} x_{16} x_{18} -  x_{2} x_{4} x_{6} x_{10} x_{15} x_{19}, x_{2} x_{10} x_{12} x_{14} x_{16} x_{18} -  x_{3} x_{4} x_{6} x_{7} x_{15} x_{19}, x_{3} x_{10} x_{12} x_{14} x_{16} x_{18} -  x_{2} x_{4} x_{6} x_{7} x_{15} x_{19}, x_{2} x_{3} x_{6} x_{15} x_{16} x_{18} -  x_{4} x_{7} x_{10} x_{12} x_{14} x_{19}, x_{3} x_{4} x_{6} x_{15} x_{16} x_{18} -  x_{2} x_{7} x_{10} x_{12} x_{14}
 x_{19}, x_{2} x_{4} x_{7} x_{15} x_{16} x_{18} -  x_{3} x_{6} x_{10} x_{12} x_{14} x_{19}, x_{3} x_{4} x_{7} x_{15} x_{16} x_{18} -  x_{2} x_{6} x_{10} x_{12} x_{14} x_{19}, x_{3} x_{4} x_{8} x_{15} x_{16} x_{18} -  x_{6} x_{7} x_{9} x_{11} x_{13} x_{19}, x_{3} x_{6} x_{8} x_{15} x_{16} x_{18} -  x_{4} x_{7} x_{9} x_{11} x_{13} x_{19}, x_{3} x_{4} x_{9} x_{15} x_{16} x_{18} -  x_{6} x_{7} x_{8} x_{11} x_{13} x_{19}, x_{4} x_{7} x_{9} x_{15} x_{16} x_{18} -  x_{3} x_{6} x_{8} x_{11} x_{13} x_{19}, x_{3} x_{4} x_{10} x_{15} x_{16} x_{18} -  x_{2} x_{6} x_{7} x_{12} x_{14} x_{19}, x_{3} x_{4} x_{11} x_{15} x_{16} x_{18} -  x_{6} x_{7} x_{8} x_{9} x_{13} x_{19}, x_{4} x_{7} x_{11} x_{15} x_{16} x_{18} -  x_{3} x_{6} x_{8} x_{9} x_{13} x_{19}, x_{3} x_{8} x_{11} x_{15} x_{16} x_{18} -  x_{4} x_{6} x_{7} x_{9} x_{13} x_{19}, x_{4} x_{8} x_{11} x_{15} x_{16} x_{18} -  x_{3} x_{6} x_{7} x_{9} x_{13} x_{19}, x_{6} x_{8} x_{11} x_{15} x_{16} x_{18} -  x_{3} x_{4} x_{7} x_{9} x_{13} x_{19}, x_{7} x_{8} x_{11} x_{15} x_
{16} x_{18}
 -  x_{3} x_{4} x_{6} x_{9} x_{13} x_{19}, x_{2} x_{3} x_{12} x_{15} x_{16} x_{18} -  x_{4} x_{6} x_{7} x_{10} x_{14} x_{19}, x_{2} x_{4} x_{12} x_{15} x_{16} x_{18} -  x_{3} x_{6} x_{7} x_{10} x_{14} x_{19}, x_{3} x_{4} x_{12} x_{15} x_{16} x_{18} -  x_{2} x_{6} x_{7} x_{10} x_{14} x_{19}, x_{3} x_{4} x_{13} x_{15} x_{16} x_{18} -  x_{6} x_{7} x_{8} x_{9} x_{11} x_{19}, x_{3} x_{6} x_{13} x_{15} x_{16} x_{18} -  x_{4} x_{7} x_{8} x_{9} x_{11} x_{19}, x_{3} x_{9} x_{13} x_{15} x_{16} x_{18} -  x_{4} x_{6} x_{7} x_{8} x_{11} x_{19}, x_{4} x_{9} x_{13} x_{15} x_{16} x_{18} -  x_{3} x_{6} x_{7} x_{8} x_{11} x_{19}, x_{6} x_{9} x_{13} x_{15} x_{16} x_{18} -  x_{3} x_{4} x_{7} x_{8} x_{11} x_{19}, x_{7} x_{9} x_{13} x_{15} x_{16} x_{18} -  x_{3} x_{4} x_{6} x_{8} x_{11} x_{19}, x_{3} x_{6} x_{14} x_{15} x_{16} x_{18} -  x_{2} x_{4} x_{7} x_{10} x_{12} x_{19}, x_{4} x_{7} x_{14} x_{15} x_{16} x_{18} -  x_{2} x_{3} x_{6} x_{10} x_{12} x_{19}, x_{3} x_{10} x_{14} x_{15} x_{16} x_{18} -  x_{2} x_{4} x_{6} x_{7} x_{12}
 x_{19}, x_{4} x_{10} x_{14} x_{15} x_{16} x_{18} -  x_{2} x_{3} x_{6} x_{7} x_{12} x_{19}, x_{4} x_{8} x_{9} x_{11} x_{17} x_{18} -  x_{3} x_{5} x_{10} x_{12} x_{13} x_{19}, x_{3} x_{8} x_{10} x_{11} x_{17} x_{18} -  x_{4} x_{5} x_{9} x_{12} x_{13} x_{19}, x_{5} x_{8} x_{10} x_{11} x_{17} x_{18} -  x_{3} x_{4} x_{9} x_{12} x_{13} x_{19}, x_{4} x_{9} x_{10} x_{11} x_{17} x_{18} -  x_{3} x_{5} x_{8} x_{12} x_{13} x_{19}, x_{3} x_{8} x_{9} x_{12} x_{17} x_{18} -  x_{4} x_{5} x_{10} x_{11} x_{13} x_{19}, x_{4} x_{8} x_{9} x_{12} x_{17} x_{18} -  x_{3} x_{5} x_{10} x_{11} x_{13} x_{19}, x_{3} x_{8} x_{10} x_{12} x_{17} x_{18} -  x_{4} x_{5} x_{9} x_{11} x_{13} x_{19}, x_{5} x_{8} x_{10} x_{12} x_{17} x_{18} -  x_{3} x_{4} x_{9} x_{11} x_{13} x_{19}, x_{4} x_{9} x_{10} x_{12} x_{17} x_{18} -  x_{3} x_{5} x_{8} x_{11} x_{13} x_{19}, x_{5} x_{9} x_{10} x_{12} x_{17} x_{18} -  x_{3} x_{4} x_{8} x_{11} x_{13} x_{19}, x_{4} x_{8} x_{11} x_{12} x_{17} x_{18} -  x_{3} x_{5} x_{9} x_{10} x_{13} x_{19}, x_{5} x_{8} x_{11} 
x_{12} x_{17} x_{18} -  x_{3} x_{4} x_{9} x_{10} x_{13} x_{19}, x_{4} x_{9} x_{11} x_{12} x_{17} x_{18} -  x_{3} x_{5} x_{8} x_{10} x_{13} x_{19}, x_{4} x_{10} x_{11} x_{12} x_{17} x_{18} -  x_{3} x_{5} x_{8} x_{9} x_{13} x_{19}, x_{5} x_{10} x_{11} x_{12} x_{17} x_{18} -  x_{3} x_{4} x_{8} x_{9} x_{13} x_{19}, x_{3} x_{8} x_{9} x_{13} x_{17} x_{18} -  x_{4} x_{5} x_{10} x_{11} x_{12} x_{19}, x_{3} x_{8} x_{10} x_{13} x_{17} x_{18} -  x_{4} x_{5} x_{9} x_{11} x_{12} x_{19}, x_{4} x_{9} x_{10} x_{13} x_{17} x_{18} -  x_{3} x_{5} x_{8} x_{11} x_{12} x_{19}, x_{5} x_{9} x_{10} x_{13} x_{17} x_{18} -  x_{3} x_{4} x_{8} x_{11} x_{12} x_{19}, x_{3} x_{8} x_{11} x_{13} x_{17} x_{18} -  x_{4} x_{5} x_{9} x_{10} x_{12} x_{19}, x_{4} x_{9} x_{11} x_{13} x_{17} x_{18} -  x_{3} x_{5} x_{8} x_{10} x_{12} x_{19}, x_{3} x_{10} x_{11} x_{13} x_{17} x_{18} -  x_{4} x_{5} x_{8} x_{9} x_{12} x_{19}, x_{4} x_{10} x_{11} x_{13} x_{17} x_{18} -  x_{3} x_{5} x_{8} x_{9} x_{12} x_{19}, x_{3} x_{8} x_{12} x_{13} x_{17} x_{18} -  x_{
4} x_{5} x_{9} x_{10} x_{11} x_{19}, x_{3} x_{9} x_{12} x_{13} x_{17} x_{18} -  x_{4} x_{5} x_{8} x_{10} x_{11} x_{19}, x_{5} x_{9} x_{12} x_{13} x_{17} x_{18} -  x_{3} x_{4} x_{8} x_{10} x_{11} x_{19}, x_{3} x_{10} x_{12} x_{13} x_{17} x_{18} -  x_{4} x_{5} x_{8} x_{9} x_{11} x_{19}, x_{5} x_{10} x_{12} x_{13} x_{17} x_{18} -  x_{3} x_{4} x_{8} x_{9} x_{11} x_{19}, x_{1} x_{2} x_{3} x_{4} x_{6} -  x_{7} x_{10} x_{11} x_{13} x_{17}, x_{1} x_{2} x_{3} x_{5} x_{6} -  x_{8} x_{10} x_{11} x_{15} x_{18}, x_{1} x_{2} x_{3} x_{4} x_{7} -  x_{6} x_{10} x_{11} x_{13} x_{17}, x_{1} x_{2} x_{4} x_{5} x_{7} -  x_{9} x_{10} x_{13} x_{16} x_{19}, x_{1} x_{2} x_{3} x_{6} x_{7} -  x_{4} x_{10} x_{11} x_{13} x_{17}, x_{1} x_{2} x_{4} x_{6} x_{7} -  x_{3} x_{10} x_{11} x_{13} x_{17}, x_{1} x_{3} x_{4} x_{6} x_{7} -  x_{8} x_{9} x_{12} x_{14} x_{17}, x_{2} x_{3} x_{4} x_{6} x_{7} -  x_{1} x_{10} x_{11} x_{13} x_{17}, x_{1} x_{2} x_{5} x_{6} x_{7} -  x_{8} x_{9} x_{12} x_{18} x_{19}, x_{1} x_{2} x_{3} x_{5} x_{8} -  x_{6} x_{10}
 x_{11} x_{15} x_{18}, x_{1} x_{3} x_{4} x_{5} x_{8} -  x_{9} x_{10} x_{14} x_{15} x_{16}, x_{1} x_{2} x_{3} x_{6} x_{8} -  x_{5} x_{10} x_{11} x_{15} x_{18}, x_{1} x_{3} x_{4} x_{6} x_{8} -  x_{7} x_{9} x_{12} x_{14} x_{17}, x_{1} x_{2} x_{5} x_{6} x_{8} -  x_{7} x_{9} x_{12} x_{18} x_{19}, x_{1} x_{3} x_{5} x_{6} x_{8} -  x_{2} x_{10} x_{11} x_{15} x_{18}, x_{2} x_{3} x_{5} x_{6} x_{8} -  x_{1} x_{10} x_{11} x_{15} x_{18}, x_{1} x_{3} x_{4} x_{7} x_{8} -  x_{6} x_{9} x_{12} x_{14} x_{17}, x_{1} x_{2} x_{5} x_{7} x_{8} -  x_{6} x_{9} x_{12} x_{18} x_{19}, x_{1} x_{2} x_{6} x_{7} x_{8} -  x_{5} x_{9} x_{12} x_{18} x_{19}, x_{1} x_{3} x_{6} x_{7} x_{8} -  x_{4} x_{9} x_{12} x_{14} x_{17}, x_{1} x_{4} x_{6} x_{7} x_{8} -  x_{3} x_{9} x_{12} x_{14} x_{17}, x_{3} x_{4} x_{6} x_{7} x_{8} -  x_{1} x_{9} x_{12} x_{14} x_{17}, x_{1} x_{5} x_{6} x_{7} x_{8} -  x_{2} x_{9} x_{12} x_{18} x_{19}, x_{2} x_{5} x_{6} x_{7} x_{8} -  x_{1} x_{9} x_{12} x_{18} x_{19}, x_{1} x_{2} x_{4} x_{5} x_{9} -  x_{7} x_{10} x_{13} x_{16}
 x_{19}, x_{1} x_{3} x_{4} x_{5} x_{9} -  x_{8} x_{10} x_{14} x_{15} x_{16}, x_{1} x_{3} x_{4} x_{6} x_{9} -  x_{7} x_{8} x_{12} x_{14} x_{17}, x_{1} x_{2} x_{5} x_{6} x_{9} -  x_{7} x_{8} x_{12} x_{18} x_{19}, x_{1} x_{2} x_{4} x_{7} x_{9} -  x_{5} x_{10} x_{13} x_{16} x_{19}, x_{1} x_{3} x_{4} x_{7} x_{9} -  x_{6} x_{8} x_{12} x_{14} x_{17}, x_{1} x_{2} x_{5} x_{7} x_{9} -  x_{6} x_{8} x_{12} x_{18} x_{19}, x_{1} x_{4} x_{5} x_{7} x_{9} -  x_{2} x_{10} x_{13} x_{16} x_{19}, x_{2} x_{4} x_{5} x_{7} x_{9} -  x_{1} x_{10} x_{13} x_{16} x_{19}, x_{1} x_{2} x_{6} x_{7} x_{9} -  x_{5} x_{8} x_{12} x_{18} x_{19}, x_{1} x_{3} x_{6} x_{7} x_{9} -  x_{4} x_{8} x_{12} x_{14} x_{17}, x_{1} x_{4} x_{6} x_{7} x_{9} -  x_{3} x_{8} x_{12} x_{14} x_{17}, x_{3} x_{4} x_{6} x_{7} x_{9} -  x_{1} x_{8} x_{12} x_{14} x_{17}, x_{1} x_{5} x_{6} x_{7} x_{9} -  x_{2} x_{8} x_{12} x_{18} x_{19}, x_{2} x_{5} x_{6} x_{7} x_{9} -  x_{1} x_{8} x_{12} x_{18} x_{19}, x_{1} x_{3} x_{4} x_{8} x_{9} -  x_{6} x_{7} x_{12} x_{14} x_{17}, x_{1}
x_{2} x_{5} x_{8} x_{9} -  x_{6} x_{7} x_{12} x_{18} x_{19}, x_{1} x_{3} x_{5} x_{8} x_{9} -  x_{4} x_{10} x_{14} x_{15} x_{16}, x_{1} x_{4} x_{5} x_{8} x_{9} -  x_{3} x_{10} x_{14} x_{15} x_{16}, x_{3} x_{4} x_{5} x_{8} x_{9} -  x_{1} x_{10} x_{14} x_{15} x_{16}, x_{1} x_{2} x_{6} x_{8} x_{9} -  x_{5} x_{7} x_{12} x_{18} x_{19}, x_{1} x_{3} x_{6} x_{8} x_{9} -  x_{4} x_{7} x_{12} x_{14} x_{17}, x_{1} x_{4} x_{6} x_{8} x_{9} -  x_{3} x_{7} x_{12} x_{14} x_{17}, x_{3} x_{4} x_{6} x_{8} x_{9} -  x_{1} x_{7} x_{12} x_{14} x_{17}, x_{1} x_{5} x_{6} x_{8} x_{9} -  x_{2} x_{7} x_{12} x_{18} x_{19}, x_{2} x_{5} x_{6} x_{8} x_{9} -  x_{1} x_{7} x_{12} x_{18} x_{19}, x_{1} x_{2} x_{7} x_{8} x_{9} -  x_{5} x_{6} x_{12} x_{18} x_{19}, x_{1} x_{3} x_{7} x_{8} x_{9} -  x_{4} x_{6} x_{12} x_{14} x_{17}, x_{1} x_{4} x_{7} x_{8} x_{9} -  x_{3} x_{6} x_{12} x_{14} x_{17}, x_{3} x_{4} x_{7} x_{8} x_{9} -  x_{1} x_{6} x_{12} x_{14} x_{17}, x_{1} x_{5} x_{7} x_{8} x_{9} -  x_{2} x_{6} x_{12} x_{18} x_{19}, x_{2} x_{5} x_{7} x_{
8} x_{9} -  x_{1} x_{6} x_{12} x_{18} x_{19}, x_{1} x_{6} x_{7} x_{8} x_{9} -  x_{2} x_{5} x_{12} x_{18} x_{19}, x_{2} x_{6} x_{7} x_{8} x_{9} -  x_{1} x_{5} x_{12} x_{18} x_{19}, x_{3} x_{6} x_{7} x_{8} x_{9} -  x_{1} x_{4} x_{12} x_{14} x_{17}, x_{4} x_{6} x_{7} x_{8} x_{9} -  x_{1} x_{3} x_{12} x_{14} x_{17}, x_{5} x_{6} x_{7} x_{8} x_{9} -  x_{1} x_{2} x_{12} x_{18} x_{19}, x_{1} x_{2} x_{3} x_{4} x_{10} -  x_{6} x_{7} x_{11} x_{13} x_{17}, x_{1} x_{2} x_{3} x_{5} x_{10} -  x_{6} x_{8} x_{11} x_{15} x_{18}, x_{1} x_{2} x_{4} x_{5} x_{10} -  x_{7} x_{9} x_{13} x_{16} x_{19}, x_{1} x_{3} x_{4} x_{5} x_{10} -  x_{8} x_{9} x_{14} x_{15} x_{16}, x_{1} x_{2} x_{3} x_{6} x_{10} -  x_{5} x_{8} x_{11} x_{15} x_{18}, x_{1} x_{2} x_{4} x_{6} x_{10} -  x_{3} x_{7} x_{11} x_{13} x_{17}, x_{1} x_{3} x_{4} x_{6} x_{10} -  x_{2} x_{7} x_{11} x_{13} x_{17}, x_{2} x_{3} x_{4} x_{6} x_{10} -  x_{1} x_{7} x_{11} x_{13} x_{17}, x_{1} x_{2} x_{5} x_{6} x_{10} -  x_{3} x_{8} x_{11} x_{15} x_{18}, x_{1} x_{3} x_{5} x_{6} x_{10}-
  x_{2} x_{8} x_{11} x_{15} x_{18}, x_{2} x_{3} x_{5} x_{6} x_{10} -  x_{1} x_{8} x_{11} x_{15} x_{18}, x_{1} x_{2} x_{3} x_{7} x_{10} -  x_{4} x_{6} x_{11} x_{13} x_{17}, x_{1} x_{2} x_{4} x_{7} x_{10} -  x_{5} x_{9} x_{13} x_{16} x_{19}, x_{1} x_{3} x_{4} x_{7} x_{10} -  x_{2} x_{6} x_{11} x_{13} x_{17}, x_{2} x_{3} x_{4} x_{7} x_{10} -  x_{1} x_{6} x_{11} x_{13} x_{17}, x_{1} x_{2} x_{5} x_{7} x_{10} -  x_{4} x_{9} x_{13} x_{16} x_{19}, x_{1} x_{4} x_{5} x_{7} x_{10} -  x_{2} x_{9} x_{13} x_{16} x_{19}, x_{2} x_{4} x_{5} x_{7} x_{10} -  x_{1} x_{9} x_{13} x_{16} x_{19}, x_{1} x_{2} x_{6} x_{7} x_{10} -  x_{3} x_{4} x_{11} x_{13} x_{17}, x_{1} x_{3} x_{6} x_{7} x_{10} -  x_{2} x_{4} x_{11} x_{13} x_{17}, x_{2} x_{3} x_{6} x_{7} x_{10} -  x_{1} x_{4} x_{11} x_{13} x_{17}, x_{1} x_{4} x_{6} x_{7} x_{10} -  x_{2} x_{3} x_{11} x_{13} x_{17}, x_{2} x_{4} x_{6} x_{7} x_{10} -  x_{1} x_{3} x_{11} x_{13} x_{17}, x_{3} x_{4} x_{6} x_{7} x_{10} -  x_{1} x_{2} x_{11} x_{13} x_{17}, x_{1} x_{2} x_{3} x_{8} x_{10} -  
x_{5} x_{6} x_{11} x_{15} x_{18}, x_{1} x_{3} x_{4} x_{8} x_{10} -  x_{5} x_{9} x_{14} x_{15} x_{16}, x_{1} x_{2} x_{5} x_{8} x_{10} -  x_{3} x_{6} x_{11} x_{15} x_{18}, x_{1} x_{3} x_{5} x_{8} x_{10} -  x_{2} x_{6} x_{11} x_{15} x_{18}, x_{2} x_{3} x_{5} x_{8} x_{10} -  x_{1} x_{6} x_{11} x_{15} x_{18}, x_{1} x_{4} x_{5} x_{8} x_{10} -  x_{3} x_{9} x_{14} x_{15} x_{16}, x_{3} x_{4} x_{5} x_{8} x_{10} -  x_{1} x_{9} x_{14} x_{15} x_{16}, x_{1} x_{2} x_{6} x_{8} x_{10} -  x_{3} x_{5} x_{11} x_{15} x_{18}, x_{1} x_{3} x_{6} x_{8} x_{10} -  x_{2} x_{5} x_{11} x_{15} x_{18}, x_{2} x_{3} x_{6} x_{8} x_{10} -  x_{1} x_{5} x_{11} x_{15} x_{18}, x_{1} x_{5} x_{6} x_{8} x_{10} -  x_{2} x_{3} x_{11} x_{15} x_{18}, x_{2} x_{5} x_{6} x_{8} x_{10} -  x_{1} x_{3} x_{11} x_{15} x_{18}, x_{3} x_{5} x_{6} x_{8} x_{10} -  x_{1} x_{2} x_{11} x_{15} x_{18}, x_{1} x_{2} x_{4} x_{9} x_{10} -  x_{5} x_{7} x_{13} x_{16} x_{19}, x_{1} x_{3} x_{4} x_{9} x_{10} -  x_{5} x_{8} x_{14} x_{15} x_{16}, x_{1} x_{2} x_{5} x_{9} x_{10} -  x_{
4} x_{7} x_{13} x_{16} x_{19}, x_{1} x_{3} x_{5} x_{9} x_{10} -  x_{4} x_{8} x_{14} x_{15} x_{16}, x_{1} x_{4} x_{5} x_{9} x_{10} -  x_{2} x_{7} x_{13} x_{16} x_{19}, x_{2} x_{4} x_{5} x_{9} x_{10} -  x_{1} x_{7} x_{13} x_{16} x_{19}, x_{3} x_{4} x_{5} x_{9} x_{10} -  x_{1} x_{8} x_{14} x_{15} x_{16}, x_{1} x_{2} x_{7} x_{9} x_{10} -  x_{4} x_{5} x_{13} x_{16} x_{19}, x_{1} x_{4} x_{7} x_{9} x_{10} -  x_{2} x_{5} x_{13} x_{16} x_{19}, x_{2} x_{4} x_{7} x_{9} x_{10} -  x_{1} x_{5} x_{13} x_{16} x_{19}, x_{1} x_{5} x_{7} x_{9} x_{10} -  x_{2} x_{4} x_{13} x_{16} x_{19}, x_{2} x_{5} x_{7} x_{9} x_{10} -  x_{1} x_{4} x_{13} x_{16} x_{19}, x_{4} x_{5} x_{7} x_{9} x_{10} -  x_{1} x_{2} x_{13} x_{16} x_{19}, x_{1} x_{3} x_{8} x_{9} x_{10} -  x_{4} x_{5} x_{14} x_{15} x_{16}, x_{1} x_{4} x_{8} x_{9} x_{10} -  x_{3} x_{5} x_{14} x_{15} x_{16}, x_{3} x_{4} x_{8} x_{9} x_{10} -  x_{1} x_{5} x_{14} x_{15} x_{16}, x_{1} x_{5} x_{8} x_{9} x_{10} -  x_{3} x_{4} x_{14} x_{15} x_{16}, x_{3} x_{5} x_{8} x_{9} x_{10} -  x_{1}x_
{4} x_{14} x_{15} x_{16}, x_{4} x_{5} x_{8} x_{9} x_{10} -  x_{1} x_{3} x_{14} x_{15} x_{16}, x_{1} x_{2} x_{3} x_{4} x_{11} -  x_{6} x_{7} x_{10} x_{13} x_{17}, x_{1} x_{2} x_{3} x_{5} x_{11} -  x_{6} x_{8} x_{10} x_{15} x_{18}, x_{1} x_{2} x_{4} x_{5} x_{11} -  x_{3} x_{12} x_{13} x_{15} x_{16}, x_{1} x_{3} x_{4} x_{5} x_{11} -  x_{2} x_{12} x_{13} x_{15} x_{16}, x_{1} x_{2} x_{3} x_{6} x_{11} -  x_{5} x_{8} x_{10} x_{15} x_{18}, x_{1} x_{3} x_{4} x_{6} x_{11} -  x_{2} x_{7} x_{10} x_{13} x_{17}, x_{1} x_{2} x_{5} x_{6} x_{11} -  x_{3} x_{8} x_{10} x_{15} x_{18}, x_{1} x_{2} x_{3} x_{7} x_{11} -  x_{8} x_{12} x_{16} x_{17} x_{18}, x_{1} x_{2} x_{4} x_{7} x_{11} -  x_{3} x_{6} x_{10} x_{13} x_{17}, x_{1} x_{3} x_{4} x_{7} x_{11} -  x_{2} x_{6} x_{10} x_{13} x_{17}, x_{2} x_{3} x_{4} x_{7} x_{11} -  x_{1} x_{6} x_{10} x_{13} x_{17}, x_{1} x_{4} x_{5} x_{7} x_{11} -  x_{8} x_{12} x_{14} x_{16} x_{19}, x_{1} x_{2} x_{6} x_{7} x_{11} -  x_{3} x_{4} x_{10} x_{13} x_{17}, x_{1} x_{3} x_{6} x_{7} x_{11} -  x_{2} 
x_{4} x_{10} x_{13} x_{17}, x_{2} x_{3} x_{6} x_{7} x_{11} -  x_{1} x_{4} x_{10} x_{13} x_{17}, x_{1} x_{4} x_{6} x_{7} x_{11} -  x_{2} x_{3} x_{10} x_{13} x_{17}, x_{3} x_{4} x_{6} x_{7} x_{11} -  x_{1} x_{2} x_{10} x_{13} x_{17}, x_{1} x_{5} x_{6} x_{7} x_{11} -  x_{10} x_{13} x_{14} x_{18} x_{19}, x_{1} x_{2} x_{3} x_{8} x_{11} -  x_{7} x_{12} x_{16} x_{17} x_{18}, x_{1} x_{2} x_{5} x_{8} x_{11} -  x_{3} x_{6} x_{10} x_{15} x_{18}, x_{1} x_{3} x_{5} x_{8} x_{11} -  x_{2} x_{6} x_{10} x_{15} x_{18}, x_{2} x_{3} x_{5} x_{8} x_{11} -  x_{1} x_{6} x_{10} x_{15} x_{18}, x_{1} x_{4} x_{5} x_{8} x_{11} -  x_{7} x_{12} x_{14} x_{16} x_{19}, x_{1} x_{2} x_{6} x_{8} x_{11} -  x_{3} x_{5} x_{10} x_{15} x_{18}, x_{1} x_{3} x_{6} x_{8} x_{11} -  x_{2} x_{5} x_{10} x_{15} x_{18}, x_{2} x_{3} x_{6} x_{8} x_{11} -  x_{1} x_{5} x_{10} x_{15} x_{18}, x_{1} x_{4} x_{6} x_{8} x_{11} -  x_{10} x_{14} x_{15} x_{17} x_{19}, x_{1} x_{5} x_{6} x_{8} x_{11} -  x_{2} x_{3} x_{10} x_{15} x_{18}, x_{2} x_{5} x_{6} x_{8} x_{11} -  x_{
1} x_{3} x_{10} x_{15} x_{18}, x_{1} x_{2} x_{7} x_{8} x_{11} -  x_{3} x_{12} x_{16} x_{17} x_{18}, x_{1} x_{3} x_{7} x_{8} x_{11} -  x_{2} x_{12} x_{16} x_{17} x_{18}, x_{1} x_{4} x_{7} x_{8} x_{11} -  x_{5} x_{12} x_{14} x_{16} x_{19}, x_{1} x_{5} x_{7} x_{8} x_{11} -  x_{4} x_{12} x_{14} x_{16} x_{19}, x_{1} x_{2} x_{3} x_{10} x_{11} -  x_{5} x_{6} x_{8} x_{15} x_{18}, x_{1} x_{2} x_{4} x_{10} x_{11} -  x_{3} x_{6} x_{7} x_{13} x_{17}, x_{1} x_{3} x_{4} x_{10} x_{11} -  x_{2} x_{6} x_{7} x_{13} x_{17}, x_{2} x_{3} x_{4} x_{10} x_{11} -  x_{1} x_{6} x_{7} x_{13} x_{17}, x_{1} x_{2} x_{5} x_{10} x_{11} -  x_{3} x_{6} x_{8} x_{15} x_{18}, x_{1} x_{3} x_{5} x_{10} x_{11} -  x_{2} x_{6} x_{8} x_{15} x_{18}, x_{2} x_{3} x_{5} x_{10} x_{11} -  x_{1} x_{6} x_{8} x_{15} x_{18}, x_{1} x_{2} x_{6} x_{10} x_{11} -  x_{3} x_{5} x_{8} x_{15} x_{18}, x_{1} x_{3} x_{6} x_{10} x_{11} -  x_{2} x_{5} x_{8} x_{15} x_{18}, x_{2} x_{3} x_{6} x_{10} x_{11} -  x_{1} x_{5} x_{8} x_{15} x_{18}, x_{1} x_{4} x_{6} x_{10} x_{11} -  x_
{8} x_{14} x_{15} x_{17} x_{19}, x_{3} x_{4} x_{6} x_{10} x_{11} -  x_{1} x_{2} x_{7} x_{13} x_{17}, x_{1} x_{5} x_{6} x_{10} x_{11} -  x_{7} x_{13} x_{14} x_{18} x_{19}, x_{2} x_{5} x_{6} x_{10} x_{11} -  x_{1} x_{3} x_{8} x_{15} x_{18}, x_{1} x_{2} x_{7} x_{10} x_{11} -  x_{3} x_{4} x_{6} x_{13} x_{17}, x_{1} x_{4} x_{7} x_{10} x_{11} -  x_{2} x_{3} x_{6} x_{13} x_{17}, x_{2} x_{4} x_{7} x_{10} x_{11} -  x_{1} x_{3} x_{6} x_{13} x_{17}, x_{1} x_{5} x_{7} x_{10} x_{11} -  x_{6} x_{13} x_{14} x_{18} x_{19}, x_{1} x_{6} x_{7} x_{10} x_{11} -  x_{5} x_{13} x_{14} x_{18} x_{19}, x_{2} x_{6} x_{7} x_{10} x_{11} -  x_{1} x_{3} x_{4} x_{13} x_{17}, x_{4} x_{6} x_{7} x_{10} x_{11} -  x_{1} x_{2} x_{3} x_{13} x_{17}, x_{1} x_{3} x_{8} x_{10} x_{11} -  x_{2} x_{5} x_{6} x_{15} x_{18}, x_{1} x_{4} x_{8} x_{10} x_{11} -  x_{6} x_{14} x_{15} x_{17} x_{19}, x_{1} x_{5} x_{8} x_{10} x_{11} -  x_{2} x_{3} x_{6} x_{15} x_{18}, x_{3} x_{5} x_{8} x_{10} x_{11} -  x_{1} x_{2} x_{6} x_{15} x_{18}, x_{1} x_{6} x_{8} x_{10} x_{11}
 -  x_{4} x_{14} x_{15} x_{17} x_{19}, x_{3} x_{6} x_{8} x_{10} x_{11} -  x_{1} x_{2} x_{5} x_{15} x_{18}, x_{5} x_{6} x_{8} x_{10} x_{11} -  x_{1} x_{2} x_{3} x_{15} x_{18}, x_{1} x_{2} x_{3} x_{4} x_{12} -  x_{5} x_{11} x_{13} x_{15} x_{16}, x_{1} x_{2} x_{3} x_{5} x_{12} -  x_{4} x_{11} x_{13} x_{15} x_{16}, x_{1} x_{2} x_{4} x_{5} x_{12} -  x_{3} x_{11} x_{13} x_{15} x_{16}, x_{1} x_{3} x_{4} x_{5} x_{12} -  x_{2} x_{11} x_{13} x_{15} x_{16}, x_{1} x_{2} x_{4} x_{6} x_{12} -  x_{9} x_{13} x_{15} x_{17} x_{19}, x_{1} x_{3} x_{4} x_{6} x_{12} -  x_{7} x_{8} x_{9} x_{14} x_{17}, x_{1} x_{2} x_{5} x_{6} x_{12} -  x_{7} x_{8} x_{9} x_{18} x_{19}, x_{1} x_{3} x_{5} x_{6} x_{12} -  x_{9} x_{13} x_{14} x_{15} x_{18}, x_{1} x_{2} x_{3} x_{7} x_{12} -  x_{8} x_{11} x_{16} x_{17} x_{18}, x_{1} x_{3} x_{4} x_{7} x_{12} -  x_{6} x_{8} x_{9} x_{14} x_{17}, x_{1} x_{2} x_{5} x_{7} x_{12} -  x_{6} x_{8} x_{9} x_{18} x_{19}, x_{1} x_{4} x_{5} x_{7} x_{12} -  x_{8} x_{11} x_{14} x_{16} x_{19}, x_{1} x_{2} x_{6} x_{7} x_{
12} -  x_{5} x_{8} x_{9} x_{18} x_{19}, x_{1} x_{3} x_{6} x_{7} x_{12} -  x_{4} x_{8} x_{9} x_{14} x_{17}, x_{1} x_{4} x_{6} x_{7} x_{12} -  x_{3} x_{8} x_{9} x_{14} x_{17}, x_{3} x_{4} x_{6} x_{7} x_{12} -  x_{1} x_{8} x_{9} x_{14} x_{17}, x_{1} x_{2} x_{3} x_{8} x_{12} -  x_{7} x_{11} x_{16} x_{17} x_{18}, x_{1} x_{3} x_{4} x_{8} x_{12} -  x_{6} x_{7} x_{9} x_{14} x_{17}, x_{1} x_{2} x_{5} x_{8} x_{12} -  x_{6} x_{7} x_{9} x_{18} x_{19}, x_{1} x_{4} x_{5} x_{8} x_{12} -  x_{7} x_{11} x_{14} x_{16} x_{19}, x_{1} x_{2} x_{6} x_{8} x_{12} -  x_{5} x_{7} x_{9} x_{18} x_{19}, x_{1} x_{3} x_{6} x_{8} x_{12} -  x_{4} x_{7} x_{9} x_{14} x_{17}, x_{1} x_{5} x_{6} x_{8} x_{12} -  x_{2} x_{7} x_{9} x_{18} x_{19}, x_{2} x_{5} x_{6} x_{8} x_{12} -  x_{1} x_{7} x_{9} x_{18} x_{19}, x_{1} x_{2} x_{7} x_{8} x_{12} -  x_{5} x_{6} x_{9} x_{18} x_{19}, x_{1} x_{3} x_{7} x_{8} x_{12} -  x_{2} x_{11} x_{16} x_{17} x_{18}, x_{1} x_{4} x_{7} x_{8} x_{12} -  x_{5} x_{11} x_{14} x_{16} x_{19}, x_{3} x_{4} x_{7} x_{8} x_{12} -  x_{
1} x_{6} x_{9} x_{14} x_{17}, x_{1} x_{5} x_{7} x_{8} x_{12} -  x_{2} x_{6} x_{9} x_{18} x_{19}, x_{2} x_{5} x_{7} x_{8} x_{12} -  x_{1} x_{6} x_{9} x_{18} x_{19}, x_{1} x_{6} x_{7} x_{8} x_{12} -  x_{2} x_{5} x_{9} x_{18} x_{19}, x_{2} x_{6} x_{7} x_{8} x_{12} -  x_{1} x_{5} x_{9} x_{18} x_{19}, x_{3} x_{6} x_{7} x_{8} x_{12} -  x_{1} x_{4} x_{9} x_{14} x_{17}, x_{1} x_{2} x_{4} x_{9} x_{12} -  x_{6} x_{13} x_{15} x_{17} x_{19}, x_{1} x_{3} x_{4} x_{9} x_{12} -  x_{6} x_{7} x_{8} x_{14} x_{17}, x_{1} x_{2} x_{5} x_{9} x_{12} -  x_{6} x_{7} x_{8} x_{18} x_{19}, x_{1} x_{3} x_{5} x_{9} x_{12} -  x_{6} x_{13} x_{14} x_{15} x_{18}, x_{1} x_{2} x_{6} x_{9} x_{12} -  x_{5} x_{7} x_{8} x_{18} x_{19}, x_{1} x_{3} x_{6} x_{9} x_{12} -  x_{5} x_{13} x_{14} x_{15} x_{18}, x_{1} x_{4} x_{6} x_{9} x_{12} -  x_{2} x_{13} x_{15} x_{17} x_{19}, x_{3} x_{4} x_{6} x_{9} x_{12} -  x_{1} x_{7} x_{8} x_{14} x_{17}, x_{1} x_{5} x_{6} x_{9} x_{12} -  x_{2} x_{7} x_{8} x_{18} x_{19}, x_{2} x_{5} x_{6} x_{9} x_{12} -  x_{1} x_{7} x_
{8} x_{18} x_{19}, x_{1} x_{2} x_{7} x_{9} x_{12} -  x_{5} x_{6} x_{8} x_{18} x_{19}, x_{1} x_{4} x_{7} x_{9} x_{12} -  x_{3} x_{6} x_{8} x_{14} x_{17}, x_{1} x_{5} x_{7} x_{9} x_{12} -  x_{2} x_{6} x_{8} x_{18} x_{19}, x_{2} x_{5} x_{7} x_{9} x_{12} -  x_{1} x_{6} x_{8} x_{18} x_{19}, x_{1} x_{6} x_{7} x_{9} x_{12} -  x_{2} x_{5} x_{8} x_{18} x_{19}, x_{2} x_{6} x_{7} x_{9} x_{12} -  x_{1} x_{5} x_{8} x_{18} x_{19}, x_{4} x_{6} x_{7} x_{9} x_{12} -  x_{1} x_{3} x_{8} x_{14} x_{17}, x_{1} x_{3} x_{8} x_{9} x_{12} -  x_{4} x_{6} x_{7} x_{14} x_{17}, x_{1} x_{4} x_{8} x_{9} x_{12} -  x_{3} x_{6} x_{7} x_{14} x_{17}, x_{3} x_{4} x_{8} x_{9} x_{12} -  x_{1} x_{6} x_{7} x_{14} x_{17}, x_{1} x_{5} x_{8} x_{9} x_{12} -  x_{2} x_{6} x_{7} x_{18} x_{19}, x_{1} x_{6} x_{8} x_{9} x_{12} -  x_{2} x_{5} x_{7} x_{18} x_{19}, x_{3} x_{6} x_{8} x_{9} x_{12} -  x_{1} x_{4} x_{7} x_{14} x_{17}, x_{5} x_{6} x_{8} x_{9} x_{12} -  x_{1} x_{2} x_{7} x_{18} x_{19}, x_{1} x_{7} x_{8} x_{9} x_{12} -  x_{2} x_{5} x_{6} x_{18} x_{19},
x_{4} x_{7} x_{8} x_{9} x_{12} -  x_{1} x_{3} x_{6} x_{14} x_{17}, x_{5} x_{7} x_{8} x_{9} x_{12} -  x_{1} x_{2} x_{6} x_{18} x_{19}, x_{6} x_{7} x_{8} x_{9} x_{12} -  x_{1} x_{2} x_{5} x_{18} x_{19}, x_{1} x_{2} x_{3} x_{11} x_{12} -  x_{7} x_{8} x_{16} x_{17} x_{18}, x_{1} x_{2} x_{4} x_{11} x_{12} -  x_{3} x_{5} x_{13} x_{15} x_{16}, x_{1} x_{3} x_{4} x_{11} x_{12} -  x_{2} x_{5} x_{13} x_{15} x_{16}, x_{2} x_{3} x_{4} x_{11} x_{12} -  x_{1} x_{5} x_{13} x_{15} x_{16}, x_{1} x_{2} x_{5} x_{11} x_{12} -  x_{3} x_{4} x_{13} x_{15} x_{16}, x_{1} x_{3} x_{5} x_{11} x_{12} -  x_{2} x_{4} x_{13} x_{15} x_{16}, x_{2} x_{3} x_{5} x_{11} x_{12} -  x_{1} x_{4} x_{13} x_{15} x_{16}, x_{1} x_{4} x_{5} x_{11} x_{12} -  x_{7} x_{8} x_{14} x_{16} x_{19}, x_{2} x_{4} x_{5} x_{11} x_{12} -  x_{1} x_{3} x_{13} x_{15} x_{16}, x_{3} x_{4} x_{5} x_{11} x_{12} -  x_{1} x_{2} x_{13} x_{15} x_{16}, x_{1} x_{2} x_{7} x_{11} x_{12} -  x_{3} x_{8} x_{16} x_{17} x_{18}, x_{1} x_{4} x_{7} x_{11} x_{12} -  x_{5} x_{8} x_{14} x_{16} x_{
19}, x_{1} x_{5} x_{7} x_{11} x_{12} -  x_{4} x_{8} x_{14} x_{16} x_{19}, x_{4} x_{5} x_{7} x_{11} x_{12} -  x_{1} x_{8} x_{14} x_{16} x_{19}, x_{1} x_{3} x_{8} x_{11} x_{12} -  x_{2} x_{7} x_{16} x_{17} x_{18}, x_{1} x_{4} x_{8} x_{11} x_{12} -  x_{5} x_{7} x_{14} x_{16} x_{19}, x_{1} x_{5} x_{8} x_{11} x_{12} -  x_{4} x_{7} x_{14} x_{16} x_{19}, x_{4} x_{5} x_{8} x_{11} x_{12} -  x_{1} x_{7} x_{14} x_{16} x_{19}, x_{1} x_{7} x_{8} x_{11} x_{12} -  x_{4} x_{5} x_{14} x_{16} x_{19}, x_{4} x_{7} x_{8} x_{11} x_{12} -  x_{1} x_{5} x_{14} x_{16} x_{19}, x_{5} x_{7} x_{8} x_{11} x_{12} -  x_{1} x_{4} x_{14} x_{16} x_{19}, x_{1} x_{2} x_{3} x_{4} x_{13} -  x_{6} x_{7} x_{10} x_{11} x_{17}, x_{1} x_{2} x_{3} x_{5} x_{13} -  x_{4} x_{11} x_{12} x_{15} x_{16}, x_{1} x_{2} x_{4} x_{5} x_{13} -  x_{7} x_{9} x_{10} x_{16} x_{19}, x_{1} x_{3} x_{4} x_{5} x_{13} -  x_{2} x_{11} x_{12} x_{15} x_{16}, x_{1} x_{2} x_{3} x_{6} x_{13} -  x_{4} x_{7} x_{10} x_{11} x_{17}, x_{1} x_{2} x_{4} x_{6} x_{13} -  x_{9} x_{12} x_{15} x_
{17} x_{19}, x_{1} x_{3} x_{4} x_{6} x_{13} -  x_{2} x_{7} x_{10} x_{11} x_{17}, x_{2} x_{3} x_{4} x_{6} x_{13} -  x_{1} x_{7} x_{10} x_{11} x_{17}, x_{1} x_{3} x_{5} x_{6} x_{13} -  x_{9} x_{12} x_{14} x_{15} x_{18}, x_{1} x_{2} x_{4} x_{7} x_{13} -  x_{5} x_{9} x_{10} x_{16} x_{19}, x_{1} x_{3} x_{4} x_{7} x_{13} -  x_{2} x_{6} x_{10} x_{11} x_{17}, x_{1} x_{2} x_{5} x_{7} x_{13} -  x_{4} x_{9} x_{10} x_{16} x_{19}, x_{1} x_{2} x_{6} x_{7} x_{13} -  x_{3} x_{4} x_{10} x_{11} x_{17}, x_{1} x_{3} x_{6} x_{7} x_{13} -  x_{2} x_{4} x_{10} x_{11} x_{17}, x_{1} x_{4} x_{6} x_{7} x_{13} -  x_{2} x_{3} x_{10} x_{11} x_{17}, x_{2} x_{4} x_{6} x_{7} x_{13} -  x_{1} x_{3} x_{10} x_{11} x_{17}, x_{3} x_{4} x_{6} x_{7} x_{13} -  x_{1} x_{2} x_{10} x_{11} x_{17}, x_{1} x_{5} x_{6} x_{7} x_{13} -  x_{10} x_{11} x_{14} x_{18} x_{19}, x_{1} x_{2} x_{4} x_{9} x_{13} -  x_{6} x_{12} x_{15} x_{17} x_{19}, x_{1} x_{2} x_{5} x_{9} x_{13} -  x_{4} x_{7} x_{10} x_{16} x_{19}, x_{1} x_{3} x_{5} x_{9} x_{13} -  x_{6} x_{12} x_{14}x_
{15} x_{18}, x_{1} x_{4} x_{5} x_{9} x_{13} -  x_{2} x_{7} x_{10} x_{16} x_{19}, x_{2} x_{4} x_{5} x_{9} x_{13} -  x_{1} x_{7} x_{10} x_{16} x_{19}, x_{1} x_{2} x_{6} x_{9} x_{13} -  x_{4} x_{12} x_{15} x_{17} x_{19}, x_{1} x_{3} x_{6} x_{9} x_{13} -  x_{5} x_{12} x_{14} x_{15} x_{18}, x_{1} x_{4} x_{6} x_{9} x_{13} -  x_{2} x_{12} x_{15} x_{17} x_{19}, x_{1} x_{5} x_{6} x_{9} x_{13} -  x_{3} x_{12} x_{14} x_{15} x_{18}, x_{1} x_{2} x_{7} x_{9} x_{13} -  x_{4} x_{5} x_{10} x_{16} x_{19}, x_{1} x_{3} x_{7} x_{9} x_{13} -  x_{10} x_{14} x_{16} x_{17} x_{18}, x_{1} x_{4} x_{7} x_{9} x_{13} -  x_{2} x_{5} x_{10} x_{16} x_{19}, x_{2} x_{4} x_{7} x_{9} x_{13} -  x_{1} x_{5} x_{10} x_{16} x_{19}, x_{1} x_{5} x_{7} x_{9} x_{13} -  x_{2} x_{4} x_{10} x_{16} x_{19}, x_{2} x_{5} x_{7} x_{9} x_{13} -  x_{1} x_{4} x_{10} x_{16} x_{19}, x_{1} x_{2} x_{3} x_{10} x_{13} -  x_{4} x_{6} x_{7} x_{11} x_{17}, x_{1} x_{2} x_{4} x_{10} x_{13} -  x_{5} x_{7} x_{9} x_{16} x_{19}, x_{1} x_{3} x_{4} x_{10} x_{13} -  x_{2} x_{6} x_{7} 
x_{11} x_{17}, x_{2} x_{3} x_{4} x_{10} x_{13} -  x_{1} x_{6} x_{7} x_{11} x_{17}, x_{1} x_{2} x_{5} x_{10} x_{13} -  x_{4} x_{7} x_{9} x_{16} x_{19}, x_{1} x_{4} x_{5} x_{10} x_{13} -  x_{2} x_{7} x_{9} x_{16} x_{19}, x_{2} x_{4} x_{5} x_{10} x_{13} -  x_{1} x_{7} x_{9} x_{16} x_{19}, x_{1} x_{2} x_{6} x_{10} x_{13} -  x_{3} x_{4} x_{7} x_{11} x_{17}, x_{1} x_{3} x_{6} x_{10} x_{13} -  x_{2} x_{4} x_{7} x_{11} x_{17}, x_{2} x_{3} x_{6} x_{10} x_{13} -  x_{1} x_{4} x_{7} x_{11} x_{17}, x_{1} x_{5} x_{6} x_{10} x_{13} -  x_{7} x_{11} x_{14} x_{18} x_{19}, x_{1} x_{2} x_{7} x_{10} x_{13} -  x_{4} x_{5} x_{9} x_{16} x_{19}, x_{1} x_{3} x_{7} x_{10} x_{13} -  x_{9} x_{14} x_{16} x_{17} x_{18}, x_{1} x_{4} x_{7} x_{10} x_{13} -  x_{2} x_{5} x_{9} x_{16} x_{19}, x_{2} x_{4} x_{7} x_{10} x_{13} -  x_{1} x_{5} x_{9} x_{16} x_{19}, x_{3} x_{4} x_{7} x_{10} x_{13} -  x_{1} x_{2} x_{6} x_{11} x_{17}, x_{1} x_{5} x_{7} x_{10} x_{13} -  x_{6} x_{11} x_{14} x_{18} x_{19}, x_{2} x_{5} x_{7} x_{10} x_{13} -  x_{1} x_{4}x_{9}
 x_{16} x_{19}, x_{1} x_{6} x_{7} x_{10} x_{13} -  x_{5} x_{11} x_{14} x_{18} x_{19}, x_{2} x_{6} x_{7} x_{10} x_{13} -  x_{1} x_{3} x_{4} x_{11} x_{17}, x_{3} x_{6} x_{7} x_{10} x_{13} -  x_{1} x_{2} x_{4} x_{11} x_{17}, x_{1} x_{3} x_{9} x_{10} x_{13} -  x_{7} x_{14} x_{16} x_{17} x_{18}, x_{1} x_{4} x_{9} x_{10} x_{13} -  x_{2} x_{5} x_{7} x_{16} x_{19}, x_{1} x_{5} x_{9} x_{10} x_{13} -  x_{2} x_{4} x_{7} x_{16} x_{19}, x_{4} x_{5} x_{9} x_{10} x_{13} -  x_{1} x_{2} x_{7} x_{16} x_{19}, x_{1} x_{7} x_{9} x_{10} x_{13} -  x_{2} x_{4} x_{5} x_{16} x_{19}, x_{4} x_{7} x_{9} x_{10} x_{13} -  x_{1} x_{2} x_{5} x_{16} x_{19}, x_{5} x_{7} x_{9} x_{10} x_{13} -  x_{1} x_{2} x_{4} x_{16} x_{19}, x_{1} x_{2} x_{3} x_{11} x_{13} -  x_{4} x_{6} x_{7} x_{10} x_{17}, x_{1} x_{2} x_{4} x_{11} x_{13} -  x_{3} x_{6} x_{7} x_{10} x_{17}, x_{1} x_{3} x_{4} x_{11} x_{13} -  x_{2} x_{6} x_{7} x_{10} x_{17}, x_{2} x_{3} x_{4} x_{11} x_{13} -  x_{1} x_{6} x_{7} x_{10} x_{17}, x_{1} x_{2} x_{5} x_{11} x_{13} -  x_{3} x_{4}x_{12}
 x_{15} x_{16}, x_{1} x_{3} x_{5} x_{11} x_{13} -  x_{2} x_{4} x_{12} x_{15} x_{16}, x_{2} x_{3} x_{5} x_{11} x_{13} -  x_{1} x_{4} x_{12} x_{15} x_{16}, x_{1} x_{4} x_{5} x_{11} x_{13} -  x_{2} x_{3} x_{12} x_{15} x_{16}, x_{2} x_{4} x_{5} x_{11} x_{13} -  x_{1} x_{3} x_{12} x_{15} x_{16}, x_{3} x_{4} x_{5} x_{11} x_{13} -  x_{1} x_{2} x_{12} x_{15} x_{16}, x_{1} x_{2} x_{6} x_{11} x_{13} -  x_{3} x_{4} x_{7} x_{10} x_{17}, x_{1} x_{3} x_{6} x_{11} x_{13} -  x_{2} x_{4} x_{7} x_{10} x_{17}, x_{2} x_{3} x_{6} x_{11} x_{13} -  x_{1} x_{4} x_{7} x_{10} x_{17}, x_{1} x_{4} x_{6} x_{11} x_{13} -  x_{2} x_{3} x_{7} x_{10} x_{17}, x_{3} x_{4} x_{6} x_{11} x_{13} -  x_{1} x_{2} x_{7} x_{10} x_{17}, x_{1} x_{5} x_{6} x_{11} x_{13} -  x_{7} x_{10} x_{14} x_{18} x_{19}, x_{1} x_{2} x_{7} x_{11} x_{13} -  x_{3} x_{4} x_{6} x_{10} x_{17}, x_{1} x_{3} x_{7} x_{11} x_{13} -  x_{2} x_{4} x_{6} x_{10} x_{17}, x_{1} x_{4} x_{7} x_{11} x_{13} -  x_{2} x_{3} x_{6} x_{10} x_{17}, x_{2} x_{4} x_{7} x_{11} x_{13} -  x_{1} x_{3} x_
{6} x_{10} x_{17}, x_{3} x_{4} x_{7} x_{11} x_{13} -  x_{1} x_{2} x_{6} x_{10} x_{17}, x_{1} x_{5} x_{7} x_{11} x_{13} -  x_{6} x_{10} x_{14} x_{18} x_{19}, x_{1} x_{6} x_{7} x_{11} x_{13} -  x_{5} x_{10} x_{14} x_{18} x_{19}, x_{2} x_{6} x_{7} x_{11} x_{13} -  x_{1} x_{3} x_{4} x_{10} x_{17}, x_{3} x_{6} x_{7} x_{11} x_{13} -  x_{1} x_{2} x_{4} x_{10} x_{17}, x_{4} x_{6} x_{7} x_{11} x_{13} -  x_{1} x_{2} x_{3} x_{10} x_{17}, x_{5} x_{6} x_{7} x_{11} x_{13} -  x_{1} x_{10} x_{14} x_{18} x_{19}, x_{1} x_{3} x_{10} x_{11} x_{13} -  x_{2} x_{4} x_{6} x_{7} x_{17}, x_{1} x_{4} x_{10} x_{11} x_{13} -  x_{2} x_{3} x_{6} x_{7} x_{17}, x_{3} x_{4} x_{10} x_{11} x_{13} -  x_{1} x_{2} x_{6} x_{7} x_{17}, x_{1} x_{5} x_{10} x_{11} x_{13} -  x_{6} x_{7} x_{14} x_{18} x_{19}, x_{1} x_{6} x_{10} x_{11} x_{13} -  x_{5} x_{7} x_{14} x_{18} x_{19}, x_{3} x_{6} x_{10} x_{11} x_{13} -  x_{1} x_{2} x_{4} x_{7} x_{17}, x_{5} x_{6} x_{10} x_{11} x_{13} -  x_{1} x_{7} x_{14} x_{18} x_{19}, x_{1} x_{7} x_{10} x_{11} x_{13} - x_{5} 
x_{6} x_{14} x_{18} x_{19}, x_{4} x_{7} x_{10} x_{11} x_{13} -  x_{1} x_{2} x_{3} x_{6} x_{17}, x_{5} x_{7} x_{10} x_{11} x_{13} -  x_{1} x_{6} x_{14} x_{18} x_{19}, x_{6} x_{7} x_{10} x_{11} x_{13} -  x_{1} x_{5} x_{14} x_{18} x_{19}, x_{1} x_{2} x_{3} x_{12} x_{13} -  x_{4} x_{5} x_{11} x_{15} x_{16}, x_{1} x_{2} x_{4} x_{12} x_{13} -  x_{6} x_{9} x_{15} x_{17} x_{19}, x_{1} x_{3} x_{4} x_{12} x_{13} -  x_{2} x_{5} x_{11} x_{15} x_{16}, x_{2} x_{3} x_{4} x_{12} x_{13} -  x_{1} x_{5} x_{11} x_{15} x_{16}, x_{1} x_{2} x_{5} x_{12} x_{13} -  x_{3} x_{4} x_{11} x_{15} x_{16}, x_{1} x_{3} x_{5} x_{12} x_{13} -  x_{6} x_{9} x_{14} x_{15} x_{18}, x_{2} x_{3} x_{5} x_{12} x_{13} -  x_{1} x_{4} x_{11} x_{15} x_{16}, x_{1} x_{4} x_{5} x_{12} x_{13} -  x_{2} x_{3} x_{11} x_{15} x_{16}, x_{2} x_{4} x_{5} x_{12} x_{13} -  x_{1} x_{3} x_{11} x_{15} x_{16}, x_{3} x_{4} x_{5} x_{12} x_{13} -  x_{1} x_{2} x_{11} x_{15} x_{16}, x_{1} x_{2} x_{6} x_{12} x_{13} -  x_{4} x_{9} x_{15} x_{17} x_{19}, x_{1} x_{3} x_{6} x_{12} x_{
13} -  x_{5} x_{9} x_{14} x_{15} x_{18}, x_{1} x_{5} x_{6} x_{12} x_{13} -  x_{3} x_{9} x_{14} x_{15} x_{18}, x_{3} x_{5} x_{6} x_{12} x_{13} -  x_{1} x_{9} x_{14} x_{15} x_{18}, x_{1} x_{3} x_{9} x_{12} x_{13} -  x_{5} x_{6} x_{14} x_{15} x_{18}, x_{1} x_{4} x_{9} x_{12} x_{13} -  x_{2} x_{6} x_{15} x_{17} x_{19}, x_{1} x_{5} x_{9} x_{12} x_{13} -  x_{3} x_{6} x_{14} x_{15} x_{18}, x_{3} x_{5} x_{9} x_{12} x_{13} -  x_{1} x_{6} x_{14} x_{15} x_{18}, x_{1} x_{6} x_{9} x_{12} x_{13} -  x_{2} x_{4} x_{15} x_{17} x_{19}, x_{3} x_{6} x_{9} x_{12} x_{13} -  x_{1} x_{5} x_{14} x_{15} x_{18}, x_{5} x_{6} x_{9} x_{12} x_{13} -  x_{1} x_{3} x_{14} x_{15} x_{18}, x_{1} x_{3} x_{11} x_{12} x_{13} -  x_{2} x_{4} x_{5} x_{15} x_{16}, x_{1} x_{4} x_{11} x_{12} x_{13} -  x_{2} x_{3} x_{5} x_{15} x_{16}, x_{3} x_{4} x_{11} x_{12} x_{13} -  x_{1} x_{2} x_{5} x_{15} x_{16}, x_{1} x_{5} x_{11} x_{12} x_{13} -  x_{2} x_{3} x_{4} x_{15} x_{16}, x_{3} x_{5} x_{11} x_{12} x_{13} -  x_{1} x_{2} x_{4} x_{15} x_{16}, x_{4} x_{5} x_{
11} x_{12} x_{13} -  x_{1} x_{2} x_{3} x_{15} x_{16}, x_{1} x_{3} x_{4} x_{6} x_{14} -  x_{7} x_{8} x_{9} x_{12} x_{17}, x_{1} x_{3} x_{5} x_{6} x_{14} -  x_{9} x_{12} x_{13} x_{15} x_{18}, x_{1} x_{3} x_{4} x_{7} x_{14} -  x_{6} x_{8} x_{9} x_{12} x_{17}, x_{1} x_{4} x_{5} x_{7} x_{14} -  x_{8} x_{11} x_{12} x_{16} x_{19}, x_{1} x_{3} x_{6} x_{7} x_{14} -  x_{4} x_{8} x_{9} x_{12} x_{17}, x_{1} x_{4} x_{6} x_{7} x_{14} -  x_{3} x_{8} x_{9} x_{12} x_{17}, x_{3} x_{4} x_{6} x_{7} x_{14} -  x_{1} x_{8} x_{9} x_{12} x_{17}, x_{1} x_{5} x_{6} x_{7} x_{14} -  x_{10} x_{11} x_{13} x_{18} x_{19}, x_{1} x_{3} x_{4} x_{8} x_{14} -  x_{6} x_{7} x_{9} x_{12} x_{17}, x_{1} x_{3} x_{5} x_{8} x_{14} -  x_{4} x_{9} x_{10} x_{15} x_{16}, x_{1} x_{4} x_{5} x_{8} x_{14} -  x_{7} x_{11} x_{12} x_{16} x_{19}, x_{1} x_{3} x_{6} x_{8} x_{14} -  x_{4} x_{7} x_{9} x_{12} x_{17}, x_{1} x_{4} x_{6} x_{8} x_{14} -  x_{10} x_{11} x_{15} x_{17} x_{19}, x_{3} x_{4} x_{6} x_{8} x_{14} -  x_{1} x_{7} x_{9} x_{12} x_{17}, x_{1} x_{4} x_{7} 
x_{8} x_{14} -  x_{5} x_{11} x_{12} x_{16} x_{19}, x_{1} x_{5} x_{7} x_{8} x_{14} -  x_{4} x_{11} x_{12} x_{16} x_{19}, x_{1} x_{6} x_{7} x_{8} x_{14} -  x_{3} x_{4} x_{9} x_{12} x_{17}, x_{4} x_{6} x_{7} x_{8} x_{14} -  x_{1} x_{3} x_{9} x_{12} x_{17}, x_{1} x_{3} x_{4} x_{9} x_{14} -  x_{6} x_{7} x_{8} x_{12} x_{17}, x_{1} x_{3} x_{5} x_{9} x_{14} -  x_{6} x_{12} x_{13} x_{15} x_{18}, x_{1} x_{4} x_{5} x_{9} x_{14} -  x_{3} x_{8} x_{10} x_{15} x_{16}, x_{1} x_{3} x_{6} x_{9} x_{14} -  x_{5} x_{12} x_{13} x_{15} x_{18}, x_{1} x_{5} x_{6} x_{9} x_{14} -  x_{3} x_{12} x_{13} x_{15} x_{18}, x_{1} x_{3} x_{7} x_{9} x_{14} -  x_{10} x_{13} x_{16} x_{17} x_{18}, x_{1} x_{4} x_{7} x_{9} x_{14} -  x_{3} x_{6} x_{8} x_{12} x_{17}, x_{3} x_{4} x_{7} x_{9} x_{14} -  x_{1} x_{6} x_{8} x_{12} x_{17}, x_{1} x_{6} x_{7} x_{9} x_{14} -  x_{3} x_{4} x_{8} x_{12} x_{17}, x_{3} x_{6} x_{7} x_{9} x_{14} -  x_{1} x_{4} x_{8} x_{12} x_{17}, x_{1} x_{3} x_{8} x_{9} x_{14} -  x_{4} x_{6} x_{7} x_{12} x_{17}, x_{1} x_{4} x_{8} x_{9}
 x_{14} -  x_{3} x_{6} x_{7} x_{12} x_{17}, x_{3} x_{4} x_{8} x_{9} x_{14} -  x_{1} x_{6} x_{7} x_{12} x_{17}, x_{1} x_{5} x_{8} x_{9} x_{14} -  x_{3} x_{4} x_{10} x_{15} x_{16}, x_{3} x_{5} x_{8} x_{9} x_{14} -  x_{1} x_{4} x_{10} x_{15} x_{16}, x_{4} x_{5} x_{8} x_{9} x_{14} -  x_{1} x_{3} x_{10} x_{15} x_{16}, x_{1} x_{6} x_{8} x_{9} x_{14} -  x_{3} x_{4} x_{7} x_{12} x_{17}, x_{3} x_{6} x_{8} x_{9} x_{14} -  x_{1} x_{4} x_{7} x_{12} x_{17}, x_{1} x_{7} x_{8} x_{9} x_{14} -  x_{3} x_{4} x_{6} x_{12} x_{17}, x_{4} x_{7} x_{8} x_{9} x_{14} -  x_{1} x_{3} x_{6} x_{12} x_{17}, x_{6} x_{7} x_{8} x_{9} x_{14} -  x_{1} x_{3} x_{4} x_{12} x_{17}, x_{1} x_{3} x_{4} x_{10} x_{14} -  x_{5} x_{8} x_{9} x_{15} x_{16}, x_{1} x_{3} x_{5} x_{10} x_{14} -  x_{4} x_{8} x_{9} x_{15} x_{16}, x_{1} x_{4} x_{5} x_{10} x_{14} -  x_{3} x_{8} x_{9} x_{15} x_{16}, x_{1} x_{4} x_{6} x_{10} x_{14} -  x_{8} x_{11} x_{15} x_{17} x_{19}, x_{1} x_{5} x_{6} x_{10} x_{14} -  x_{7} x_{11} x_{13} x_{18} x_{19}, x_{1} x_{3} x_{7} x_{10} x_{
14} -  x_{9} x_{13} x_{16} x_{17} x_{18}, x_{1} x_{5} x_{7} x_{10} x_{14} -  x_{6} x_{11} x_{13} x_{18} x_{19}, x_{1} x_{6} x_{7} x_{10} x_{14} -  x_{5} x_{11} x_{13} x_{18} x_{19}, x_{1} x_{3} x_{8} x_{10} x_{14} -  x_{4} x_{5} x_{9} x_{15} x_{16}, x_{1} x_{4} x_{8} x_{10} x_{14} -  x_{6} x_{11} x_{15} x_{17} x_{19}, x_{3} x_{4} x_{8} x_{10} x_{14} -  x_{1} x_{5} x_{9} x_{15} x_{16}, x_{1} x_{5} x_{8} x_{10} x_{14} -  x_{3} x_{4} x_{9} x_{15} x_{16}, x_{3} x_{5} x_{8} x_{10} x_{14} -  x_{1} x_{4} x_{9} x_{15} x_{16}, x_{4} x_{5} x_{8} x_{10} x_{14} -  x_{1} x_{3} x_{9} x_{15} x_{16}, x_{1} x_{6} x_{8} x_{10} x_{14} -  x_{4} x_{11} x_{15} x_{17} x_{19}, x_{1} x_{3} x_{9} x_{10} x_{14} -  x_{7} x_{13} x_{16} x_{17} x_{18}, x_{1} x_{4} x_{9} x_{10} x_{14} -  x_{3} x_{5} x_{8} x_{15} x_{16}, x_{3} x_{4} x_{9} x_{10} x_{14} -  x_{1} x_{5} x_{8} x_{15} x_{16}, x_{1} x_{5} x_{9} x_{10} x_{14} -  x_{3} x_{4} x_{8} x_{15} x_{16}, x_{3} x_{5} x_{9} x_{10} x_{14} -  x_{1} x_{4} x_{8} x_{15} x_{16}, x_{4} x_{5} x_{9} x_
{10} x_{14} -  x_{1} x_{3} x_{8} x_{15} x_{16}, x_{1} x_{7} x_{9} x_{10} x_{14} -  x_{3} x_{13} x_{16} x_{17} x_{18}, x_{1} x_{8} x_{9} x_{10} x_{14} -  x_{3} x_{4} x_{5} x_{15} x_{16}, x_{3} x_{8} x_{9} x_{10} x_{14} -  x_{1} x_{4} x_{5} x_{15} x_{16}, x_{4} x_{8} x_{9} x_{10} x_{14} -  x_{1} x_{3} x_{5} x_{15} x_{16}, x_{5} x_{8} x_{9} x_{10} x_{14} -  x_{1} x_{3} x_{4} x_{15} x_{16}, x_{1} x_{4} x_{5} x_{11} x_{14} -  x_{7} x_{8} x_{12} x_{16} x_{19}, x_{1} x_{5} x_{6} x_{11} x_{14} -  x_{7} x_{10} x_{13} x_{18} x_{19}, x_{1} x_{4} x_{7} x_{11} x_{14} -  x_{5} x_{8} x_{12} x_{16} x_{19}, x_{1} x_{5} x_{7} x_{11} x_{14} -  x_{6} x_{10} x_{13} x_{18} x_{19}, x_{4} x_{5} x_{7} x_{11} x_{14} -  x_{1} x_{8} x_{12} x_{16} x_{19}, x_{1} x_{6} x_{7} x_{11} x_{14} -  x_{5} x_{10} x_{13} x_{18} x_{19}, x_{5} x_{6} x_{7} x_{11} x_{14} -  x_{1} x_{10} x_{13} x_{18} x_{19}, x_{1} x_{4} x_{8} x_{11} x_{14} -  x_{6} x_{10} x_{15} x_{17} x_{19}, x_{1} x_{5} x_{8} x_{11} x_{14} -  x_{4} x_{7} x_{12} x_{16} x_{19}, x_{4} x_
{5} x_{8} x_{11} x_{14} -  x_{1} x_{7} x_{12} x_{16} x_{19}, x_{1} x_{6} x_{8} x_{11} x_{14} -  x_{4} x_{10} x_{15} x_{17} x_{19}, x_{1} x_{7} x_{8} x_{11} x_{14} -  x_{4} x_{5} x_{12} x_{16} x_{19}, x_{4} x_{7} x_{8} x_{11} x_{14} -  x_{1} x_{5} x_{12} x_{16} x_{19}, x_{5} x_{7} x_{8} x_{11} x_{14} -  x_{1} x_{4} x_{12} x_{16} x_{19}, x_{1} x_{4} x_{10} x_{11} x_{14} -  x_{6} x_{8} x_{15} x_{17} x_{19}, x_{1} x_{5} x_{10} x_{11} x_{14} -  x_{6} x_{7} x_{13} x_{18} x_{19}, x_{1} x_{6} x_{10} x_{11} x_{14} -  x_{5} x_{7} x_{13} x_{18} x_{19}, x_{5} x_{6} x_{10} x_{11} x_{14} -  x_{1} x_{7} x_{13} x_{18} x_{19}, x_{1} x_{7} x_{10} x_{11} x_{14} -  x_{5} x_{6} x_{13} x_{18} x_{19}, x_{5} x_{7} x_{10} x_{11} x_{14} -  x_{1} x_{6} x_{13} x_{18} x_{19}, x_{6} x_{7} x_{10} x_{11} x_{14} -  x_{1} x_{5} x_{13} x_{18} x_{19}, x_{1} x_{8} x_{10} x_{11} x_{14} -  x_{4} x_{6} x_{15} x_{17} x_{19}, x_{4} x_{8} x_{10} x_{11} x_{14} -  x_{1} x_{6} x_{15} x_{17} x_{19}, x_{6} x_{8} x_{10} x_{11} x_{14} -  x_{1} x_{4} x_{15} 
x_{17} x_{19}, x_{1} x_{3} x_{4} x_{12} x_{14} -  x_{6} x_{7} x_{8} x_{9} x_{17}, x_{1} x_{3} x_{5} x_{12} x_{14} -  x_{6} x_{9} x_{13} x_{15} x_{18}, x_{1} x_{4} x_{5} x_{12} x_{14} -  x_{7} x_{8} x_{11} x_{16} x_{19}, x_{1} x_{3} x_{6} x_{12} x_{14} -  x_{5} x_{9} x_{13} x_{15} x_{18}, x_{1} x_{4} x_{6} x_{12} x_{14} -  x_{3} x_{7} x_{8} x_{9} x_{17}, x_{3} x_{4} x_{6} x_{12} x_{14} -  x_{1} x_{7} x_{8} x_{9} x_{17}, x_{1} x_{5} x_{6} x_{12} x_{14} -  x_{3} x_{9} x_{13} x_{15} x_{18}, x_{3} x_{5} x_{6} x_{12} x_{14} -  x_{1} x_{9} x_{13} x_{15} x_{18}, x_{1} x_{3} x_{7} x_{12} x_{14} -  x_{4} x_{6} x_{8} x_{9} x_{17}, x_{1} x_{4} x_{7} x_{12} x_{14} -  x_{5} x_{8} x_{11} x_{16} x_{19}, x_{3} x_{4} x_{7} x_{12} x_{14} -  x_{1} x_{6} x_{8} x_{9} x_{17}, x_{1} x_{5} x_{7} x_{12} x_{14} -  x_{4} x_{8} x_{11} x_{16} x_{19}, x_{4} x_{5} x_{7} x_{12} x_{14} -  x_{1} x_{8} x_{11} x_{16} x_{19}, x_{1} x_{6} x_{7} x_{12} x_{14} -  x_{3} x_{4} x_{8} x_{9} x_{17}, x_{3} x_{6} x_{7} x_{12} x_{14} -  x_{1} x_{4} x_{8} x_
{9} x_{17}, x_{4} x_{6} x_{7} x_{12} x_{14} -  x_{1} x_{3} x_{8} x_{9} x_{17}, x_{1} x_{3} x_{8} x_{12} x_{14} -  x_{4} x_{6} x_{7} x_{9} x_{17}, x_{1} x_{4} x_{8} x_{12} x_{14} -  x_{5} x_{7} x_{11} x_{16} x_{19}, x_{3} x_{4} x_{8} x_{12} x_{14} -  x_{1} x_{6} x_{7} x_{9} x_{17}, x_{1} x_{5} x_{8} x_{12} x_{14} -  x_{4} x_{7} x_{11} x_{16} x_{19}, x_{4} x_{5} x_{8} x_{12} x_{14} -  x_{1} x_{7} x_{11} x_{16} x_{19}, x_{1} x_{6} x_{8} x_{12} x_{14} -  x_{3} x_{4} x_{7} x_{9} x_{17}, x_{3} x_{6} x_{8} x_{12} x_{14} -  x_{1} x_{4} x_{7} x_{9} x_{17}, x_{1} x_{7} x_{8} x_{12} x_{14} -  x_{4} x_{5} x_{11} x_{16} x_{19}, x_{4} x_{7} x_{8} x_{12} x_{14} -  x_{1} x_{5} x_{11} x_{16} x_{19}, x_{5} x_{7} x_{8} x_{12} x_{14} -  x_{1} x_{4} x_{11} x_{16} x_{19}, x_{6} x_{7} x_{8} x_{12} x_{14} -  x_{1} x_{3} x_{4} x_{9} x_{17}, x_{1} x_{3} x_{9} x_{12} x_{14} -  x_{5} x_{6} x_{13} x_{15} x_{18}, x_{1} x_{4} x_{9} x_{12} x_{14} -  x_{3} x_{6} x_{7} x_{8} x_{17}, x_{3} x_{4} x_{9} x_{12} x_{14} -  x_{1} x_{6} x_{7} x_{8} 
x_{17}, x_{1} x_{5} x_{9} x_{12} x_{14} -  x_{3} x_{6} x_{13} x_{15} x_{18}, x_{3} x_{5} x_{9} x_{12} x_{14} -  x_{1} x_{6} x_{13} x_{15} x_{18}, x_{1} x_{6} x_{9} x_{12} x_{14} -  x_{3} x_{5} x_{13} x_{15} x_{18}, x_{3} x_{6} x_{9} x_{12} x_{14} -  x_{1} x_{5} x_{13} x_{15} x_{18}, x_{5} x_{6} x_{9} x_{12} x_{14} -  x_{1} x_{3} x_{13} x_{15} x_{18}, x_{1} x_{7} x_{9} x_{12} x_{14} -  x_{3} x_{4} x_{6} x_{8} x_{17}, x_{4} x_{7} x_{9} x_{12} x_{14} -  x_{1} x_{3} x_{6} x_{8} x_{17}, x_{6} x_{7} x_{9} x_{12} x_{14} -  x_{1} x_{3} x_{4} x_{8} x_{17}, x_{1} x_{8} x_{9} x_{12} x_{14} -  x_{3} x_{4} x_{6} x_{7} x_{17}, x_{3} x_{8} x_{9} x_{12} x_{14} -  x_{1} x_{4} x_{6} x_{7} x_{17}, x_{4} x_{8} x_{9} x_{12} x_{14} -  x_{1} x_{3} x_{6} x_{7} x_{17}, x_{6} x_{8} x_{9} x_{12} x_{14} -  x_{1} x_{3} x_{4} x_{7} x_{17}, x_{7} x_{8} x_{9} x_{12} x_{14} -  x_{1} x_{3} x_{4} x_{6} x_{17}, x_{1} x_{4} x_{11} x_{12} x_{14} -  x_{5} x_{7} x_{8} x_{16} x_{19}, x_{1} x_{5} x_{11} x_{12} x_{14} -  x_{4} x_{7} x_{8} x_{16}x_{19}
, x_{4} x_{5} x_{11} x_{12} x_{14} -  x_{1} x_{7} x_{8} x_{16} x_{19}, x_{1} x_{7} x_{11} x_{12} x_{14} -  x_{4} x_{5} x_{8} x_{16} x_{19}, x_{4} x_{7} x_{11} x_{12} x_{14} -  x_{1} x_{5} x_{8} x_{16} x_{19}, x_{5} x_{7} x_{11} x_{12} x_{14} -  x_{1} x_{4} x_{8} x_{16} x_{19}, x_{1} x_{8} x_{11} x_{12} x_{14} -  x_{4} x_{5} x_{7} x_{16} x_{19}, x_{4} x_{8} x_{11} x_{12} x_{14} -  x_{1} x_{5} x_{7} x_{16} x_{19}, x_{5} x_{8} x_{11} x_{12} x_{14} -  x_{1} x_{4} x_{7} x_{16} x_{19}, x_{7} x_{8} x_{11} x_{12} x_{14} -  x_{1} x_{4} x_{5} x_{16} x_{19}, x_{1} x_{3} x_{5} x_{13} x_{14} -  x_{6} x_{9} x_{12} x_{15} x_{18}, x_{1} x_{3} x_{6} x_{13} x_{14} -  x_{5} x_{9} x_{12} x_{15} x_{18}, x_{1} x_{5} x_{6} x_{13} x_{14} -  x_{7} x_{10} x_{11} x_{18} x_{19}, x_{3} x_{5} x_{6} x_{13} x_{14} -  x_{1} x_{9} x_{12} x_{15} x_{18}, x_{1} x_{5} x_{7} x_{13} x_{14} -  x_{6} x_{10} x_{11} x_{18} x_{19}, x_{1} x_{6} x_{7} x_{13} x_{14} -  x_{5} x_{10} x_{11} x_{18} x_{19}, x_{5} x_{6} x_{7} x_{13} x_{14} -  x_{1} x_{10} x_{
11} x_{18} x_{19}, x_{1} x_{3} x_{9} x_{13} x_{14} -  x_{7} x_{10} x_{16} x_{17} x_{18}, x_{1} x_{5} x_{9} x_{13} x_{14} -  x_{3} x_{6} x_{12} x_{15} x_{18}, x_{3} x_{5} x_{9} x_{13} x_{14} -  x_{1} x_{6} x_{12} x_{15} x_{18}, x_{1} x_{6} x_{9} x_{13} x_{14} -  x_{3} x_{5} x_{12} x_{15} x_{18}, x_{3} x_{6} x_{9} x_{13} x_{14} -  x_{1} x_{5} x_{12} x_{15} x_{18}, x_{5} x_{6} x_{9} x_{13} x_{14} -  x_{1} x_{3} x_{12} x_{15} x_{18}, x_{1} x_{7} x_{9} x_{13} x_{14} -  x_{3} x_{10} x_{16} x_{17} x_{18}, x_{1} x_{3} x_{10} x_{13} x_{14} -  x_{7} x_{9} x_{16} x_{17} x_{18}, x_{1} x_{5} x_{10} x_{13} x_{14} -  x_{6} x_{7} x_{11} x_{18} x_{19}, x_{1} x_{6} x_{10} x_{13} x_{14} -  x_{5} x_{7} x_{11} x_{18} x_{19}, x_{5} x_{6} x_{10} x_{13} x_{14} -  x_{1} x_{7} x_{11} x_{18} x_{19}, x_{1} x_{7} x_{10} x_{13} x_{14} -  x_{5} x_{6} x_{11} x_{18} x_{19}, x_{5} x_{7} x_{10} x_{13} x_{14} -  x_{1} x_{6} x_{11} x_{18} x_{19}, x_{6} x_{7} x_{10} x_{13} x_{14} -  x_{1} x_{5} x_{11} x_{18} x_{19}, x_{1} x_{9} x_{10} x_{13}x_{
14} -  x_{3} x_{7} x_{16} x_{17} x_{18}, x_{3} x_{9} x_{10} x_{13} x_{14} -  x_{1} x_{7} x_{16} x_{17} x_{18}, x_{7} x_{9} x_{10} x_{13} x_{14} -  x_{1} x_{3} x_{16} x_{17} x_{18}, x_{1} x_{5} x_{11} x_{13} x_{14} -  x_{6} x_{7} x_{10} x_{18} x_{19}, x_{1} x_{6} x_{11} x_{13} x_{14} -  x_{5} x_{7} x_{10} x_{18} x_{19}, x_{5} x_{6} x_{11} x_{13} x_{14} -  x_{1} x_{7} x_{10} x_{18} x_{19}, x_{1} x_{7} x_{11} x_{13} x_{14} -  x_{5} x_{6} x_{10} x_{18} x_{19}, x_{5} x_{7} x_{11} x_{13} x_{14} -  x_{1} x_{6} x_{10} x_{18} x_{19}, x_{6} x_{7} x_{11} x_{13} x_{14} -  x_{1} x_{5} x_{10} x_{18} x_{19}, x_{1} x_{10} x_{11} x_{13} x_{14} -  x_{5} x_{6} x_{7} x_{18} x_{19}, x_{5} x_{10} x_{11} x_{13} x_{14} -  x_{1} x_{6} x_{7} x_{18} x_{19}, x_{6} x_{10} x_{11} x_{13} x_{14} -  x_{1} x_{5} x_{7} x_{18} x_{19}, x_{7} x_{10} x_{11} x_{13} x_{14} -  x_{1} x_{5} x_{6} x_{18} x_{19}, x_{1} x_{3} x_{12} x_{13} x_{14} -  x_{5} x_{6} x_{9} x_{15} x_{18}, x_{1} x_{5} x_{12} x_{13} x_{14} -  x_{3} x_{6} x_{9} x_{15} x_{18}, x_{3}
 x_{5} x_{12} x_{13} x_{14} -  x_{1} x_{6} x_{9} x_{15} x_{18}, x_{1} x_{6} x_{12} x_{13} x_{14} -  x_{3} x_{5} x_{9} x_{15} x_{18}, x_{3} x_{6} x_{12} x_{13} x_{14} -  x_{1} x_{5} x_{9} x_{15} x_{18}, x_{5} x_{6} x_{12} x_{13} x_{14} -  x_{1} x_{3} x_{9} x_{15} x_{18}, x_{1} x_{9} x_{12} x_{13} x_{14} -  x_{3} x_{5} x_{6} x_{15} x_{18}, x_{3} x_{9} x_{12} x_{13} x_{14} -  x_{1} x_{5} x_{6} x_{15} x_{18}, x_{5} x_{9} x_{12} x_{13} x_{14} -  x_{1} x_{3} x_{6} x_{15} x_{18}, x_{6} x_{9} x_{12} x_{13} x_{14} -  x_{1} x_{3} x_{5} x_{15} x_{18}, x_{1} x_{11} x_{12} x_{13} x_{14} -  x_{15} x_{16} x_{17} x_{18} x_{19}, x_{1} x_{2} x_{3} x_{4} x_{15} -  x_{5} x_{11} x_{12} x_{13} x_{16}, x_{1} x_{2} x_{3} x_{5} x_{15} -  x_{6} x_{8} x_{10} x_{11} x_{18}, x_{1} x_{2} x_{4} x_{5} x_{15} -  x_{3} x_{11} x_{12} x_{13} x_{16}, x_{1} x_{3} x_{4} x_{5} x_{15} -  x_{8} x_{9} x_{10} x_{14} x_{16}, x_{1} x_{2} x_{3} x_{6} x_{15} -  x_{5} x_{8} x_{10} x_{11} x_{18}, x_{1} x_{2} x_{4} x_{6} x_{15} -  x_{9} x_{12} x_{13} x_{17} 
x_{19}, x_{1} x_{2} x_{5} x_{6} x_{15} -  x_{3} x_{8} x_{10} x_{11} x_{18}, x_{1} x_{3} x_{5} x_{6} x_{15} -  x_{9} x_{12} x_{13} x_{14} x_{18}, x_{2} x_{3} x_{5} x_{6} x_{15} -  x_{1} x_{8} x_{10} x_{11} x_{18}, x_{1} x_{3} x_{4} x_{8} x_{15} -  x_{5} x_{9} x_{10} x_{14} x_{16}, x_{1} x_{2} x_{5} x_{8} x_{15} -  x_{3} x_{6} x_{10} x_{11} x_{18}, x_{1} x_{3} x_{5} x_{8} x_{15} -  x_{2} x_{6} x_{10} x_{11} x_{18}, x_{1} x_{2} x_{6} x_{8} x_{15} -  x_{3} x_{5} x_{10} x_{11} x_{18}, x_{1} x_{3} x_{6} x_{8} x_{15} -  x_{2} x_{5} x_{10} x_{11} x_{18}, x_{1} x_{4} x_{6} x_{8} x_{15} -  x_{10} x_{11} x_{14} x_{17} x_{19}, x_{1} x_{5} x_{6} x_{8} x_{15} -  x_{2} x_{3} x_{10} x_{11} x_{18}, x_{2} x_{5} x_{6} x_{8} x_{15} -  x_{1} x_{3} x_{10} x_{11} x_{18}, x_{3} x_{5} x_{6} x_{8} x_{15} -  x_{1} x_{2} x_{10} x_{11} x_{18}, x_{1} x_{2} x_{4} x_{9} x_{15} -  x_{6} x_{12} x_{13} x_{17} x_{19}, x_{1} x_{3} x_{4} x_{9} x_{15} -  x_{5} x_{8} x_{10} x_{14} x_{16}, x_{1} x_{3} x_{5} x_{9} x_{15} -  x_{6} x_{12} x_{13} x_{14}
 x_{18}, x_{1} x_{4} x_{5} x_{9} x_{15} -  x_{3} x_{8} x_{10} x_{14} x_{16}, x_{3} x_{4} x_{5} x_{9} x_{15} -  x_{1} x_{8} x_{10} x_{14} x_{16}, x_{1} x_{2} x_{6} x_{9} x_{15} -  x_{4} x_{12} x_{13} x_{17} x_{19}, x_{1} x_{3} x_{6} x_{9} x_{15} -  x_{5} x_{12} x_{13} x_{14} x_{18}, x_{1} x_{4} x_{6} x_{9} x_{15} -  x_{2} x_{12} x_{13} x_{17} x_{19}, x_{1} x_{5} x_{6} x_{9} x_{15} -  x_{3} x_{12} x_{13} x_{14} x_{18}, x_{1} x_{2} x_{8} x_{9} x_{15} -  x_{10} x_{16} x_{17} x_{18} x_{19}, x_{1} x_{3} x_{8} x_{9} x_{15} -  x_{4} x_{5} x_{10} x_{14} x_{16}, x_{1} x_{4} x_{8} x_{9} x_{15} -  x_{3} x_{5} x_{10} x_{14} x_{16}, x_{3} x_{4} x_{8} x_{9} x_{15} -  x_{1} x_{5} x_{10} x_{14} x_{16}, x_{1} x_{5} x_{8} x_{9} x_{15} -  x_{3} x_{4} x_{10} x_{14} x_{16}, x_{3} x_{5} x_{8} x_{9} x_{15} -  x_{1} x_{4} x_{10} x_{14} x_{16}, x_{1} x_{2} x_{3} x_{10} x_{15} -  x_{5} x_{6} x_{8} x_{11} x_{18}, x_{1} x_{3} x_{4} x_{10} x_{15} -  x_{5} x_{8} x_{9} x_{14} x_{16}, x_{1} x_{2} x_{5} x_{10} x_{15} -  x_{3} x_{6} x_{8}x_{
11} x_{18}, x_{1} x_{3} x_{5} x_{10} x_{15} -  x_{2} x_{6} x_{8} x_{11} x_{18}, x_{2} x_{3} x_{5} x_{10} x_{15} -  x_{1} x_{6} x_{8} x_{11} x_{18}, x_{1} x_{4} x_{5} x_{10} x_{15} -  x_{3} x_{8} x_{9} x_{14} x_{16}, x_{3} x_{4} x_{5} x_{10} x_{15} -  x_{1} x_{8} x_{9} x_{14} x_{16}, x_{1} x_{2} x_{6} x_{10} x_{15} -  x_{3} x_{5} x_{8} x_{11} x_{18}, x_{1} x_{3} x_{6} x_{10} x_{15} -  x_{2} x_{5} x_{8} x_{11} x_{18}, x_{2} x_{3} x_{6} x_{10} x_{15} -  x_{1} x_{5} x_{8} x_{11} x_{18}, x_{1} x_{4} x_{6} x_{10} x_{15} -  x_{8} x_{11} x_{14} x_{17} x_{19}, x_{1} x_{2} x_{8} x_{10} x_{15} -  x_{9} x_{16} x_{17} x_{18} x_{19}, x_{1} x_{3} x_{8} x_{10} x_{15} -  x_{2} x_{5} x_{6} x_{11} x_{18}, x_{1} x_{4} x_{8} x_{10} x_{15} -  x_{6} x_{11} x_{14} x_{17} x_{19}, x_{3} x_{4} x_{8} x_{10} x_{15} -  x_{1} x_{5} x_{9} x_{14} x_{16}, x_{1} x_{5} x_{8} x_{10} x_{15} -  x_{2} x_{3} x_{6} x_{11} x_{18}, x_{2} x_{5} x_{8} x_{10} x_{15} -  x_{1} x_{3} x_{6} x_{11} x_{18}, x_{3} x_{5} x_{8} x_{10} x_{15} -  x_{1} x_{2} x_{6} 
x_{11} x_{18}, x_{1} x_{6} x_{8} x_{10} x_{15} -  x_{4} x_{11} x_{14} x_{17} x_{19}, x_{2} x_{6} x_{8} x_{10} x_{15} -  x_{1} x_{3} x_{5} x_{11} x_{18}, x_{3} x_{6} x_{8} x_{10} x_{15} -  x_{1} x_{2} x_{5} x_{11} x_{18}, x_{1} x_{2} x_{9} x_{10} x_{15} -  x_{8} x_{16} x_{17} x_{18} x_{19}, x_{1} x_{4} x_{9} x_{10} x_{15} -  x_{3} x_{5} x_{8} x_{14} x_{16}, x_{1} x_{5} x_{9} x_{10} x_{15} -  x_{3} x_{4} x_{8} x_{14} x_{16}, x_{4} x_{5} x_{9} x_{10} x_{15} -  x_{1} x_{3} x_{8} x_{14} x_{16}, x_{1} x_{8} x_{9} x_{10} x_{15} -  x_{2} x_{16} x_{17} x_{18} x_{19}, x_{4} x_{8} x_{9} x_{10} x_{15} -  x_{1} x_{3} x_{5} x_{14} x_{16}, x_{5} x_{8} x_{9} x_{10} x_{15} -  x_{1} x_{3} x_{4} x_{14} x_{16}, x_{1} x_{2} x_{3} x_{11} x_{15} -  x_{5} x_{6} x_{8} x_{10} x_{18}, x_{1} x_{2} x_{4} x_{11} x_{15} -  x_{3} x_{5} x_{12} x_{13} x_{16}, x_{1} x_{3} x_{4} x_{11} x_{15} -  x_{2} x_{5} x_{12} x_{13} x_{16}, x_{2} x_{3} x_{4} x_{11} x_{15} -  x_{1} x_{5} x_{12} x_{13} x_{16}, x_{1} x_{2} x_{5} x_{11} x_{15} -  x_{3} x_{6} 
x_{8} x_{10} x_{18}, x_{1} x_{3} x_{5} x_{11} x_{15} -  x_{2} x_{6} x_{8} x_{10} x_{18}, x_{2} x_{3} x_{5} x_{11} x_{15} -  x_{1} x_{6} x_{8} x_{10} x_{18}, x_{1} x_{4} x_{5} x_{11} x_{15} -  x_{2} x_{3} x_{12} x_{13} x_{16}, x_{2} x_{4} x_{5} x_{11} x_{15} -  x_{1} x_{3} x_{12} x_{13} x_{16}, x_{3} x_{4} x_{5} x_{11} x_{15} -  x_{1} x_{2} x_{12} x_{13} x_{16}, x_{1} x_{2} x_{6} x_{11} x_{15} -  x_{3} x_{5} x_{8} x_{10} x_{18}, x_{1} x_{3} x_{6} x_{11} x_{15} -  x_{2} x_{5} x_{8} x_{10} x_{18}, x_{2} x_{3} x_{6} x_{11} x_{15} -  x_{1} x_{5} x_{8} x_{10} x_{18}, x_{1} x_{4} x_{6} x_{11} x_{15} -  x_{8} x_{10} x_{14} x_{17} x_{19}, x_{1} x_{5} x_{6} x_{11} x_{15} -  x_{2} x_{3} x_{8} x_{10} x_{18}, x_{2} x_{5} x_{6} x_{11} x_{15} -  x_{1} x_{3} x_{8} x_{10} x_{18}, x_{1} x_{2} x_{8} x_{11} x_{15} -  x_{3} x_{5} x_{6} x_{10} x_{18}, x_{1} x_{3} x_{8} x_{11} x_{15} -  x_{2} x_{5} x_{6} x_{10} x_{18}, x_{1} x_{4} x_{8} x_{11} x_{15} -  x_{6} x_{10} x_{14} x_{17} x_{19}, x_{1} x_{5} x_{8} x_{11} x_{15} -  x_{2}x_{
3} x_{6} x_{10} x_{18}, x_{2} x_{5} x_{8} x_{11} x_{15} -  x_{1} x_{3} x_{6} x_{10} x_{18}, x_{3} x_{5} x_{8} x_{11} x_{15} -  x_{1} x_{2} x_{6} x_{10} x_{18}, x_{1} x_{6} x_{8} x_{11} x_{15} -  x_{4} x_{10} x_{14} x_{17} x_{19}, x_{2} x_{6} x_{8} x_{11} x_{15} -  x_{1} x_{3} x_{5} x_{10} x_{18}, x_{3} x_{6} x_{8} x_{11} x_{15} -  x_{1} x_{2} x_{5} x_{10} x_{18}, x_{4} x_{6} x_{8} x_{11} x_{15} -  x_{1} x_{10} x_{14} x_{17} x_{19}, x_{5} x_{6} x_{8} x_{11} x_{15} -  x_{1} x_{2} x_{3} x_{10} x_{18}, x_{1} x_{2} x_{10} x_{11} x_{15} -  x_{3} x_{5} x_{6} x_{8} x_{18}, x_{1} x_{4} x_{10} x_{11} x_{15} -  x_{6} x_{8} x_{14} x_{17} x_{19}, x_{1} x_{5} x_{10} x_{11} x_{15} -  x_{2} x_{3} x_{6} x_{8} x_{18}, x_{2} x_{5} x_{10} x_{11} x_{15} -  x_{1} x_{3} x_{6} x_{8} x_{18}, x_{1} x_{6} x_{10} x_{11} x_{15} -  x_{4} x_{8} x_{14} x_{17} x_{19}, x_{2} x_{6} x_{10} x_{11} x_{15} -  x_{1} x_{3} x_{5} x_{8} x_{18}, x_{4} x_{6} x_{10} x_{11} x_{15} -  x_{1} x_{8} x_{14} x_{17} x_{19}, x_{1} x_{8} x_{10} x_{11} x_{15} -  x_
{4} x_{6} x_{14} x_{17} x_{19}, x_{4} x_{8} x_{10} x_{11} x_{15} -  x_{1} x_{6} x_{14} x_{17} x_{19}, x_{5} x_{8} x_{10} x_{11} x_{15} -  x_{1} x_{2} x_{3} x_{6} x_{18}, x_{6} x_{8} x_{10} x_{11} x_{15} -  x_{1} x_{4} x_{14} x_{17} x_{19}, x_{1} x_{2} x_{3} x_{12} x_{15} -  x_{4} x_{5} x_{11} x_{13} x_{16}, x_{1} x_{2} x_{4} x_{12} x_{15} -  x_{6} x_{9} x_{13} x_{17} x_{19}, x_{1} x_{3} x_{4} x_{12} x_{15} -  x_{2} x_{5} x_{11} x_{13} x_{16}, x_{2} x_{3} x_{4} x_{12} x_{15} -  x_{1} x_{5} x_{11} x_{13} x_{16}, x_{1} x_{2} x_{5} x_{12} x_{15} -  x_{3} x_{4} x_{11} x_{13} x_{16}, x_{1} x_{3} x_{5} x_{12} x_{15} -  x_{6} x_{9} x_{13} x_{14} x_{18}, x_{2} x_{3} x_{5} x_{12} x_{15} -  x_{1} x_{4} x_{11} x_{13} x_{16}, x_{1} x_{4} x_{5} x_{12} x_{15} -  x_{2} x_{3} x_{11} x_{13} x_{16}, x_{2} x_{4} x_{5} x_{12} x_{15} -  x_{1} x_{3} x_{11} x_{13} x_{16}, x_{3} x_{4} x_{5} x_{12} x_{15} -  x_{1} x_{2} x_{11} x_{13} x_{16}, x_{1} x_{2} x_{6} x_{12} x_{15} -  x_{4} x_{9} x_{13} x_{17} x_{19}, x_{1} x_{3} x_{6} x_{12} 
x_{15} -  x_{5} x_{9} x_{13} x_{14} x_{18}, x_{1} x_{4} x_{6} x_{12} x_{15} -  x_{2} x_{9} x_{13} x_{17} x_{19}, x_{2} x_{4} x_{6} x_{12} x_{15} -  x_{1} x_{9} x_{13} x_{17} x_{19}, x_{1} x_{2} x_{9} x_{12} x_{15} -  x_{4} x_{6} x_{13} x_{17} x_{19}, x_{1} x_{4} x_{9} x_{12} x_{15} -  x_{2} x_{6} x_{13} x_{17} x_{19}, x_{2} x_{4} x_{9} x_{12} x_{15} -  x_{1} x_{6} x_{13} x_{17} x_{19}, x_{1} x_{5} x_{9} x_{12} x_{15} -  x_{3} x_{6} x_{13} x_{14} x_{18}, x_{1} x_{6} x_{9} x_{12} x_{15} -  x_{2} x_{4} x_{13} x_{17} x_{19}, x_{2} x_{6} x_{9} x_{12} x_{15} -  x_{1} x_{4} x_{13} x_{17} x_{19}, x_{4} x_{6} x_{9} x_{12} x_{15} -  x_{1} x_{2} x_{13} x_{17} x_{19}, x_{1} x_{2} x_{11} x_{12} x_{15} -  x_{3} x_{4} x_{5} x_{13} x_{16}, x_{1} x_{4} x_{11} x_{12} x_{15} -  x_{2} x_{3} x_{5} x_{13} x_{16}, x_{2} x_{4} x_{11} x_{12} x_{15} -  x_{1} x_{3} x_{5} x_{13} x_{16}, x_{1} x_{5} x_{11} x_{12} x_{15} -  x_{2} x_{3} x_{4} x_{13} x_{16}, x_{2} x_{5} x_{11} x_{12} x_{15} -  x_{1} x_{3} x_{4} x_{13} x_{16}, x_{4} x_{5} x_
{11} x_{12} x_{15} -  x_{1} x_{2} x_{3} x_{13} x_{16}, x_{1} x_{2} x_{4} x_{13} x_{15} -  x_{6} x_{9} x_{12} x_{17} x_{19}, x_{1} x_{3} x_{4} x_{13} x_{15} -  x_{2} x_{5} x_{11} x_{12} x_{16}, x_{1} x_{2} x_{5} x_{13} x_{15} -  x_{3} x_{4} x_{11} x_{12} x_{16}, x_{1} x_{3} x_{5} x_{13} x_{15} -  x_{6} x_{9} x_{12} x_{14} x_{18}, x_{1} x_{2} x_{6} x_{13} x_{15} -  x_{4} x_{9} x_{12} x_{17} x_{19}, x_{1} x_{3} x_{6} x_{13} x_{15} -  x_{5} x_{9} x_{12} x_{14} x_{18}, x_{1} x_{4} x_{6} x_{13} x_{15} -  x_{2} x_{9} x_{12} x_{17} x_{19}, x_{2} x_{4} x_{6} x_{13} x_{15} -  x_{1} x_{9} x_{12} x_{17} x_{19}, x_{1} x_{5} x_{6} x_{13} x_{15} -  x_{3} x_{9} x_{12} x_{14} x_{18}, x_{3} x_{5} x_{6} x_{13} x_{15} -  x_{1} x_{9} x_{12} x_{14} x_{18}, x_{1} x_{2} x_{9} x_{13} x_{15} -  x_{4} x_{6} x_{12} x_{17} x_{19}, x_{1} x_{3} x_{9} x_{13} x_{15} -  x_{5} x_{6} x_{12} x_{14} x_{18}, x_{1} x_{4} x_{9} x_{13} x_{15} -  x_{2} x_{6} x_{12} x_{17} x_{19}, x_{2} x_{4} x_{9} x_{13} x_{15} -  x_{1} x_{6} x_{12} x_{17} x_{19}, x_{
1} x_{5} x_{9} x_{13} x_{15} -  x_{3} x_{6} x_{12} x_{14} x_{18}, x_{3} x_{5} x_{9} x_{13} x_{15} -  x_{1} x_{6} x_{12} x_{14} x_{18}, x_{1} x_{6} x_{9} x_{13} x_{15} -  x_{2} x_{4} x_{12} x_{17} x_{19}, x_{2} x_{6} x_{9} x_{13} x_{15} -  x_{1} x_{4} x_{12} x_{17} x_{19}, x_{3} x_{6} x_{9} x_{13} x_{15} -  x_{1} x_{5} x_{12} x_{14} x_{18}, x_{4} x_{6} x_{9} x_{13} x_{15} -  x_{1} x_{2} x_{12} x_{17} x_{19}, x_{5} x_{6} x_{9} x_{13} x_{15} -  x_{1} x_{3} x_{12} x_{14} x_{18}, x_{1} x_{2} x_{11} x_{13} x_{15} -  x_{3} x_{4} x_{5} x_{12} x_{16}, x_{1} x_{3} x_{11} x_{13} x_{15} -  x_{2} x_{4} x_{5} x_{12} x_{16}, x_{1} x_{4} x_{11} x_{13} x_{15} -  x_{2} x_{3} x_{5} x_{12} x_{16}, x_{2} x_{4} x_{11} x_{13} x_{15} -  x_{1} x_{3} x_{5} x_{12} x_{16}, x_{3} x_{4} x_{11} x_{13} x_{15} -  x_{1} x_{2} x_{5} x_{12} x_{16}, x_{1} x_{5} x_{11} x_{13} x_{15} -  x_{2} x_{3} x_{4} x_{12} x_{16}, x_{2} x_{5} x_{11} x_{13} x_{15} -  x_{1} x_{3} x_{4} x_{12} x_{16}, x_{3} x_{5} x_{11} x_{13} x_{15} -  x_{1} x_{2} x_{4} x_{12} 
x_{16}, x_{1} x_{2} x_{12} x_{13} x_{15} -  x_{4} x_{6} x_{9} x_{17} x_{19}, x_{1} x_{3} x_{12} x_{13} x_{15} -  x_{5} x_{6} x_{9} x_{14} x_{18}, x_{1} x_{4} x_{12} x_{13} x_{15} -  x_{2} x_{6} x_{9} x_{17} x_{19}, x_{2} x_{4} x_{12} x_{13} x_{15} -  x_{1} x_{6} x_{9} x_{17} x_{19}, x_{3} x_{4} x_{12} x_{13} x_{15} -  x_{1} x_{2} x_{5} x_{11} x_{16}, x_{1} x_{5} x_{12} x_{13} x_{15} -  x_{3} x_{6} x_{9} x_{14} x_{18}, x_{2} x_{5} x_{12} x_{13} x_{15} -  x_{1} x_{3} x_{4} x_{11} x_{16}, x_{3} x_{5} x_{12} x_{13} x_{15} -  x_{1} x_{6} x_{9} x_{14} x_{18}, x_{1} x_{6} x_{12} x_{13} x_{15} -  x_{2} x_{4} x_{9} x_{17} x_{19}, x_{2} x_{6} x_{12} x_{13} x_{15} -  x_{1} x_{4} x_{9} x_{17} x_{19}, x_{3} x_{6} x_{12} x_{13} x_{15} -  x_{1} x_{5} x_{9} x_{14} x_{18}, x_{1} x_{9} x_{12} x_{13} x_{15} -  x_{2} x_{4} x_{6} x_{17} x_{19}, x_{4} x_{9} x_{12} x_{13} x_{15} -  x_{1} x_{2} x_{6} x_{17} x_{19}, x_{5} x_{9} x_{12} x_{13} x_{15} -  x_{1} x_{3} x_{6} x_{14} x_{18}, x_{6} x_{9} x_{12} x_{13} x_{15} -  x_{1} x_{2} x_
{4} x_{17} x_{19}, x_{1} x_{11} x_{12} x_{13} x_{15} -  x_{14} x_{16} x_{17} x_{18} x_{19}, x_{4} x_{11} x_{12} x_{13} x_{15} -  x_{1} x_{2} x_{3} x_{5} x_{16}, x_{5} x_{11} x_{12} x_{13} x_{15} -  x_{1} x_{2} x_{3} x_{4} x_{16}, x_{1} x_{3} x_{4} x_{14} x_{15} -  x_{5} x_{8} x_{9} x_{10} x_{16}, x_{1} x_{3} x_{5} x_{14} x_{15} -  x_{6} x_{9} x_{12} x_{13} x_{18}, x_{1} x_{4} x_{5} x_{14} x_{15} -  x_{3} x_{8} x_{9} x_{10} x_{16}, x_{1} x_{3} x_{6} x_{14} x_{15} -  x_{5} x_{9} x_{12} x_{13} x_{18}, x_{1} x_{4} x_{6} x_{14} x_{15} -  x_{8} x_{10} x_{11} x_{17} x_{19}, x_{1} x_{5} x_{6} x_{14} x_{15} -  x_{3} x_{9} x_{12} x_{13} x_{18}, x_{3} x_{5} x_{6} x_{14} x_{15} -  x_{1} x_{9} x_{12} x_{13} x_{18}, x_{1} x_{4} x_{8} x_{14} x_{15} -  x_{6} x_{10} x_{11} x_{17} x_{19}, x_{1} x_{5} x_{8} x_{14} x_{15} -  x_{3} x_{4} x_{9} x_{10} x_{16}, x_{1} x_{6} x_{8} x_{14} x_{15} -  x_{4} x_{10} x_{11} x_{17} x_{19}, x_{4} x_{6} x_{8} x_{14} x_{15} -  x_{1} x_{10} x_{11} x_{17} x_{19}, x_{1} x_{3} x_{9} x_{14} x_{15} - 
 x_{5} x_{6} x_{12} x_{13} x_{18}, x_{1} x_{4} x_{9} x_{14} x_{15} -  x_{3} x_{5} x_{8} x_{10} x_{16}, x_{3} x_{4} x_{9} x_{14} x_{15} -  x_{1} x_{5} x_{8} x_{10} x_{16}, x_{1} x_{5} x_{9} x_{14} x_{15} -  x_{3} x_{6} x_{12} x_{13} x_{18}, x_{3} x_{5} x_{9} x_{14} x_{15} -  x_{1} x_{6} x_{12} x_{13} x_{18}, x_{4} x_{5} x_{9} x_{14} x_{15} -  x_{1} x_{3} x_{8} x_{10} x_{16}, x_{1} x_{6} x_{9} x_{14} x_{15} -  x_{3} x_{5} x_{12} x_{13} x_{18}, x_{3} x_{6} x_{9} x_{14} x_{15} -  x_{1} x_{5} x_{12} x_{13} x_{18}, x_{5} x_{6} x_{9} x_{14} x_{15} -  x_{1} x_{3} x_{12} x_{13} x_{18}, x_{1} x_{8} x_{9} x_{14} x_{15} -  x_{3} x_{4} x_{5} x_{10} x_{16}, x_{4} x_{8} x_{9} x_{14} x_{15} -  x_{1} x_{3} x_{5} x_{10} x_{16}, x_{5} x_{8} x_{9} x_{14} x_{15} -  x_{1} x_{3} x_{4} x_{10} x_{16}, x_{1} x_{3} x_{10} x_{14} x_{15} -  x_{4} x_{5} x_{8} x_{9} x_{16}, x_{1} x_{4} x_{10} x_{14} x_{15} -  x_{6} x_{8} x_{11} x_{17} x_{19}, x_{3} x_{4} x_{10} x_{14} x_{15} -  x_{1} x_{5} x_{8} x_{9} x_{16}, x_{1} x_{5} x_{10} x_{14} x_{
15} -  x_{3} x_{4} x_{8} x_{9} x_{16}, x_{3} x_{5} x_{10} x_{14} x_{15} -  x_{1} x_{4} x_{8} x_{9} x_{16}, x_{4} x_{5} x_{10} x_{14} x_{15} -  x_{1} x_{3} x_{8} x_{9} x_{16}, x_{1} x_{6} x_{10} x_{14} x_{15} -  x_{4} x_{8} x_{11} x_{17} x_{19}, x_{4} x_{6} x_{10} x_{14} x_{15} -  x_{1} x_{8} x_{11} x_{17} x_{19}, x_{1} x_{8} x_{10} x_{14} x_{15} -  x_{4} x_{6} x_{11} x_{17} x_{19}, x_{4} x_{8} x_{10} x_{14} x_{15} -  x_{1} x_{6} x_{11} x_{17} x_{19}, x_{5} x_{8} x_{10} x_{14} x_{15} -  x_{1} x_{3} x_{4} x_{9} x_{16}, x_{6} x_{8} x_{10} x_{14} x_{15} -  x_{1} x_{4} x_{11} x_{17} x_{19}, x_{1} x_{9} x_{10} x_{14} x_{15} -  x_{3} x_{4} x_{5} x_{8} x_{16}, x_{4} x_{9} x_{10} x_{14} x_{15} -  x_{1} x_{3} x_{5} x_{8} x_{16}, x_{5} x_{9} x_{10} x_{14} x_{15} -  x_{1} x_{3} x_{4} x_{8} x_{16}, x_{8} x_{9} x_{10} x_{14} x_{15} -  x_{1} x_{3} x_{4} x_{5} x_{16}, x_{1} x_{4} x_{11} x_{14} x_{15} -  x_{6} x_{8} x_{10} x_{17} x_{19}, x_{1} x_{6} x_{11} x_{14} x_{15} -  x_{4} x_{8} x_{10} x_{17} x_{19}, x_{1} x_{8}x_{11} 
x_{14} x_{15} -  x_{4} x_{6} x_{10} x_{17} x_{19}, x_{4} x_{8} x_{11} x_{14} x_{15} -  x_{1} x_{6} x_{10} x_{17} x_{19}, x_{6} x_{8} x_{11} x_{14} x_{15} -  x_{1} x_{4} x_{10} x_{17} x_{19}, x_{1} x_{10} x_{11} x_{14} x_{15} -  x_{4} x_{6} x_{8} x_{17} x_{19}, x_{4} x_{10} x_{11} x_{14} x_{15} -  x_{1} x_{6} x_{8} x_{17} x_{19}, x_{6} x_{10} x_{11} x_{14} x_{15} -  x_{1} x_{4} x_{8} x_{17} x_{19}, x_{8} x_{10} x_{11} x_{14} x_{15} -  x_{1} x_{4} x_{6} x_{17} x_{19}, x_{1} x_{3} x_{12} x_{14} x_{15} -  x_{5} x_{6} x_{9} x_{13} x_{18}, x_{1} x_{5} x_{12} x_{14} x_{15} -  x_{3} x_{6} x_{9} x_{13} x_{18}, x_{3} x_{5} x_{12} x_{14} x_{15} -  x_{1} x_{6} x_{9} x_{13} x_{18}, x_{1} x_{6} x_{12} x_{14} x_{15} -  x_{3} x_{5} x_{9} x_{13} x_{18}, x_{3} x_{6} x_{12} x_{14} x_{15} -  x_{1} x_{5} x_{9} x_{13} x_{18}, x_{1} x_{9} x_{12} x_{14} x_{15} -  x_{3} x_{5} x_{6} x_{13} x_{18}, x_{5} x_{9} x_{12} x_{14} x_{15} -  x_{1} x_{3} x_{6} x_{13} x_{18}, x_{6} x_{9} x_{12} x_{14} x_{15} -  x_{1} x_{3} x_{5} x_{13} x_{18}, 
x_{1} x_{11} x_{12} x_{14} x_{15} -  x_{13} x_{16} x_{17} x_{18} x_{19}, x_{1} x_{5} x_{13} x_{14} x_{15} -  x_{3} x_{6} x_{9} x_{12} x_{18}, x_{1} x_{6} x_{13} x_{14} x_{15} -  x_{3} x_{5} x_{9} x_{12} x_{18}, x_{5} x_{6} x_{13} x_{14} x_{15} -  x_{1} x_{3} x_{9} x_{12} x_{18}, x_{1} x_{9} x_{13} x_{14} x_{15} -  x_{3} x_{5} x_{6} x_{12} x_{18}, x_{5} x_{9} x_{13} x_{14} x_{15} -  x_{1} x_{3} x_{6} x_{12} x_{18}, x_{6} x_{9} x_{13} x_{14} x_{15} -  x_{1} x_{3} x_{5} x_{12} x_{18}, x_{1} x_{11} x_{13} x_{14} x_{15} -  x_{12} x_{16} x_{17} x_{18} x_{19}, x_{1} x_{12} x_{13} x_{14} x_{15} -  x_{11} x_{16} x_{17} x_{18} x_{19}, x_{5} x_{12} x_{13} x_{14} x_{15} -  x_{1} x_{3} x_{6} x_{9} x_{18}, x_{6} x_{12} x_{13} x_{14} x_{15} -  x_{1} x_{3} x_{5} x_{9} x_{18}, x_{9} x_{12} x_{13} x_{14} x_{15} -  x_{1} x_{3} x_{5} x_{6} x_{18}, x_{11} x_{12} x_{13} x_{14} x_{15} -  x_{1} x_{16} x_{17} x_{18} x_{19}, x_{1} x_{2} x_{4} x_{5} x_{16} -  x_{7} x_{9} x_{10} x_{13} x_{19}, x_{1} x_{2} x_{3} x_{7} x_{16} -  x_{8} x_{
11} x_{12} x_{17} x_{18}, x_{1} x_{2} x_{4} x_{7} x_{16} -  x_{5} x_{9} x_{10} x_{13} x_{19}, x_{1} x_{2} x_{5} x_{7} x_{16} -  x_{4} x_{9} x_{10} x_{13} x_{19}, x_{1} x_{4} x_{5} x_{7} x_{16} -  x_{8} x_{11} x_{12} x_{14} x_{19}, x_{2} x_{4} x_{5} x_{7} x_{16} -  x_{1} x_{9} x_{10} x_{13} x_{19}, x_{1} x_{2} x_{3} x_{8} x_{16} -  x_{7} x_{11} x_{12} x_{17} x_{18}, x_{1} x_{4} x_{5} x_{8} x_{16} -  x_{7} x_{11} x_{12} x_{14} x_{19}, x_{1} x_{2} x_{7} x_{8} x_{16} -  x_{3} x_{11} x_{12} x_{17} x_{18}, x_{1} x_{3} x_{7} x_{8} x_{16} -  x_{2} x_{11} x_{12} x_{17} x_{18}, x_{1} x_{4} x_{7} x_{8} x_{16} -  x_{5} x_{11} x_{12} x_{14} x_{19}, x_{1} x_{5} x_{7} x_{8} x_{16} -  x_{4} x_{11} x_{12} x_{14} x_{19}, x_{1} x_{2} x_{5} x_{9} x_{16} -  x_{4} x_{7} x_{10} x_{13} x_{19}, x_{1} x_{4} x_{5} x_{9} x_{16} -  x_{2} x_{7} x_{10} x_{13} x_{19}, x_{1} x_{2} x_{7} x_{9} x_{16} -  x_{4} x_{5} x_{10} x_{13} x_{19}, x_{1} x_{3} x_{7} x_{9} x_{16} -  x_{10} x_{13} x_{14} x_{17} x_{18}, x_{1} x_{4} x_{7} x_{9} x_{16}-  x_{
2} x_{5} x_{10} x_{13} x_{19}, x_{1} x_{5} x_{7} x_{9} x_{16} -  x_{2} x_{4} x_{10} x_{13} x_{19}, x_{2} x_{5} x_{7} x_{9} x_{16} -  x_{1} x_{4} x_{10} x_{13} x_{19}, x_{4} x_{5} x_{7} x_{9} x_{16} -  x_{1} x_{2} x_{10} x_{13} x_{19}, x_{1} x_{2} x_{8} x_{9} x_{16} -  x_{10} x_{15} x_{17} x_{18} x_{19}, x_{1} x_{2} x_{4} x_{10} x_{16} -  x_{5} x_{7} x_{9} x_{13} x_{19}, x_{1} x_{2} x_{5} x_{10} x_{16} -  x_{4} x_{7} x_{9} x_{13} x_{19}, x_{1} x_{4} x_{5} x_{10} x_{16} -  x_{2} x_{7} x_{9} x_{13} x_{19}, x_{2} x_{4} x_{5} x_{10} x_{16} -  x_{1} x_{7} x_{9} x_{13} x_{19}, x_{1} x_{2} x_{7} x_{10} x_{16} -  x_{4} x_{5} x_{9} x_{13} x_{19}, x_{1} x_{3} x_{7} x_{10} x_{16} -  x_{9} x_{13} x_{14} x_{17} x_{18}, x_{1} x_{4} x_{7} x_{10} x_{16} -  x_{2} x_{5} x_{9} x_{13} x_{19}, x_{2} x_{4} x_{7} x_{10} x_{16} -  x_{1} x_{5} x_{9} x_{13} x_{19}, x_{1} x_{2} x_{8} x_{10} x_{16} -  x_{9} x_{15} x_{17} x_{18} x_{19}, x_{1} x_{2} x_{9} x_{10} x_{16} -  x_{8} x_{15} x_{17} x_{18} x_{19}, x_{1} x_{3} x_{9} x_{10} x_{16} -
  x_{7} x_{13} x_{14} x_{17} x_{18}, x_{1} x_{4} x_{9} x_{10} x_{16} -  x_{2} x_{5} x_{7} x_{13} x_{19}, x_{1} x_{5} x_{9} x_{10} x_{16} -  x_{2} x_{4} x_{7} x_{13} x_{19}, x_{2} x_{5} x_{9} x_{10} x_{16} -  x_{1} x_{4} x_{7} x_{13} x_{19}, x_{4} x_{5} x_{9} x_{10} x_{16} -  x_{1} x_{2} x_{7} x_{13} x_{19}, x_{1} x_{7} x_{9} x_{10} x_{16} -  x_{2} x_{4} x_{5} x_{13} x_{19}, x_{2} x_{7} x_{9} x_{10} x_{16} -  x_{1} x_{4} x_{5} x_{13} x_{19}, x_{4} x_{7} x_{9} x_{10} x_{16} -  x_{1} x_{2} x_{5} x_{13} x_{19}, x_{1} x_{8} x_{9} x_{10} x_{16} -  x_{2} x_{15} x_{17} x_{18} x_{19}, x_{1} x_{2} x_{3} x_{11} x_{16} -  x_{7} x_{8} x_{12} x_{17} x_{18}, x_{1} x_{4} x_{5} x_{11} x_{16} -  x_{7} x_{8} x_{12} x_{14} x_{19}, x_{1} x_{2} x_{7} x_{11} x_{16} -  x_{3} x_{8} x_{12} x_{17} x_{18}, x_{1} x_{3} x_{7} x_{11} x_{16} -  x_{2} x_{8} x_{12} x_{17} x_{18}, x_{2} x_{3} x_{7} x_{11} x_{16} -  x_{1} x_{8} x_{12} x_{17} x_{18}, x_{1} x_{4} x_{7} x_{11} x_{16} -  x_{5} x_{8} x_{12} x_{14} x_{19}, x_{1} x_{5} x_{7} x_{11} x_
{16} -  x_{4} x_{8} x_{12} x_{14} x_{19}, x_{4} x_{5} x_{7} x_{11} x_{16} -  x_{1} x_{8} x_{12} x_{14} x_{19}, x_{1} x_{2} x_{8} x_{11} x_{16} -  x_{3} x_{7} x_{12} x_{17} x_{18}, x_{1} x_{3} x_{8} x_{11} x_{16} -  x_{2} x_{7} x_{12} x_{17} x_{18}, x_{2} x_{3} x_{8} x_{11} x_{16} -  x_{1} x_{7} x_{12} x_{17} x_{18}, x_{1} x_{4} x_{8} x_{11} x_{16} -  x_{5} x_{7} x_{12} x_{14} x_{19}, x_{1} x_{5} x_{8} x_{11} x_{16} -  x_{4} x_{7} x_{12} x_{14} x_{19}, x_{4} x_{5} x_{8} x_{11} x_{16} -  x_{1} x_{7} x_{12} x_{14} x_{19}, x_{1} x_{7} x_{8} x_{11} x_{16} -  x_{4} x_{5} x_{12} x_{14} x_{19}, x_{2} x_{7} x_{8} x_{11} x_{16} -  x_{1} x_{3} x_{12} x_{17} x_{18}, x_{3} x_{7} x_{8} x_{11} x_{16} -  x_{1} x_{2} x_{12} x_{17} x_{18}, x_{4} x_{7} x_{8} x_{11} x_{16} -  x_{1} x_{5} x_{12} x_{14} x_{19}, x_{5} x_{7} x_{8} x_{11} x_{16} -  x_{1} x_{4} x_{12} x_{14} x_{19}, x_{1} x_{2} x_{3} x_{12} x_{16} -  x_{7} x_{8} x_{11} x_{17} x_{18}, x_{1} x_{4} x_{5} x_{12} x_{16} -  x_{7} x_{8} x_{11} x_{14} x_{19}, x_{1} x_{2} x_{
7} x_{12} x_{16} -  x_{3} x_{8} x_{11} x_{17} x_{18}, x_{1} x_{3} x_{7} x_{12} x_{16} -  x_{2} x_{8} x_{11} x_{17} x_{18}, x_{2} x_{3} x_{7} x_{12} x_{16} -  x_{1} x_{8} x_{11} x_{17} x_{18}, x_{1} x_{4} x_{7} x_{12} x_{16} -  x_{5} x_{8} x_{11} x_{14} x_{19}, x_{1} x_{2} x_{8} x_{12} x_{16} -  x_{3} x_{7} x_{11} x_{17} x_{18}, x_{1} x_{3} x_{8} x_{12} x_{16} -  x_{2} x_{7} x_{11} x_{17} x_{18}, x_{2} x_{3} x_{8} x_{12} x_{16} -  x_{1} x_{7} x_{11} x_{17} x_{18}, x_{1} x_{5} x_{8} x_{12} x_{16} -  x_{4} x_{7} x_{11} x_{14} x_{19}, x_{1} x_{7} x_{8} x_{12} x_{16} -  x_{4} x_{5} x_{11} x_{14} x_{19}, x_{2} x_{7} x_{8} x_{12} x_{16} -  x_{1} x_{3} x_{11} x_{17} x_{18}, x_{3} x_{7} x_{8} x_{12} x_{16} -  x_{1} x_{2} x_{11} x_{17} x_{18}, x_{1} x_{2} x_{11} x_{12} x_{16} -  x_{3} x_{7} x_{8} x_{17} x_{18}, x_{1} x_{3} x_{11} x_{12} x_{16} -  x_{2} x_{7} x_{8} x_{17} x_{18}, x_{2} x_{3} x_{11} x_{12} x_{16} -  x_{1} x_{7} x_{8} x_{17} x_{18}, x_{1} x_{4} x_{11} x_{12} x_{16} -  x_{5} x_{7} x_{8} x_{14} x_{19}, x_{
1} x_{5} x_{11} x_{12} x_{16} -  x_{4} x_{7} x_{8} x_{14} x_{19}, x_{4} x_{5} x_{11} x_{12} x_{16} -  x_{1} x_{7} x_{8} x_{14} x_{19}, x_{1} x_{7} x_{11} x_{12} x_{16} -  x_{4} x_{5} x_{8} x_{14} x_{19}, x_{2} x_{7} x_{11} x_{12} x_{16} -  x_{1} x_{3} x_{8} x_{17} x_{18}, x_{4} x_{7} x_{11} x_{12} x_{16} -  x_{1} x_{5} x_{8} x_{14} x_{19}, x_{1} x_{8} x_{11} x_{12} x_{16} -  x_{4} x_{5} x_{7} x_{14} x_{19}, x_{3} x_{8} x_{11} x_{12} x_{16} -  x_{1} x_{2} x_{7} x_{17} x_{18}, x_{5} x_{8} x_{11} x_{12} x_{16} -  x_{1} x_{4} x_{7} x_{14} x_{19}, x_{7} x_{8} x_{11} x_{12} x_{16} -  x_{1} x_{4} x_{5} x_{14} x_{19}, x_{1} x_{2} x_{4} x_{13} x_{16} -  x_{5} x_{7} x_{9} x_{10} x_{19}, x_{1} x_{2} x_{5} x_{13} x_{16} -  x_{4} x_{7} x_{9} x_{10} x_{19}, x_{1} x_{4} x_{5} x_{13} x_{16} -  x_{2} x_{7} x_{9} x_{10} x_{19}, x_{2} x_{4} x_{5} x_{13} x_{16} -  x_{1} x_{7} x_{9} x_{10} x_{19}, x_{1} x_{2} x_{7} x_{13} x_{16} -  x_{4} x_{5} x_{9} x_{10} x_{19}, x_{1} x_{3} x_{7} x_{13} x_{16} -  x_{9} x_{10} x_{14} x_{17} x_{
18}, x_{1} x_{4} x_{7} x_{13} x_{16} -  x_{2} x_{5} x_{9} x_{10} x_{19}, x_{2} x_{4} x_{7} x_{13} x_{16} -  x_{1} x_{5} x_{9} x_{10} x_{19}, x_{1} x_{5} x_{7} x_{13} x_{16} -  x_{2} x_{4} x_{9} x_{10} x_{19}, x_{2} x_{5} x_{7} x_{13} x_{16} -  x_{1} x_{4} x_{9} x_{10} x_{19}, x_{1} x_{2} x_{9} x_{13} x_{16} -  x_{4} x_{5} x_{7} x_{10} x_{19}, x_{1} x_{3} x_{9} x_{13} x_{16} -  x_{7} x_{10} x_{14} x_{17} x_{18}, x_{1} x_{4} x_{9} x_{13} x_{16} -  x_{2} x_{5} x_{7} x_{10} x_{19}, x_{1} x_{5} x_{9} x_{13} x_{16} -  x_{2} x_{4} x_{7} x_{10} x_{19}, x_{2} x_{5} x_{9} x_{13} x_{16} -  x_{1} x_{4} x_{7} x_{10} x_{19}, x_{4} x_{5} x_{9} x_{13} x_{16} -  x_{1} x_{2} x_{7} x_{10} x_{19}, x_{1} x_{7} x_{9} x_{13} x_{16} -  x_{2} x_{4} x_{5} x_{10} x_{19}, x_{2} x_{7} x_{9} x_{13} x_{16} -  x_{1} x_{4} x_{5} x_{10} x_{19}, x_{3} x_{7} x_{9} x_{13} x_{16} -  x_{1} x_{10} x_{14} x_{17} x_{18}, x_{4} x_{7} x_{9} x_{13} x_{16} -  x_{1} x_{2} x_{5} x_{10} x_{19}, x_{5} x_{7} x_{9} x_{13} x_{16} -  x_{1} x_{2} x_{4} x_{10} x_{
19}, x_{1} x_{2} x_{10} x_{13} x_{16} -  x_{4} x_{5} x_{7} x_{9} x_{19}, x_{1} x_{3} x_{10} x_{13} x_{16} -  x_{7} x_{9} x_{14} x_{17} x_{18}, x_{1} x_{5} x_{10} x_{13} x_{16} -  x_{2} x_{4} x_{7} x_{9} x_{19}, x_{2} x_{5} x_{10} x_{13} x_{16} -  x_{1} x_{4} x_{7} x_{9} x_{19}, x_{1} x_{7} x_{10} x_{13} x_{16} -  x_{2} x_{4} x_{5} x_{9} x_{19}, x_{2} x_{7} x_{10} x_{13} x_{16} -  x_{1} x_{4} x_{5} x_{9} x_{19}, x_{3} x_{7} x_{10} x_{13} x_{16} -  x_{1} x_{9} x_{14} x_{17} x_{18}, x_{1} x_{9} x_{10} x_{13} x_{16} -  x_{2} x_{4} x_{5} x_{7} x_{19}, x_{3} x_{9} x_{10} x_{13} x_{16} -  x_{1} x_{7} x_{14} x_{17} x_{18}, x_{5} x_{9} x_{10} x_{13} x_{16} -  x_{1} x_{2} x_{4} x_{7} x_{19}, x_{7} x_{9} x_{10} x_{13} x_{16} -  x_{1} x_{2} x_{4} x_{5} x_{19}, x_{1} x_{11} x_{12} x_{13} x_{16} -  x_{14} x_{15} x_{17} x_{18} x_{19}, x_{1} x_{4} x_{5} x_{14} x_{16} -  x_{7} x_{8} x_{11} x_{12} x_{19}, x_{1} x_{3} x_{7} x_{14} x_{16} -  x_{9} x_{10} x_{13} x_{17} x_{18}, x_{1} x_{4} x_{7} x_{14} x_{16} -  x_{5} x_{8} x_{11}
 x_{12} x_{19}, x_{1} x_{5} x_{7} x_{14} x_{16} -  x_{4} x_{8} x_{11} x_{12} x_{19}, x_{4} x_{5} x_{7} x_{14} x_{16} -  x_{1} x_{8} x_{11} x_{12} x_{19}, x_{1} x_{4} x_{8} x_{14} x_{16} -  x_{5} x_{7} x_{11} x_{12} x_{19}, x_{1} x_{5} x_{8} x_{14} x_{16} -  x_{4} x_{7} x_{11} x_{12} x_{19}, x_{4} x_{5} x_{8} x_{14} x_{16} -  x_{1} x_{7} x_{11} x_{12} x_{19}, x_{1} x_{7} x_{8} x_{14} x_{16} -  x_{4} x_{5} x_{11} x_{12} x_{19}, x_{4} x_{7} x_{8} x_{14} x_{16} -  x_{1} x_{5} x_{11} x_{12} x_{19}, x_{5} x_{7} x_{8} x_{14} x_{16} -  x_{1} x_{4} x_{11} x_{12} x_{19}, x_{1} x_{3} x_{9} x_{14} x_{16} -  x_{7} x_{10} x_{13} x_{17} x_{18}, x_{1} x_{7} x_{9} x_{14} x_{16} -  x_{3} x_{10} x_{13} x_{17} x_{18}, x_{3} x_{7} x_{9} x_{14} x_{16} -  x_{1} x_{10} x_{13} x_{17} x_{18}, x_{1} x_{3} x_{10} x_{14} x_{16} -  x_{7} x_{9} x_{13} x_{17} x_{18}, x_{1} x_{7} x_{10} x_{14} x_{16} -  x_{3} x_{9} x_{13} x_{17} x_{18}, x_{3} x_{7} x_{10} x_{14} x_{16} -  x_{1} x_{9} x_{13} x_{17} x_{18}, x_{1} x_{9} x_{10} x_{14} x_{16} -  
x_{3} x_{7} x_{13} x_{17} x_{18}, x_{3} x_{9} x_{10} x_{14} x_{16} -  x_{1} x_{7} x_{13} x_{17} x_{18}, x_{7} x_{9} x_{10} x_{14} x_{16} -  x_{1} x_{3} x_{13} x_{17} x_{18}, x_{1} x_{5} x_{11} x_{14} x_{16} -  x_{4} x_{7} x_{8} x_{12} x_{19}, x_{1} x_{7} x_{11} x_{14} x_{16} -  x_{4} x_{5} x_{8} x_{12} x_{19}, x_{5} x_{7} x_{11} x_{14} x_{16} -  x_{1} x_{4} x_{8} x_{12} x_{19}, x_{1} x_{8} x_{11} x_{14} x_{16} -  x_{4} x_{5} x_{7} x_{12} x_{19}, x_{5} x_{8} x_{11} x_{14} x_{16} -  x_{1} x_{4} x_{7} x_{12} x_{19}, x_{7} x_{8} x_{11} x_{14} x_{16} -  x_{1} x_{4} x_{5} x_{12} x_{19}, x_{1} x_{4} x_{12} x_{14} x_{16} -  x_{5} x_{7} x_{8} x_{11} x_{19}, x_{1} x_{5} x_{12} x_{14} x_{16} -  x_{4} x_{7} x_{8} x_{11} x_{19}, x_{4} x_{5} x_{12} x_{14} x_{16} -  x_{1} x_{7} x_{8} x_{11} x_{19}, x_{1} x_{7} x_{12} x_{14} x_{16} -  x_{4} x_{5} x_{8} x_{11} x_{19}, x_{4} x_{7} x_{12} x_{14} x_{16} -  x_{1} x_{5} x_{8} x_{11} x_{19}, x_{1} x_{8} x_{12} x_{14} x_{16} -  x_{4} x_{5} x_{7} x_{11} x_{19}, x_{5} x_{8} x_{12} x_{
14} x_{16} -  x_{1} x_{4} x_{7} x_{11} x_{19}, x_{7} x_{8} x_{12} x_{14} x_{16} -  x_{1} x_{4} x_{5} x_{11} x_{19}, x_{1} x_{11} x_{12} x_{14} x_{16} -  x_{13} x_{15} x_{17} x_{18} x_{19}, x_{5} x_{11} x_{12} x_{14} x_{16} -  x_{1} x_{4} x_{7} x_{8} x_{19}, x_{7} x_{11} x_{12} x_{14} x_{16} -  x_{1} x_{4} x_{5} x_{8} x_{19}, x_{8} x_{11} x_{12} x_{14} x_{16} -  x_{1} x_{4} x_{5} x_{7} x_{19}, x_{1} x_{3} x_{13} x_{14} x_{16} -  x_{7} x_{9} x_{10} x_{17} x_{18}, x_{1} x_{7} x_{13} x_{14} x_{16} -  x_{3} x_{9} x_{10} x_{17} x_{18}, x_{1} x_{9} x_{13} x_{14} x_{16} -  x_{3} x_{7} x_{10} x_{17} x_{18}, x_{3} x_{9} x_{13} x_{14} x_{16} -  x_{1} x_{7} x_{10} x_{17} x_{18}, x_{7} x_{9} x_{13} x_{14} x_{16} -  x_{1} x_{3} x_{10} x_{17} x_{18}, x_{1} x_{10} x_{13} x_{14} x_{16} -  x_{3} x_{7} x_{9} x_{17} x_{18}, x_{3} x_{10} x_{13} x_{14} x_{16} -  x_{1} x_{7} x_{9} x_{17} x_{18}, x_{7} x_{10} x_{13} x_{14} x_{16} -  x_{1} x_{3} x_{9} x_{17} x_{18}, x_{9} x_{10} x_{13} x_{14} x_{16} -  x_{1} x_{3} x_{7} x_{17} x_{18}
, x_{1} x_{11} x_{13} x_{14} x_{16} -  x_{12} x_{15} x_{17} x_{18} x_{19}, x_{1} x_{12} x_{13} x_{14} x_{16} -  x_{11} x_{15} x_{17} x_{18} x_{19}, x_{11} x_{12} x_{13} x_{14} x_{16} -  x_{1} x_{15} x_{17} x_{18} x_{19}, x_{1} x_{2} x_{8} x_{15} x_{16} -  x_{9} x_{10} x_{17} x_{18} x_{19}, x_{1} x_{2} x_{9} x_{15} x_{16} -  x_{8} x_{10} x_{17} x_{18} x_{19}, x_{1} x_{8} x_{9} x_{15} x_{16} -  x_{2} x_{10} x_{17} x_{18} x_{19}, x_{2} x_{8} x_{9} x_{15} x_{16} -  x_{1} x_{10} x_{17} x_{18} x_{19}, x_{1} x_{2} x_{10} x_{15} x_{16} -  x_{8} x_{9} x_{17} x_{18} x_{19}, x_{1} x_{8} x_{10} x_{15} x_{16} -  x_{2} x_{9} x_{17} x_{18} x_{19}, x_{2} x_{8} x_{10} x_{15} x_{16} -  x_{1} x_{9} x_{17} x_{18} x_{19}, x_{1} x_{9} x_{10} x_{15} x_{16} -  x_{2} x_{8} x_{17} x_{18} x_{19}, x_{2} x_{9} x_{10} x_{15} x_{16} -  x_{1} x_{8} x_{17} x_{18} x_{19}, x_{8} x_{9} x_{10} x_{15} x_{16} -  x_{1} x_{2} x_{17} x_{18} x_{19}, x_{1} x_{11} x_{12} x_{15} x_{16} -  x_{13} x_{14} x_{17} x_{18} x_{19}, x_{1} x_{11} x_{13} x_{15} x_{
16} -  x_{12} x_{14} x_{17} x_{18} x_{19}, x_{1} x_{12} x_{13} x_{15} x_{16} -  x_{11} x_{14} x_{17} x_{18} x_{19}, x_{11} x_{12} x_{13} x_{15} x_{16} -  x_{1} x_{14} x_{17} x_{18} x_{19}, x_{1} x_{11} x_{14} x_{15} x_{16} -  x_{12} x_{13} x_{17} x_{18} x_{19}, x_{1} x_{12} x_{14} x_{15} x_{16} -  x_{11} x_{13} x_{17} x_{18} x_{19}, x_{11} x_{12} x_{14} x_{15} x_{16} -  x_{1} x_{13} x_{17} x_{18} x_{19}, x_{1} x_{13} x_{14} x_{15} x_{16} -  x_{11} x_{12} x_{17} x_{18} x_{19}, x_{11} x_{13} x_{14} x_{15} x_{16} -  x_{1} x_{12} x_{17} x_{18} x_{19}, x_{12} x_{13} x_{14} x_{15} x_{16} -  x_{1} x_{11} x_{17} x_{18} x_{19}, x_{1} x_{2} x_{4} x_{6} x_{17} -  x_{9} x_{12} x_{13} x_{15} x_{19}, x_{1} x_{2} x_{3} x_{7} x_{17} -  x_{8} x_{11} x_{12} x_{16} x_{18}, x_{1} x_{2} x_{3} x_{8} x_{17} -  x_{7} x_{11} x_{12} x_{16} x_{18}, x_{1} x_{4} x_{6} x_{8} x_{17} -  x_{10} x_{11} x_{14} x_{15} x_{19}, x_{1} x_{2} x_{7} x_{8} x_{17} -  x_{3} x_{11} x_{12} x_{16} x_{18}, x_{1} x_{3} x_{7} x_{8} x_{17} -  x_{2} x_{11}x_{
12} x_{16} x_{18}, x_{1} x_{2} x_{4} x_{9} x_{17} -  x_{6} x_{12} x_{13} x_{15} x_{19}, x_{1} x_{2} x_{6} x_{9} x_{17} -  x_{4} x_{12} x_{13} x_{15} x_{19}, x_{1} x_{4} x_{6} x_{9} x_{17} -  x_{2} x_{12} x_{13} x_{15} x_{19}, x_{1} x_{3} x_{7} x_{9} x_{17} -  x_{10} x_{13} x_{14} x_{16} x_{18}, x_{1} x_{2} x_{8} x_{9} x_{17} -  x_{10} x_{15} x_{16} x_{18} x_{19}, x_{1} x_{4} x_{6} x_{10} x_{17} -  x_{8} x_{11} x_{14} x_{15} x_{19}, x_{1} x_{3} x_{7} x_{10} x_{17} -  x_{9} x_{13} x_{14} x_{16} x_{18}, x_{1} x_{2} x_{8} x_{10} x_{17} -  x_{9} x_{15} x_{16} x_{18} x_{19}, x_{1} x_{4} x_{8} x_{10} x_{17} -  x_{6} x_{11} x_{14} x_{15} x_{19}, x_{1} x_{6} x_{8} x_{10} x_{17} -  x_{4} x_{11} x_{14} x_{15} x_{19}, x_{1} x_{2} x_{9} x_{10} x_{17} -  x_{8} x_{15} x_{16} x_{18} x_{19}, x_{1} x_{3} x_{9} x_{10} x_{17} -  x_{7} x_{13} x_{14} x_{16} x_{18}, x_{1} x_{7} x_{9} x_{10} x_{17} -  x_{3} x_{13} x_{14} x_{16} x_{18}, x_{1} x_{8} x_{9} x_{10} x_{17} -  x_{2} x_{15} x_{16} x_{18} x_{19}, x_{1} x_{2} x_{3} x_{11}x_{
17} -  x_{7} x_{8} x_{12} x_{16} x_{18}, x_{1} x_{4} x_{6} x_{11} x_{17} -  x_{8} x_{10} x_{14} x_{15} x_{19}, x_{1} x_{2} x_{7} x_{11} x_{17} -  x_{3} x_{8} x_{12} x_{16} x_{18}, x_{1} x_{3} x_{7} x_{11} x_{17} -  x_{2} x_{8} x_{12} x_{16} x_{18}, x_{2} x_{3} x_{7} x_{11} x_{17} -  x_{1} x_{8} x_{12} x_{16} x_{18}, x_{1} x_{2} x_{8} x_{11} x_{17} -  x_{3} x_{7} x_{12} x_{16} x_{18}, x_{1} x_{3} x_{8} x_{11} x_{17} -  x_{2} x_{7} x_{12} x_{16} x_{18}, x_{2} x_{3} x_{8} x_{11} x_{17} -  x_{1} x_{7} x_{12} x_{16} x_{18}, x_{1} x_{4} x_{8} x_{11} x_{17} -  x_{6} x_{10} x_{14} x_{15} x_{19}, x_{1} x_{6} x_{8} x_{11} x_{17} -  x_{4} x_{10} x_{14} x_{15} x_{19}, x_{4} x_{6} x_{8} x_{11} x_{17} -  x_{1} x_{10} x_{14} x_{15} x_{19}, x_{1} x_{7} x_{8} x_{11} x_{17} -  x_{2} x_{3} x_{12} x_{16} x_{18}, x_{2} x_{7} x_{8} x_{11} x_{17} -  x_{1} x_{3} x_{12} x_{16} x_{18}, x_{3} x_{7} x_{8} x_{11} x_{17} -  x_{1} x_{2} x_{12} x_{16} x_{18}, x_{1} x_{4} x_{10} x_{11} x_{17} -  x_{6} x_{8} x_{14} x_{15} x_{19}, x_{1} x_{6} 
x_{10} x_{11} x_{17} -  x_{4} x_{8} x_{14} x_{15} x_{19}, x_{4} x_{6} x_{10} x_{11} x_{17} -  x_{1} x_{8} x_{14} x_{15} x_{19}, x_{1} x_{8} x_{10} x_{11} x_{17} -  x_{4} x_{6} x_{14} x_{15} x_{19}, x_{4} x_{8} x_{10} x_{11} x_{17} -  x_{1} x_{6} x_{14} x_{15} x_{19}, x_{6} x_{8} x_{10} x_{11} x_{17} -  x_{1} x_{4} x_{14} x_{15} x_{19}, x_{1} x_{2} x_{3} x_{12} x_{17} -  x_{7} x_{8} x_{11} x_{16} x_{18}, x_{1} x_{2} x_{4} x_{12} x_{17} -  x_{6} x_{9} x_{13} x_{15} x_{19}, x_{1} x_{2} x_{6} x_{12} x_{17} -  x_{4} x_{9} x_{13} x_{15} x_{19}, x_{1} x_{4} x_{6} x_{12} x_{17} -  x_{2} x_{9} x_{13} x_{15} x_{19}, x_{2} x_{4} x_{6} x_{12} x_{17} -  x_{1} x_{9} x_{13} x_{15} x_{19}, x_{1} x_{2} x_{7} x_{12} x_{17} -  x_{3} x_{8} x_{11} x_{16} x_{18}, x_{1} x_{3} x_{7} x_{12} x_{17} -  x_{2} x_{8} x_{11} x_{16} x_{18}, x_{2} x_{3} x_{7} x_{12} x_{17} -  x_{1} x_{8} x_{11} x_{16} x_{18}, x_{1} x_{2} x_{8} x_{12} x_{17} -  x_{3} x_{7} x_{11} x_{16} x_{18}, x_{1} x_{3} x_{8} x_{12} x_{17} -  x_{2} x_{7} x_{11} x_{16}x_{
18}, x_{2} x_{3} x_{8} x_{12} x_{17} -  x_{1} x_{7} x_{11} x_{16} x_{18}, x_{1} x_{7} x_{8} x_{12} x_{17} -  x_{2} x_{3} x_{11} x_{16} x_{18}, x_{2} x_{7} x_{8} x_{12} x_{17} -  x_{1} x_{3} x_{11} x_{16} x_{18}, x_{3} x_{7} x_{8} x_{12} x_{17} -  x_{1} x_{2} x_{11} x_{16} x_{18}, x_{1} x_{2} x_{9} x_{12} x_{17} -  x_{4} x_{6} x_{13} x_{15} x_{19}, x_{1} x_{4} x_{9} x_{12} x_{17} -  x_{2} x_{6} x_{13} x_{15} x_{19}, x_{2} x_{4} x_{9} x_{12} x_{17} -  x_{1} x_{6} x_{13} x_{15} x_{19}, x_{1} x_{6} x_{9} x_{12} x_{17} -  x_{2} x_{4} x_{13} x_{15} x_{19}, x_{2} x_{6} x_{9} x_{12} x_{17} -  x_{1} x_{4} x_{13} x_{15} x_{19}, x_{4} x_{6} x_{9} x_{12} x_{17} -  x_{1} x_{2} x_{13} x_{15} x_{19}, x_{1} x_{2} x_{11} x_{12} x_{17} -  x_{3} x_{7} x_{8} x_{16} x_{18}, x_{1} x_{3} x_{11} x_{12} x_{17} -  x_{2} x_{7} x_{8} x_{16} x_{18}, x_{2} x_{3} x_{11} x_{12} x_{17} -  x_{1} x_{7} x_{8} x_{16} x_{18}, x_{1} x_{7} x_{11} x_{12} x_{17} -  x_{2} x_{3} x_{8} x_{16} x_{18}, x_{2} x_{7} x_{11} x_{12} x_{17} -  x_{1} x_{3} x_{8}
 x_{16} x_{18}, x_{1} x_{8} x_{11} x_{12} x_{17} -  x_{2} x_{3} x_{7} x_{16} x_{18}, x_{3} x_{8} x_{11} x_{12} x_{17} -  x_{1} x_{2} x_{7} x_{16} x_{18}, x_{7} x_{8} x_{11} x_{12} x_{17} -  x_{1} x_{2} x_{3} x_{16} x_{18}, x_{1} x_{2} x_{4} x_{13} x_{17} -  x_{6} x_{9} x_{12} x_{15} x_{19}, x_{1} x_{2} x_{6} x_{13} x_{17} -  x_{4} x_{9} x_{12} x_{15} x_{19}, x_{1} x_{4} x_{6} x_{13} x_{17} -  x_{2} x_{9} x_{12} x_{15} x_{19}, x_{2} x_{4} x_{6} x_{13} x_{17} -  x_{1} x_{9} x_{12} x_{15} x_{19}, x_{1} x_{3} x_{7} x_{13} x_{17} -  x_{9} x_{10} x_{14} x_{16} x_{18}, x_{1} x_{2} x_{9} x_{13} x_{17} -  x_{4} x_{6} x_{12} x_{15} x_{19}, x_{1} x_{3} x_{9} x_{13} x_{17} -  x_{7} x_{10} x_{14} x_{16} x_{18}, x_{1} x_{4} x_{9} x_{13} x_{17} -  x_{2} x_{6} x_{12} x_{15} x_{19}, x_{2} x_{4} x_{9} x_{13} x_{17} -  x_{1} x_{6} x_{12} x_{15} x_{19}, x_{1} x_{6} x_{9} x_{13} x_{17} -  x_{2} x_{4} x_{12} x_{15} x_{19}, x_{2} x_{6} x_{9} x_{13} x_{17} -  x_{1} x_{4} x_{12} x_{15} x_{19}, x_{4} x_{6} x_{9} x_{13} x_{17} -  x_{1}
 x_{2} x_{12} x_{15} x_{19}, x_{1} x_{7} x_{9} x_{13} x_{17} -  x_{3} x_{10} x_{14} x_{16} x_{18}, x_{3} x_{7} x_{9} x_{13} x_{17} -  x_{1} x_{10} x_{14} x_{16} x_{18}, x_{1} x_{3} x_{10} x_{13} x_{17} -  x_{7} x_{9} x_{14} x_{16} x_{18}, x_{1} x_{7} x_{10} x_{13} x_{17} -  x_{3} x_{9} x_{14} x_{16} x_{18}, x_{3} x_{7} x_{10} x_{13} x_{17} -  x_{1} x_{9} x_{14} x_{16} x_{18}, x_{1} x_{9} x_{10} x_{13} x_{17} -  x_{3} x_{7} x_{14} x_{16} x_{18}, x_{3} x_{9} x_{10} x_{13} x_{17} -  x_{1} x_{7} x_{14} x_{16} x_{18}, x_{7} x_{9} x_{10} x_{13} x_{17} -  x_{1} x_{3} x_{14} x_{16} x_{18}, x_{1} x_{2} x_{12} x_{13} x_{17} -  x_{4} x_{6} x_{9} x_{15} x_{19}, x_{1} x_{4} x_{12} x_{13} x_{17} -  x_{2} x_{6} x_{9} x_{15} x_{19}, x_{2} x_{4} x_{12} x_{13} x_{17} -  x_{1} x_{6} x_{9} x_{15} x_{19}, x_{1} x_{6} x_{12} x_{13} x_{17} -  x_{2} x_{4} x_{9} x_{15} x_{19}, x_{2} x_{6} x_{12} x_{13} x_{17} -  x_{1} x_{4} x_{9} x_{15} x_{19}, x_{1} x_{9} x_{12} x_{13} x_{17} -  x_{2} x_{4} x_{6} x_{15} x_{19}, x_{4} x_{9} x_{12} x_
{13} x_{17} -  x_{1} x_{2} x_{6} x_{15} x_{19}, x_{6} x_{9} x_{12} x_{13} x_{17} -  x_{1} x_{2} x_{4} x_{15} x_{19}, x_{1} x_{11} x_{12} x_{13} x_{17} -  x_{14} x_{15} x_{16} x_{18} x_{19}, x_{1} x_{4} x_{6} x_{14} x_{17} -  x_{8} x_{10} x_{11} x_{15} x_{19}, x_{1} x_{3} x_{7} x_{14} x_{17} -  x_{9} x_{10} x_{13} x_{16} x_{18}, x_{1} x_{4} x_{8} x_{14} x_{17} -  x_{6} x_{10} x_{11} x_{15} x_{19}, x_{1} x_{6} x_{8} x_{14} x_{17} -  x_{4} x_{10} x_{11} x_{15} x_{19}, x_{4} x_{6} x_{8} x_{14} x_{17} -  x_{1} x_{10} x_{11} x_{15} x_{19}, x_{1} x_{3} x_{9} x_{14} x_{17} -  x_{7} x_{10} x_{13} x_{16} x_{18}, x_{1} x_{7} x_{9} x_{14} x_{17} -  x_{3} x_{10} x_{13} x_{16} x_{18}, x_{3} x_{7} x_{9} x_{14} x_{17} -  x_{1} x_{10} x_{13} x_{16} x_{18}, x_{1} x_{3} x_{10} x_{14} x_{17} -  x_{7} x_{9} x_{13} x_{16} x_{18}, x_{1} x_{4} x_{10} x_{14} x_{17} -  x_{6} x_{8} x_{11} x_{15} x_{19}, x_{1} x_{6} x_{10} x_{14} x_{17} -  x_{4} x_{8} x_{11} x_{15} x_{19}, x_{4} x_{6} x_{10} x_{14} x_{17} -  x_{1} x_{8} x_{11} x_{15}
 x_{19}, x_{1} x_{7} x_{10} x_{14} x_{17} -  x_{3} x_{9} x_{13} x_{16} x_{18}, x_{3} x_{7} x_{10} x_{14} x_{17} -  x_{1} x_{9} x_{13} x_{16} x_{18}, x_{1} x_{8} x_{10} x_{14} x_{17} -  x_{4} x_{6} x_{11} x_{15} x_{19}, x_{4} x_{8} x_{10} x_{14} x_{17} -  x_{1} x_{6} x_{11} x_{15} x_{19}, x_{6} x_{8} x_{10} x_{14} x_{17} -  x_{1} x_{4} x_{11} x_{15} x_{19}, x_{1} x_{9} x_{10} x_{14} x_{17} -  x_{3} x_{7} x_{13} x_{16} x_{18}, x_{3} x_{9} x_{10} x_{14} x_{17} -  x_{1} x_{7} x_{13} x_{16} x_{18}, x_{7} x_{9} x_{10} x_{14} x_{17} -  x_{1} x_{3} x_{13} x_{16} x_{18}, x_{1} x_{4} x_{11} x_{14} x_{17} -  x_{6} x_{8} x_{10} x_{15} x_{19}, x_{1} x_{6} x_{11} x_{14} x_{17} -  x_{4} x_{8} x_{10} x_{15} x_{19}, x_{1} x_{8} x_{11} x_{14} x_{17} -  x_{4} x_{6} x_{10} x_{15} x_{19}, x_{4} x_{8} x_{11} x_{14} x_{17} -  x_{1} x_{6} x_{10} x_{15} x_{19}, x_{6} x_{8} x_{11} x_{14} x_{17} -  x_{1} x_{4} x_{10} x_{15} x_{19}, x_{1} x_{10} x_{11} x_{14} x_{17} -  x_{4} x_{6} x_{8} x_{15} x_{19}, x_{4} x_{10} x_{11} x_{14} x_{17} -
  x_{1} x_{6} x_{8} x_{15} x_{19}, x_{6} x_{10} x_{11} x_{14} x_{17} -  x_{1} x_{4} x_{8} x_{15} x_{19}, x_{8} x_{10} x_{11} x_{14} x_{17} -  x_{1} x_{4} x_{6} x_{15} x_{19}, x_{1} x_{11} x_{12} x_{14} x_{17} -  x_{13} x_{15} x_{16} x_{18} x_{19}, x_{1} x_{3} x_{13} x_{14} x_{17} -  x_{7} x_{9} x_{10} x_{16} x_{18}, x_{1} x_{7} x_{13} x_{14} x_{17} -  x_{3} x_{9} x_{10} x_{16} x_{18}, x_{1} x_{9} x_{13} x_{14} x_{17} -  x_{3} x_{7} x_{10} x_{16} x_{18}, x_{3} x_{9} x_{13} x_{14} x_{17} -  x_{1} x_{7} x_{10} x_{16} x_{18}, x_{7} x_{9} x_{13} x_{14} x_{17} -  x_{1} x_{3} x_{10} x_{16} x_{18}, x_{1} x_{10} x_{13} x_{14} x_{17} -  x_{3} x_{7} x_{9} x_{16} x_{18}, x_{3} x_{10} x_{13} x_{14} x_{17} -  x_{1} x_{7} x_{9} x_{16} x_{18}, x_{7} x_{10} x_{13} x_{14} x_{17} -  x_{1} x_{3} x_{9} x_{16} x_{18}, x_{9} x_{10} x_{13} x_{14} x_{17} -  x_{1} x_{3} x_{7} x_{16} x_{18}, x_{1} x_{11} x_{13} x_{14} x_{17} -  x_{12} x_{15} x_{16} x_{18} x_{19}, x_{1} x_{12} x_{13} x_{14} x_{17} -  x_{11} x_{15} x_{16} x_{18} x_{19},
x_{11} x_{12} x_{13} x_{14} x_{17} -  x_{1} x_{15} x_{16} x_{18} x_{19}, x_{1} x_{2} x_{4} x_{15} x_{17} -  x_{6} x_{9} x_{12} x_{13} x_{19}, x_{1} x_{2} x_{6} x_{15} x_{17} -  x_{4} x_{9} x_{12} x_{13} x_{19}, x_{1} x_{4} x_{6} x_{15} x_{17} -  x_{8} x_{10} x_{11} x_{14} x_{19}, x_{2} x_{4} x_{6} x_{15} x_{17} -  x_{1} x_{9} x_{12} x_{13} x_{19}, x_{1} x_{2} x_{8} x_{15} x_{17} -  x_{9} x_{10} x_{16} x_{18} x_{19}, x_{1} x_{6} x_{8} x_{15} x_{17} -  x_{4} x_{10} x_{11} x_{14} x_{19}, x_{1} x_{2} x_{9} x_{15} x_{17} -  x_{8} x_{10} x_{16} x_{18} x_{19}, x_{1} x_{4} x_{9} x_{15} x_{17} -  x_{2} x_{6} x_{12} x_{13} x_{19}, x_{2} x_{4} x_{9} x_{15} x_{17} -  x_{1} x_{6} x_{12} x_{13} x_{19}, x_{1} x_{6} x_{9} x_{15} x_{17} -  x_{2} x_{4} x_{12} x_{13} x_{19}, x_{2} x_{6} x_{9} x_{15} x_{17} -  x_{1} x_{4} x_{12} x_{13} x_{19}, x_{4} x_{6} x_{9} x_{15} x_{17} -  x_{1} x_{2} x_{12} x_{13} x_{19}, x_{1} x_{8} x_{9} x_{15} x_{17} -  x_{2} x_{10} x_{16} x_{18} x_{19}, x_{2} x_{8} x_{9} x_{15} x_{17} -  x_{1} x_{10}x_
{16} x_{18} x_{19}, x_{1} x_{2} x_{10} x_{15} x_{17} -  x_{8} x_{9} x_{16} x_{18} x_{19}, x_{1} x_{4} x_{10} x_{15} x_{17} -  x_{6} x_{8} x_{11} x_{14} x_{19}, x_{1} x_{6} x_{10} x_{15} x_{17} -  x_{4} x_{8} x_{11} x_{14} x_{19}, x_{4} x_{6} x_{10} x_{15} x_{17} -  x_{1} x_{8} x_{11} x_{14} x_{19}, x_{1} x_{8} x_{10} x_{15} x_{17} -  x_{2} x_{9} x_{16} x_{18} x_{19}, x_{2} x_{8} x_{10} x_{15} x_{17} -  x_{1} x_{9} x_{16} x_{18} x_{19}, x_{6} x_{8} x_{10} x_{15} x_{17} -  x_{1} x_{4} x_{11} x_{14} x_{19}, x_{1} x_{9} x_{10} x_{15} x_{17} -  x_{2} x_{8} x_{16} x_{18} x_{19}, x_{2} x_{9} x_{10} x_{15} x_{17} -  x_{1} x_{8} x_{16} x_{18} x_{19}, x_{8} x_{9} x_{10} x_{15} x_{17} -  x_{1} x_{2} x_{16} x_{18} x_{19}, x_{1} x_{4} x_{11} x_{15} x_{17} -  x_{6} x_{8} x_{10} x_{14} x_{19}, x_{1} x_{6} x_{11} x_{15} x_{17} -  x_{4} x_{8} x_{10} x_{14} x_{19}, x_{4} x_{6} x_{11} x_{15} x_{17} -  x_{1} x_{8} x_{10} x_{14} x_{19}, x_{1} x_{8} x_{11} x_{15} x_{17} -  x_{4} x_{6} x_{10} x_{14} x_{19}, x_{6} x_{8} x_{11} x_{
15} x_{17} -  x_{1} x_{4} x_{10} x_{14} x_{19}, x_{1} x_{10} x_{11} x_{15} x_{17} -  x_{4} x_{6} x_{8} x_{14} x_{19}, x_{4} x_{10} x_{11} x_{15} x_{17} -  x_{1} x_{6} x_{8} x_{14} x_{19}, x_{6} x_{10} x_{11} x_{15} x_{17} -  x_{1} x_{4} x_{8} x_{14} x_{19}, x_{8} x_{10} x_{11} x_{15} x_{17} -  x_{1} x_{4} x_{6} x_{14} x_{19}, x_{1} x_{2} x_{12} x_{15} x_{17} -  x_{4} x_{6} x_{9} x_{13} x_{19}, x_{1} x_{4} x_{12} x_{15} x_{17} -  x_{2} x_{6} x_{9} x_{13} x_{19}, x_{2} x_{4} x_{12} x_{15} x_{17} -  x_{1} x_{6} x_{9} x_{13} x_{19}, x_{1} x_{6} x_{12} x_{15} x_{17} -  x_{2} x_{4} x_{9} x_{13} x_{19}, x_{2} x_{6} x_{12} x_{15} x_{17} -  x_{1} x_{4} x_{9} x_{13} x_{19}, x_{4} x_{6} x_{12} x_{15} x_{17} -  x_{1} x_{2} x_{9} x_{13} x_{19}, x_{1} x_{9} x_{12} x_{15} x_{17} -  x_{2} x_{4} x_{6} x_{13} x_{19}, x_{2} x_{9} x_{12} x_{15} x_{17} -  x_{1} x_{4} x_{6} x_{13} x_{19}, x_{4} x_{9} x_{12} x_{15} x_{17} -  x_{1} x_{2} x_{6} x_{13} x_{19}, x_{6} x_{9} x_{12} x_{15} x_{17} -  x_{1} x_{2} x_{4} x_{13} x_{19}, x_{1}
x_{11} x_{12} x_{15} x_{17} -  x_{13} x_{14} x_{16} x_{18} x_{19}, x_{1} x_{2} x_{13} x_{15} x_{17} -  x_{4} x_{6} x_{9} x_{12} x_{19}, x_{1} x_{6} x_{13} x_{15} x_{17} -  x_{2} x_{4} x_{9} x_{12} x_{19}, x_{2} x_{6} x_{13} x_{15} x_{17} -  x_{1} x_{4} x_{9} x_{12} x_{19}, x_{1} x_{9} x_{13} x_{15} x_{17} -  x_{2} x_{4} x_{6} x_{12} x_{19}, x_{2} x_{9} x_{13} x_{15} x_{17} -  x_{1} x_{4} x_{6} x_{12} x_{19}, x_{6} x_{9} x_{13} x_{15} x_{17} -  x_{1} x_{2} x_{4} x_{12} x_{19}, x_{1} x_{11} x_{13} x_{15} x_{17} -  x_{12} x_{14} x_{16} x_{18} x_{19}, x_{1} x_{12} x_{13} x_{15} x_{17} -  x_{11} x_{14} x_{16} x_{18} x_{19}, x_{2} x_{12} x_{13} x_{15} x_{17} -  x_{1} x_{4} x_{6} x_{9} x_{19}, x_{6} x_{12} x_{13} x_{15} x_{17} -  x_{1} x_{2} x_{4} x_{9} x_{19}, x_{9} x_{12} x_{13} x_{15} x_{17} -  x_{1} x_{2} x_{4} x_{6} x_{19}, x_{11} x_{12} x_{13} x_{15} x_{17} -  x_{1} x_{14} x_{16} x_{18} x_{19}, x_{1} x_{4} x_{14} x_{15} x_{17} -  x_{6} x_{8} x_{10} x_{11} x_{19}, x_{1} x_{6} x_{14} x_{15} x_{17} -  x_{4} x_{8}
 x_{10} x_{11} x_{19}, x_{4} x_{6} x_{14} x_{15} x_{17} -  x_{1} x_{8} x_{10} x_{11} x_{19}, x_{1} x_{8} x_{14} x_{15} x_{17} -  x_{4} x_{6} x_{10} x_{11} x_{19}, x_{6} x_{8} x_{14} x_{15} x_{17} -  x_{1} x_{4} x_{10} x_{11} x_{19}, x_{1} x_{10} x_{14} x_{15} x_{17} -  x_{4} x_{6} x_{8} x_{11} x_{19}, x_{4} x_{10} x_{14} x_{15} x_{17} -  x_{1} x_{6} x_{8} x_{11} x_{19}, x_{6} x_{10} x_{14} x_{15} x_{17} -  x_{1} x_{4} x_{8} x_{11} x_{19}, x_{8} x_{10} x_{14} x_{15} x_{17} -  x_{1} x_{4} x_{6} x_{11} x_{19}, x_{1} x_{11} x_{14} x_{15} x_{17} -  x_{12} x_{13} x_{16} x_{18} x_{19}, x_{4} x_{11} x_{14} x_{15} x_{17} -  x_{1} x_{6} x_{8} x_{10} x_{19}, x_{6} x_{11} x_{14} x_{15} x_{17} -  x_{1} x_{4} x_{8} x_{10} x_{19}, x_{8} x_{11} x_{14} x_{15} x_{17} -  x_{1} x_{4} x_{6} x_{10} x_{19}, x_{10} x_{11} x_{14} x_{15} x_{17} -  x_{1} x_{4} x_{6} x_{8} x_{19}, x_{1} x_{12} x_{14} x_{15} x_{17} -  x_{11} x_{13} x_{16} x_{18} x_{19}, x_{11} x_{12} x_{14} x_{15} x_{17} -  x_{1} x_{13} x_{16} x_{18} x_{19}, x_{1} x_{13}
 x_{14} x_{15} x_{17} -  x_{11} x_{12} x_{16} x_{18} x_{19}, x_{11} x_{13} x_{14} x_{15} x_{17} -  x_{1} x_{12} x_{16} x_{18} x_{19}, x_{12} x_{13} x_{14} x_{15} x_{17} -  x_{1} x_{11} x_{16} x_{18} x_{19}, x_{1} x_{2} x_{3} x_{16} x_{17} -  x_{7} x_{8} x_{11} x_{12} x_{18}, x_{1} x_{2} x_{7} x_{16} x_{17} -  x_{3} x_{8} x_{11} x_{12} x_{18}, x_{1} x_{3} x_{7} x_{16} x_{17} -  x_{9} x_{10} x_{13} x_{14} x_{18}, x_{2} x_{3} x_{7} x_{16} x_{17} -  x_{1} x_{8} x_{11} x_{12} x_{18}, x_{1} x_{2} x_{8} x_{16} x_{17} -  x_{9} x_{10} x_{15} x_{18} x_{19}, x_{1} x_{3} x_{8} x_{16} x_{17} -  x_{2} x_{7} x_{11} x_{12} x_{18}, x_{2} x_{3} x_{8} x_{16} x_{17} -  x_{1} x_{7} x_{11} x_{12} x_{18}, x_{1} x_{7} x_{8} x_{16} x_{17} -  x_{2} x_{3} x_{11} x_{12} x_{18}, x_{2} x_{7} x_{8} x_{16} x_{17} -  x_{1} x_{3} x_{11} x_{12} x_{18}, x_{3} x_{7} x_{8} x_{16} x_{17} -  x_{1} x_{2} x_{11} x_{12} x_{18}, x_{1} x_{2} x_{9} x_{16} x_{17} -  x_{8} x_{10} x_{15} x_{18} x_{19}, x_{1} x_{7} x_{9} x_{16} x_{17} -  x_{3} x_{10} x_{13}
x_{14} x_{18}, x_{1} x_{8} x_{9} x_{16} x_{17} -  x_{2} x_{10} x_{15} x_{18} x_{19}, x_{2} x_{8} x_{9} x_{16} x_{17} -  x_{1} x_{10} x_{15} x_{18} x_{19}, x_{1} x_{2} x_{10} x_{16} x_{17} -  x_{8} x_{9} x_{15} x_{18} x_{19}, x_{1} x_{3} x_{10} x_{16} x_{17} -  x_{7} x_{9} x_{13} x_{14} x_{18}, x_{1} x_{7} x_{10} x_{16} x_{17} -  x_{3} x_{9} x_{13} x_{14} x_{18}, x_{3} x_{7} x_{10} x_{16} x_{17} -  x_{1} x_{9} x_{13} x_{14} x_{18}, x_{1} x_{8} x_{10} x_{16} x_{17} -  x_{2} x_{9} x_{15} x_{18} x_{19}, x_{2} x_{8} x_{10} x_{16} x_{17} -  x_{1} x_{9} x_{15} x_{18} x_{19}, x_{1} x_{9} x_{10} x_{16} x_{17} -  x_{2} x_{8} x_{15} x_{18} x_{19}, x_{2} x_{9} x_{10} x_{16} x_{17} -  x_{1} x_{8} x_{15} x_{18} x_{19}, x_{7} x_{9} x_{10} x_{16} x_{17} -  x_{1} x_{3} x_{13} x_{14} x_{18}, x_{8} x_{9} x_{10} x_{16} x_{17} -  x_{1} x_{2} x_{15} x_{18} x_{19}, x_{1} x_{2} x_{11} x_{16} x_{17} -  x_{3} x_{7} x_{8} x_{12} x_{18}, x_{1} x_{7} x_{11} x_{16} x_{17} -  x_{2} x_{3} x_{8} x_{12} x_{18}, x_{2} x_{7} x_{11} x_{16} x_{
17} -  x_{1} x_{3} x_{8} x_{12} x_{18}, x_{1} x_{8} x_{11} x_{16} x_{17} -  x_{2} x_{3} x_{7} x_{12} x_{18}, x_{2} x_{8} x_{11} x_{16} x_{17} -  x_{1} x_{3} x_{7} x_{12} x_{18}, x_{7} x_{8} x_{11} x_{16} x_{17} -  x_{1} x_{2} x_{3} x_{12} x_{18}, x_{1} x_{2} x_{12} x_{16} x_{17} -  x_{3} x_{7} x_{8} x_{11} x_{18}, x_{1} x_{3} x_{12} x_{16} x_{17} -  x_{2} x_{7} x_{8} x_{11} x_{18}, x_{2} x_{3} x_{12} x_{16} x_{17} -  x_{1} x_{7} x_{8} x_{11} x_{18}, x_{1} x_{7} x_{12} x_{16} x_{17} -  x_{2} x_{3} x_{8} x_{11} x_{18}, x_{2} x_{7} x_{12} x_{16} x_{17} -  x_{1} x_{3} x_{8} x_{11} x_{18}, x_{3} x_{7} x_{12} x_{16} x_{17} -  x_{1} x_{2} x_{8} x_{11} x_{18}, x_{1} x_{8} x_{12} x_{16} x_{17} -  x_{2} x_{3} x_{7} x_{11} x_{18}, x_{2} x_{8} x_{12} x_{16} x_{17} -  x_{1} x_{3} x_{7} x_{11} x_{18}, x_{3} x_{8} x_{12} x_{16} x_{17} -  x_{1} x_{2} x_{7} x_{11} x_{18}, x_{7} x_{8} x_{12} x_{16} x_{17} -  x_{1} x_{2} x_{3} x_{11} x_{18}, x_{1} x_{11} x_{12} x_{16} x_{17} -  x_{13} x_{14} x_{15} x_{18} x_{19}, x_{2} x_{11}x_
{12} x_{16} x_{17} -  x_{1} x_{3} x_{7} x_{8} x_{18}, x_{7} x_{11} x_{12} x_{16} x_{17} -  x_{1} x_{2} x_{3} x_{8} x_{18}, x_{8} x_{11} x_{12} x_{16} x_{17} -  x_{1} x_{2} x_{3} x_{7} x_{18}, x_{1} x_{3} x_{13} x_{16} x_{17} -  x_{7} x_{9} x_{10} x_{14} x_{18}, x_{1} x_{7} x_{13} x_{16} x_{17} -  x_{3} x_{9} x_{10} x_{14} x_{18}, x_{3} x_{7} x_{13} x_{16} x_{17} -  x_{1} x_{9} x_{10} x_{14} x_{18}, x_{1} x_{9} x_{13} x_{16} x_{17} -  x_{3} x_{7} x_{10} x_{14} x_{18}, x_{7} x_{9} x_{13} x_{16} x_{17} -  x_{1} x_{3} x_{10} x_{14} x_{18}, x_{1} x_{10} x_{13} x_{16} x_{17} -  x_{3} x_{7} x_{9} x_{14} x_{18}, x_{3} x_{10} x_{13} x_{16} x_{17} -  x_{1} x_{7} x_{9} x_{14} x_{18}, x_{7} x_{10} x_{13} x_{16} x_{17} -  x_{1} x_{3} x_{9} x_{14} x_{18}, x_{9} x_{10} x_{13} x_{16} x_{17} -  x_{1} x_{3} x_{7} x_{14} x_{18}, x_{1} x_{11} x_{13} x_{16} x_{17} -  x_{12} x_{14} x_{15} x_{18} x_{19}, x_{1} x_{12} x_{13} x_{16} x_{17} -  x_{11} x_{14} x_{15} x_{18} x_{19}, x_{11} x_{12} x_{13} x_{16} x_{17} -  x_{1} x_{14} x_{
15} x_{18} x_{19}, x_{1} x_{3} x_{14} x_{16} x_{17} -  x_{7} x_{9} x_{10} x_{13} x_{18}, x_{1} x_{7} x_{14} x_{16} x_{17} -  x_{3} x_{9} x_{10} x_{13} x_{18}, x_{3} x_{7} x_{14} x_{16} x_{17} -  x_{1} x_{9} x_{10} x_{13} x_{18}, x_{1} x_{9} x_{14} x_{16} x_{17} -  x_{3} x_{7} x_{10} x_{13} x_{18}, x_{7} x_{9} x_{14} x_{16} x_{17} -  x_{1} x_{3} x_{10} x_{13} x_{18}, x_{1} x_{10} x_{14} x_{16} x_{17} -  x_{3} x_{7} x_{9} x_{13} x_{18}, x_{3} x_{10} x_{14} x_{16} x_{17} -  x_{1} x_{7} x_{9} x_{13} x_{18}, x_{7} x_{10} x_{14} x_{16} x_{17} -  x_{1} x_{3} x_{9} x_{13} x_{18}, x_{9} x_{10} x_{14} x_{16} x_{17} -  x_{1} x_{3} x_{7} x_{13} x_{18}, x_{1} x_{11} x_{14} x_{16} x_{17} -  x_{12} x_{13} x_{15} x_{18} x_{19}, x_{1} x_{12} x_{14} x_{16} x_{17} -  x_{11} x_{13} x_{15} x_{18} x_{19}, x_{11} x_{12} x_{14} x_{16} x_{17} -  x_{1} x_{13} x_{15} x_{18} x_{19}, x_{1} x_{13} x_{14} x_{16} x_{17} -  x_{11} x_{12} x_{15} x_{18} x_{19}, x_{3} x_{13} x_{14} x_{16} x_{17} -  x_{1} x_{7} x_{9} x_{10} x_{18}, x_{7} x_{13}
x_{14} x_{16} x_{17} -  x_{1} x_{3} x_{9} x_{10} x_{18}, x_{9} x_{13} x_{14} x_{16} x_{17} -  x_{1} x_{3} x_{7} x_{10} x_{18}, x_{10} x_{13} x_{14} x_{16} x_{17} -  x_{1} x_{3} x_{7} x_{9} x_{18}, x_{11} x_{13} x_{14} x_{16} x_{17} -  x_{1} x_{12} x_{15} x_{18} x_{19}, x_{12} x_{13} x_{14} x_{16} x_{17} -  x_{1} x_{11} x_{15} x_{18} x_{19}, x_{1} x_{2} x_{15} x_{16} x_{17} -  x_{8} x_{9} x_{10} x_{18} x_{19}, x_{1} x_{8} x_{15} x_{16} x_{17} -  x_{2} x_{9} x_{10} x_{18} x_{19}, x_{2} x_{8} x_{15} x_{16} x_{17} -  x_{1} x_{9} x_{10} x_{18} x_{19}, x_{1} x_{9} x_{15} x_{16} x_{17} -  x_{2} x_{8} x_{10} x_{18} x_{19}, x_{2} x_{9} x_{15} x_{16} x_{17} -  x_{1} x_{8} x_{10} x_{18} x_{19}, x_{8} x_{9} x_{15} x_{16} x_{17} -  x_{1} x_{2} x_{10} x_{18} x_{19}, x_{1} x_{10} x_{15} x_{16} x_{17} -  x_{2} x_{8} x_{9} x_{18} x_{19}, x_{2} x_{10} x_{15} x_{16} x_{17} -  x_{1} x_{8} x_{9} x_{18} x_{19}, x_{8} x_{10} x_{15} x_{16} x_{17} -  x_{1} x_{2} x_{9} x_{18} x_{19}, x_{9} x_{10} x_{15} x_{16} x_{17} -  x_{1} x_{2} x_
{8} x_{18} x_{19}, x_{1} x_{11} x_{15} x_{16} x_{17} -  x_{12} x_{13} x_{14} x_{18} x_{19}, x_{1} x_{12} x_{15} x_{16} x_{17} -  x_{11} x_{13} x_{14} x_{18} x_{19}, x_{11} x_{12} x_{15} x_{16} x_{17} -  x_{1} x_{13} x_{14} x_{18} x_{19}, x_{1} x_{13} x_{15} x_{16} x_{17} -  x_{11} x_{12} x_{14} x_{18} x_{19}, x_{11} x_{13} x_{15} x_{16} x_{17} -  x_{1} x_{12} x_{14} x_{18} x_{19}, x_{12} x_{13} x_{15} x_{16} x_{17} -  x_{1} x_{11} x_{14} x_{18} x_{19}, x_{1} x_{14} x_{15} x_{16} x_{17} -  x_{11} x_{12} x_{13} x_{18} x_{19}, x_{11} x_{14} x_{15} x_{16} x_{17} -  x_{1} x_{12} x_{13} x_{18} x_{19}, x_{12} x_{14} x_{15} x_{16} x_{17} -  x_{1} x_{11} x_{13} x_{18} x_{19}, x_{13} x_{14} x_{15} x_{16} x_{17} -  x_{1} x_{11} x_{12} x_{18} x_{19}, x_{1} x_{2} x_{5} x_{6} x_{18} -  x_{7} x_{8} x_{9} x_{12} x_{19}, x_{1} x_{2} x_{5} x_{7} x_{18} -  x_{6} x_{8} x_{9} x_{12} x_{19}, x_{1} x_{2} x_{6} x_{7} x_{18} -  x_{5} x_{8} x_{9} x_{12} x_{19}, x_{1} x_{5} x_{6} x_{7} x_{18} -  x_{10} x_{11} x_{13} x_{14} x_{19}, x_{
2} x_{5} x_{6} x_{7} x_{18} -  x_{1} x_{8} x_{9} x_{12} x_{19}, x_{1} x_{2} x_{5} x_{8} x_{18} -  x_{6} x_{7} x_{9} x_{12} x_{19}, x_{1} x_{2} x_{6} x_{8} x_{18} -  x_{5} x_{7} x_{9} x_{12} x_{19}, x_{1} x_{5} x_{6} x_{8} x_{18} -  x_{2} x_{7} x_{9} x_{12} x_{19}, x_{2} x_{5} x_{6} x_{8} x_{18} -  x_{1} x_{7} x_{9} x_{12} x_{19}, x_{1} x_{2} x_{7} x_{8} x_{18} -  x_{5} x_{6} x_{9} x_{12} x_{19}, x_{1} x_{5} x_{7} x_{8} x_{18} -  x_{2} x_{6} x_{9} x_{12} x_{19}, x_{2} x_{5} x_{7} x_{8} x_{18} -  x_{1} x_{6} x_{9} x_{12} x_{19}, x_{1} x_{6} x_{7} x_{8} x_{18} -  x_{2} x_{5} x_{9} x_{12} x_{19}, x_{2} x_{6} x_{7} x_{8} x_{18} -  x_{1} x_{5} x_{9} x_{12} x_{19}, x_{5} x_{6} x_{7} x_{8} x_{18} -  x_{1} x_{2} x_{9} x_{12} x_{19}, x_{1} x_{2} x_{5} x_{9} x_{18} -  x_{6} x_{7} x_{8} x_{12} x_{19}, x_{1} x_{2} x_{6} x_{9} x_{18} -  x_{5} x_{7} x_{8} x_{12} x_{19}, x_{1} x_{5} x_{6} x_{9} x_{18} -  x_{2} x_{7} x_{8} x_{12} x_{19}, x_{2} x_{5} x_{6} x_{9} x_{18} -  x_{1} x_{7} x_{8} x_{12} x_{19}, x_{1} x_{2} x_{7} x_{
9} x_{18} -  x_{5} x_{6} x_{8} x_{12} x_{19}, x_{1} x_{5} x_{7} x_{9} x_{18} -  x_{2} x_{6} x_{8} x_{12} x_{19}, x_{2} x_{5} x_{7} x_{9} x_{18} -  x_{1} x_{6} x_{8} x_{12} x_{19}, x_{1} x_{6} x_{7} x_{9} x_{18} -  x_{2} x_{5} x_{8} x_{12} x_{19}, x_{2} x_{6} x_{7} x_{9} x_{18} -  x_{1} x_{5} x_{8} x_{12} x_{19}, x_{5} x_{6} x_{7} x_{9} x_{18} -  x_{1} x_{2} x_{8} x_{12} x_{19}, x_{1} x_{2} x_{8} x_{9} x_{18} -  x_{10} x_{15} x_{16} x_{17} x_{19}, x_{1} x_{5} x_{8} x_{9} x_{18} -  x_{2} x_{6} x_{7} x_{12} x_{19}, x_{2} x_{5} x_{8} x_{9} x_{18} -  x_{1} x_{6} x_{7} x_{12} x_{19}, x_{1} x_{6} x_{8} x_{9} x_{18} -  x_{2} x_{5} x_{7} x_{12} x_{19}, x_{2} x_{6} x_{8} x_{9} x_{18} -  x_{1} x_{5} x_{7} x_{12} x_{19}, x_{5} x_{6} x_{8} x_{9} x_{18} -  x_{1} x_{2} x_{7} x_{12} x_{19}, x_{1} x_{7} x_{8} x_{9} x_{18} -  x_{2} x_{5} x_{6} x_{12} x_{19}, x_{2} x_{7} x_{8} x_{9} x_{18} -  x_{1} x_{5} x_{6} x_{12} x_{19}, x_{5} x_{7} x_{8} x_{9} x_{18} -  x_{1} x_{2} x_{6} x_{12} x_{19}, x_{6} x_{7} x_{8} x_{9} x_{18} -  x_{
1} x_{2} x_{5} x_{12} x_{19}, x_{1} x_{5} x_{6} x_{10} x_{18} -  x_{7} x_{11} x_{13} x_{14} x_{19}, x_{1} x_{5} x_{7} x_{10} x_{18} -  x_{6} x_{11} x_{13} x_{14} x_{19}, x_{1} x_{6} x_{7} x_{10} x_{18} -  x_{5} x_{11} x_{13} x_{14} x_{19}, x_{1} x_{2} x_{8} x_{10} x_{18} -  x_{9} x_{15} x_{16} x_{17} x_{19}, x_{1} x_{2} x_{9} x_{10} x_{18} -  x_{8} x_{15} x_{16} x_{17} x_{19}, x_{1} x_{8} x_{9} x_{10} x_{18} -  x_{2} x_{15} x_{16} x_{17} x_{19}, x_{1} x_{5} x_{6} x_{11} x_{18} -  x_{7} x_{10} x_{13} x_{14} x_{19}, x_{1} x_{5} x_{7} x_{11} x_{18} -  x_{6} x_{10} x_{13} x_{14} x_{19}, x_{1} x_{6} x_{7} x_{11} x_{18} -  x_{5} x_{10} x_{13} x_{14} x_{19}, x_{5} x_{6} x_{7} x_{11} x_{18} -  x_{1} x_{10} x_{13} x_{14} x_{19}, x_{1} x_{5} x_{10} x_{11} x_{18} -  x_{6} x_{7} x_{13} x_{14} x_{19}, x_{1} x_{6} x_{10} x_{11} x_{18} -  x_{5} x_{7} x_{13} x_{14} x_{19}, x_{5} x_{6} x_{10} x_{11} x_{18} -  x_{1} x_{7} x_{13} x_{14} x_{19}, x_{1} x_{7} x_{10} x_{11} x_{18} -  x_{5} x_{6} x_{13} x_{14} x_{19}, x_{5} x_{7} x_
{10} x_{11} x_{18} -  x_{1} x_{6} x_{13} x_{14} x_{19}, x_{6} x_{7} x_{10} x_{11} x_{18} -  x_{1} x_{5} x_{13} x_{14} x_{19}, x_{1} x_{2} x_{5} x_{12} x_{18} -  x_{6} x_{7} x_{8} x_{9} x_{19}, x_{1} x_{2} x_{6} x_{12} x_{18} -  x_{5} x_{7} x_{8} x_{9} x_{19}, x_{1} x_{5} x_{6} x_{12} x_{18} -  x_{2} x_{7} x_{8} x_{9} x_{19}, x_{2} x_{5} x_{6} x_{12} x_{18} -  x_{1} x_{7} x_{8} x_{9} x_{19}, x_{1} x_{2} x_{7} x_{12} x_{18} -  x_{5} x_{6} x_{8} x_{9} x_{19}, x_{1} x_{5} x_{7} x_{12} x_{18} -  x_{2} x_{6} x_{8} x_{9} x_{19}, x_{2} x_{5} x_{7} x_{12} x_{18} -  x_{1} x_{6} x_{8} x_{9} x_{19}, x_{1} x_{6} x_{7} x_{12} x_{18} -  x_{2} x_{5} x_{8} x_{9} x_{19}, x_{2} x_{6} x_{7} x_{12} x_{18} -  x_{1} x_{5} x_{8} x_{9} x_{19}, x_{1} x_{2} x_{8} x_{12} x_{18} -  x_{5} x_{6} x_{7} x_{9} x_{19}, x_{1} x_{5} x_{8} x_{12} x_{18} -  x_{2} x_{6} x_{7} x_{9} x_{19}, x_{2} x_{5} x_{8} x_{12} x_{18} -  x_{1} x_{6} x_{7} x_{9} x_{19}, x_{1} x_{6} x_{8} x_{12} x_{18} -  x_{2} x_{5} x_{7} x_{9} x_{19}, x_{2} x_{6} x_{8} x_{12} x_
{18} -  x_{1} x_{5} x_{7} x_{9} x_{19}, x_{5} x_{6} x_{8} x_{12} x_{18} -  x_{1} x_{2} x_{7} x_{9} x_{19}, x_{1} x_{7} x_{8} x_{12} x_{18} -  x_{2} x_{5} x_{6} x_{9} x_{19}, x_{2} x_{7} x_{8} x_{12} x_{18} -  x_{1} x_{5} x_{6} x_{9} x_{19}, x_{5} x_{7} x_{8} x_{12} x_{18} -  x_{1} x_{2} x_{6} x_{9} x_{19}, x_{6} x_{7} x_{8} x_{12} x_{18} -  x_{1} x_{2} x_{5} x_{9} x_{19}, x_{1} x_{2} x_{9} x_{12} x_{18} -  x_{5} x_{6} x_{7} x_{8} x_{19}, x_{1} x_{5} x_{9} x_{12} x_{18} -  x_{2} x_{6} x_{7} x_{8} x_{19}, x_{2} x_{5} x_{9} x_{12} x_{18} -  x_{1} x_{6} x_{7} x_{8} x_{19}, x_{1} x_{6} x_{9} x_{12} x_{18} -  x_{2} x_{5} x_{7} x_{8} x_{19}, x_{2} x_{6} x_{9} x_{12} x_{18} -  x_{1} x_{5} x_{7} x_{8} x_{19}, x_{5} x_{6} x_{9} x_{12} x_{18} -  x_{1} x_{2} x_{7} x_{8} x_{19}, x_{1} x_{7} x_{9} x_{12} x_{18} -  x_{2} x_{5} x_{6} x_{8} x_{19}, x_{2} x_{7} x_{9} x_{12} x_{18} -  x_{1} x_{5} x_{6} x_{8} x_{19}, x_{5} x_{7} x_{9} x_{12} x_{18} -  x_{1} x_{2} x_{6} x_{8} x_{19}, x_{6} x_{7} x_{9} x_{12} x_{18} -  x_{1} x_{2}
 x_{5} x_{8} x_{19}, x_{1} x_{8} x_{9} x_{12} x_{18} -  x_{2} x_{5} x_{6} x_{7} x_{19}, x_{5} x_{8} x_{9} x_{12} x_{18} -  x_{1} x_{2} x_{6} x_{7} x_{19}, x_{6} x_{8} x_{9} x_{12} x_{18} -  x_{1} x_{2} x_{5} x_{7} x_{19}, x_{7} x_{8} x_{9} x_{12} x_{18} -  x_{1} x_{2} x_{5} x_{6} x_{19}, x_{1} x_{5} x_{6} x_{13} x_{18} -  x_{7} x_{10} x_{11} x_{14} x_{19}, x_{1} x_{5} x_{7} x_{13} x_{18} -  x_{6} x_{10} x_{11} x_{14} x_{19}, x_{1} x_{6} x_{7} x_{13} x_{18} -  x_{5} x_{10} x_{11} x_{14} x_{19}, x_{5} x_{6} x_{7} x_{13} x_{18} -  x_{1} x_{10} x_{11} x_{14} x_{19}, x_{1} x_{5} x_{10} x_{13} x_{18} -  x_{6} x_{7} x_{11} x_{14} x_{19}, x_{1} x_{6} x_{10} x_{13} x_{18} -  x_{5} x_{7} x_{11} x_{14} x_{19}, x_{5} x_{6} x_{10} x_{13} x_{18} -  x_{1} x_{7} x_{11} x_{14} x_{19}, x_{1} x_{7} x_{10} x_{13} x_{18} -  x_{5} x_{6} x_{11} x_{14} x_{19}, x_{5} x_{7} x_{10} x_{13} x_{18} -  x_{1} x_{6} x_{11} x_{14} x_{19}, x_{6} x_{7} x_{10} x_{13} x_{18} -  x_{1} x_{5} x_{11} x_{14} x_{19}, x_{1} x_{5} x_{11} x_{13} x_{18} - 
x_{6} x_{7} x_{10} x_{14} x_{19}, x_{1} x_{6} x_{11} x_{13} x_{18} -  x_{5} x_{7} x_{10} x_{14} x_{19}, x_{5} x_{6} x_{11} x_{13} x_{18} -  x_{1} x_{7} x_{10} x_{14} x_{19}, x_{1} x_{7} x_{11} x_{13} x_{18} -  x_{5} x_{6} x_{10} x_{14} x_{19}, x_{5} x_{7} x_{11} x_{13} x_{18} -  x_{1} x_{6} x_{10} x_{14} x_{19}, x_{6} x_{7} x_{11} x_{13} x_{18} -  x_{1} x_{5} x_{10} x_{14} x_{19}, x_{1} x_{10} x_{11} x_{13} x_{18} -  x_{5} x_{6} x_{7} x_{14} x_{19}, x_{5} x_{10} x_{11} x_{13} x_{18} -  x_{1} x_{6} x_{7} x_{14} x_{19}, x_{6} x_{10} x_{11} x_{13} x_{18} -  x_{1} x_{5} x_{7} x_{14} x_{19}, x_{7} x_{10} x_{11} x_{13} x_{18} -  x_{1} x_{5} x_{6} x_{14} x_{19}, x_{1} x_{11} x_{12} x_{13} x_{18} -  x_{14} x_{15} x_{16} x_{17} x_{19}, x_{1} x_{5} x_{6} x_{14} x_{18} -  x_{7} x_{10} x_{11} x_{13} x_{19}, x_{1} x_{5} x_{7} x_{14} x_{18} -  x_{6} x_{10} x_{11} x_{13} x_{19}, x_{1} x_{6} x_{7} x_{14} x_{18} -  x_{5} x_{10} x_{11} x_{13} x_{19}, x_{5} x_{6} x_{7} x_{14} x_{18} -  x_{1} x_{10} x_{11} x_{13} x_{19}, x_{1}x_
{5} x_{10} x_{14} x_{18} -  x_{6} x_{7} x_{11} x_{13} x_{19}, x_{1} x_{6} x_{10} x_{14} x_{18} -  x_{5} x_{7} x_{11} x_{13} x_{19}, x_{5} x_{6} x_{10} x_{14} x_{18} -  x_{1} x_{7} x_{11} x_{13} x_{19}, x_{1} x_{7} x_{10} x_{14} x_{18} -  x_{5} x_{6} x_{11} x_{13} x_{19}, x_{5} x_{7} x_{10} x_{14} x_{18} -  x_{1} x_{6} x_{11} x_{13} x_{19}, x_{6} x_{7} x_{10} x_{14} x_{18} -  x_{1} x_{5} x_{11} x_{13} x_{19}, x_{1} x_{5} x_{11} x_{14} x_{18} -  x_{6} x_{7} x_{10} x_{13} x_{19}, x_{1} x_{6} x_{11} x_{14} x_{18} -  x_{5} x_{7} x_{10} x_{13} x_{19}, x_{5} x_{6} x_{11} x_{14} x_{18} -  x_{1} x_{7} x_{10} x_{13} x_{19}, x_{1} x_{7} x_{11} x_{14} x_{18} -  x_{5} x_{6} x_{10} x_{13} x_{19}, x_{5} x_{7} x_{11} x_{14} x_{18} -  x_{1} x_{6} x_{10} x_{13} x_{19}, x_{6} x_{7} x_{11} x_{14} x_{18} -  x_{1} x_{5} x_{10} x_{13} x_{19}, x_{1} x_{10} x_{11} x_{14} x_{18} -  x_{5} x_{6} x_{7} x_{13} x_{19}, x_{5} x_{10} x_{11} x_{14} x_{18} -  x_{1} x_{6} x_{7} x_{13} x_{19}, x_{6} x_{10} x_{11} x_{14} x_{18} -  x_{1} x_{5} 
x_{7} x_{13} x_{19}, x_{7} x_{10} x_{11} x_{14} x_{18} -  x_{1} x_{5} x_{6} x_{13} x_{19}, x_{1} x_{11} x_{12} x_{14} x_{18} -  x_{13} x_{15} x_{16} x_{17} x_{19}, x_{1} x_{5} x_{13} x_{14} x_{18} -  x_{6} x_{7} x_{10} x_{11} x_{19}, x_{1} x_{6} x_{13} x_{14} x_{18} -  x_{5} x_{7} x_{10} x_{11} x_{19}, x_{5} x_{6} x_{13} x_{14} x_{18} -  x_{1} x_{7} x_{10} x_{11} x_{19}, x_{1} x_{7} x_{13} x_{14} x_{18} -  x_{5} x_{6} x_{10} x_{11} x_{19}, x_{5} x_{7} x_{13} x_{14} x_{18} -  x_{1} x_{6} x_{10} x_{11} x_{19}, x_{6} x_{7} x_{13} x_{14} x_{18} -  x_{1} x_{5} x_{10} x_{11} x_{19}, x_{1} x_{10} x_{13} x_{14} x_{18} -  x_{5} x_{6} x_{7} x_{11} x_{19}, x_{5} x_{10} x_{13} x_{14} x_{18} -  x_{1} x_{6} x_{7} x_{11} x_{19}, x_{6} x_{10} x_{13} x_{14} x_{18} -  x_{1} x_{5} x_{7} x_{11} x_{19}, x_{7} x_{10} x_{13} x_{14} x_{18} -  x_{1} x_{5} x_{6} x_{11} x_{19}, x_{1} x_{11} x_{13} x_{14} x_{18} -  x_{12} x_{15} x_{16} x_{17} x_{19}, x_{5} x_{11} x_{13} x_{14} x_{18} -  x_{1} x_{6} x_{7} x_{10} x_{19}, x_{6} x_{11} x_{
13} x_{14} x_{18} -  x_{1} x_{5} x_{7} x_{10} x_{19}, x_{7} x_{11} x_{13} x_{14} x_{18} -  x_{1} x_{5} x_{6} x_{10} x_{19}, x_{10} x_{11} x_{13} x_{14} x_{18} -  x_{1} x_{5} x_{6} x_{7} x_{19}, x_{1} x_{12} x_{13} x_{14} x_{18} -  x_{11} x_{15} x_{16} x_{17} x_{19}, x_{11} x_{12} x_{13} x_{14} x_{18} -  x_{1} x_{15} x_{16} x_{17} x_{19}, x_{1} x_{2} x_{8} x_{15} x_{18} -  x_{9} x_{10} x_{16} x_{17} x_{19}, x_{1} x_{2} x_{9} x_{15} x_{18} -  x_{8} x_{10} x_{16} x_{17} x_{19}, x_{1} x_{8} x_{9} x_{15} x_{18} -  x_{2} x_{10} x_{16} x_{17} x_{19}, x_{2} x_{8} x_{9} x_{15} x_{18} -  x_{1} x_{10} x_{16} x_{17} x_{19}, x_{1} x_{2} x_{10} x_{15} x_{18} -  x_{8} x_{9} x_{16} x_{17} x_{19}, x_{1} x_{8} x_{10} x_{15} x_{18} -  x_{2} x_{9} x_{16} x_{17} x_{19}, x_{2} x_{8} x_{10} x_{15} x_{18} -  x_{1} x_{9} x_{16} x_{17} x_{19}, x_{1} x_{9} x_{10} x_{15} x_{18} -  x_{2} x_{8} x_{16} x_{17} x_{19}, x_{2} x_{9} x_{10} x_{15} x_{18} -  x_{1} x_{8} x_{16} x_{17} x_{19}, x_{8} x_{9} x_{10} x_{15} x_{18} -  x_{1} x_{2} x_{16}
 x_{17} x_{19}, x_{1} x_{11} x_{12} x_{15} x_{18} -  x_{13} x_{14} x_{16} x_{17} x_{19}, x_{1} x_{11} x_{13} x_{15} x_{18} -  x_{12} x_{14} x_{16} x_{17} x_{19}, x_{1} x_{12} x_{13} x_{15} x_{18} -  x_{11} x_{14} x_{16} x_{17} x_{19}, x_{11} x_{12} x_{13} x_{15} x_{18} -  x_{1} x_{14} x_{16} x_{17} x_{19}, x_{1} x_{11} x_{14} x_{15} x_{18} -  x_{12} x_{13} x_{16} x_{17} x_{19}, x_{1} x_{12} x_{14} x_{15} x_{18} -  x_{11} x_{13} x_{16} x_{17} x_{19}, x_{11} x_{12} x_{14} x_{15} x_{18} -  x_{1} x_{13} x_{16} x_{17} x_{19}, x_{1} x_{13} x_{14} x_{15} x_{18} -  x_{11} x_{12} x_{16} x_{17} x_{19}, x_{11} x_{13} x_{14} x_{15} x_{18} -  x_{1} x_{12} x_{16} x_{17} x_{19}, x_{12} x_{13} x_{14} x_{15} x_{18} -  x_{1} x_{11} x_{16} x_{17} x_{19}, x_{1} x_{2} x_{8} x_{16} x_{18} -  x_{9} x_{10} x_{15} x_{17} x_{19}, x_{1} x_{2} x_{9} x_{16} x_{18} -  x_{8} x_{10} x_{15} x_{17} x_{19}, x_{1} x_{8} x_{9} x_{16} x_{18} -  x_{2} x_{10} x_{15} x_{17} x_{19}, x_{2} x_{8} x_{9} x_{16} x_{18} -  x_{1} x_{10} x_{15} x_{17} x_{19}
, x_{1} x_{2} x_{10} x_{16} x_{18} -  x_{8} x_{9} x_{15} x_{17} x_{19}, x_{1} x_{8} x_{10} x_{16} x_{18} -  x_{2} x_{9} x_{15} x_{17} x_{19}, x_{2} x_{8} x_{10} x_{16} x_{18} -  x_{1} x_{9} x_{15} x_{17} x_{19}, x_{1} x_{9} x_{10} x_{16} x_{18} -  x_{2} x_{8} x_{15} x_{17} x_{19}, x_{2} x_{9} x_{10} x_{16} x_{18} -  x_{1} x_{8} x_{15} x_{17} x_{19}, x_{8} x_{9} x_{10} x_{16} x_{18} -  x_{1} x_{2} x_{15} x_{17} x_{19}, x_{1} x_{11} x_{12} x_{16} x_{18} -  x_{13} x_{14} x_{15} x_{17} x_{19}, x_{1} x_{11} x_{13} x_{16} x_{18} -  x_{12} x_{14} x_{15} x_{17} x_{19}, x_{1} x_{12} x_{13} x_{16} x_{18} -  x_{11} x_{14} x_{15} x_{17} x_{19}, x_{11} x_{12} x_{13} x_{16} x_{18} -  x_{1} x_{14} x_{15} x_{17} x_{19}, x_{1} x_{11} x_{14} x_{16} x_{18} -  x_{12} x_{13} x_{15} x_{17} x_{19}, x_{1} x_{12} x_{14} x_{16} x_{18} -  x_{11} x_{13} x_{15} x_{17} x_{19}, x_{11} x_{12} x_{14} x_{16} x_{18} -  x_{1} x_{13} x_{15} x_{17} x_{19}, x_{1} x_{13} x_{14} x_{16} x_{18} -  x_{11} x_{12} x_{15} x_{17} x_{19}, x_{11} x_{13} x_{
14} x_{16} x_{18} -  x_{1} x_{12} x_{15} x_{17} x_{19}, x_{12} x_{13} x_{14} x_{16} x_{18} -  x_{1} x_{11} x_{15} x_{17} x_{19}, x_{1} x_{2} x_{15} x_{16} x_{18} -  x_{8} x_{9} x_{10} x_{17} x_{19}, x_{1} x_{8} x_{15} x_{16} x_{18} -  x_{2} x_{9} x_{10} x_{17} x_{19}, x_{2} x_{8} x_{15} x_{16} x_{18} -  x_{1} x_{9} x_{10} x_{17} x_{19}, x_{1} x_{9} x_{15} x_{16} x_{18} -  x_{2} x_{8} x_{10} x_{17} x_{19}, x_{2} x_{9} x_{15} x_{16} x_{18} -  x_{1} x_{8} x_{10} x_{17} x_{19}, x_{8} x_{9} x_{15} x_{16} x_{18} -  x_{1} x_{2} x_{10} x_{17} x_{19}, x_{1} x_{10} x_{15} x_{16} x_{18} -  x_{2} x_{8} x_{9} x_{17} x_{19}, x_{2} x_{10} x_{15} x_{16} x_{18} -  x_{1} x_{8} x_{9} x_{17} x_{19}, x_{8} x_{10} x_{15} x_{16} x_{18} -  x_{1} x_{2} x_{9} x_{17} x_{19}, x_{9} x_{10} x_{15} x_{16} x_{18} -  x_{1} x_{2} x_{8} x_{17} x_{19}, x_{1} x_{11} x_{15} x_{16} x_{18} -  x_{12} x_{13} x_{14} x_{17} x_{19}, x_{1} x_{12} x_{15} x_{16} x_{18} -  x_{11} x_{13} x_{14} x_{17} x_{19}, x_{11} x_{12} x_{15} x_{16} x_{18} -  x_{1} x_{
13} x_{14} x_{17} x_{19}, x_{1} x_{13} x_{15} x_{16} x_{18} -  x_{11} x_{12} x_{14} x_{17} x_{19}, x_{11} x_{13} x_{15} x_{16} x_{18} -  x_{1} x_{12} x_{14} x_{17} x_{19}, x_{12} x_{13} x_{15} x_{16} x_{18} -  x_{1} x_{11} x_{14} x_{17} x_{19}, x_{1} x_{14} x_{15} x_{16} x_{18} -  x_{11} x_{12} x_{13} x_{17} x_{19}, x_{11} x_{14} x_{15} x_{16} x_{18} -  x_{1} x_{12} x_{13} x_{17} x_{19}, x_{12} x_{14} x_{15} x_{16} x_{18} -  x_{1} x_{11} x_{13} x_{17} x_{19}, x_{13} x_{14} x_{15} x_{16} x_{18} -  x_{1} x_{11} x_{12} x_{17} x_{19}, x_{1} x_{2} x_{8} x_{17} x_{18} -  x_{9} x_{10} x_{15} x_{16} x_{19}, x_{1} x_{2} x_{9} x_{17} x_{18} -  x_{8} x_{10} x_{15} x_{16} x_{19}, x_{1} x_{8} x_{9} x_{17} x_{18} -  x_{2} x_{10} x_{15} x_{16} x_{19}, x_{2} x_{8} x_{9} x_{17} x_{18} -  x_{1} x_{10} x_{15} x_{16} x_{19}, x_{1} x_{2} x_{10} x_{17} x_{18} -  x_{8} x_{9} x_{15} x_{16} x_{19}, x_{1} x_{8} x_{10} x_{17} x_{18} -  x_{2} x_{9} x_{15} x_{16} x_{19}, x_{2} x_{8} x_{10} x_{17} x_{18} -  x_{1} x_{9} x_{15} x_{16} x_{
19}, x_{1} x_{9} x_{10} x_{17} x_{18} -  x_{2} x_{8} x_{15} x_{16} x_{19}, x_{2} x_{9} x_{10} x_{17} x_{18} -  x_{1} x_{8} x_{15} x_{16} x_{19}, x_{8} x_{9} x_{10} x_{17} x_{18} -  x_{1} x_{2} x_{15} x_{16} x_{19}, x_{1} x_{11} x_{12} x_{17} x_{18} -  x_{13} x_{14} x_{15} x_{16} x_{19}, x_{1} x_{11} x_{13} x_{17} x_{18} -  x_{12} x_{14} x_{15} x_{16} x_{19}, x_{1} x_{12} x_{13} x_{17} x_{18} -  x_{11} x_{14} x_{15} x_{16} x_{19}, x_{11} x_{12} x_{13} x_{17} x_{18} -  x_{1} x_{14} x_{15} x_{16} x_{19}, x_{1} x_{11} x_{14} x_{17} x_{18} -  x_{12} x_{13} x_{15} x_{16} x_{19}, x_{1} x_{12} x_{14} x_{17} x_{18} -  x_{11} x_{13} x_{15} x_{16} x_{19}, x_{11} x_{12} x_{14} x_{17} x_{18} -  x_{1} x_{13} x_{15} x_{16} x_{19}, x_{1} x_{13} x_{14} x_{17} x_{18} -  x_{11} x_{12} x_{15} x_{16} x_{19}, x_{11} x_{13} x_{14} x_{17} x_{18} -  x_{1} x_{12} x_{15} x_{16} x_{19}, x_{12} x_{13} x_{14} x_{17} x_{18} -  x_{1} x_{11} x_{15} x_{16} x_{19}, x_{1} x_{2} x_{15} x_{17} x_{18} -  x_{8} x_{9} x_{10} x_{16} x_{19}, x_{1} x_{
8} x_{15} x_{17} x_{18} -  x_{2} x_{9} x_{10} x_{16} x_{19}, x_{2} x_{8} x_{15} x_{17} x_{18} -  x_{1} x_{9} x_{10} x_{16} x_{19}, x_{1} x_{9} x_{15} x_{17} x_{18} -  x_{2} x_{8} x_{10} x_{16} x_{19}, x_{2} x_{9} x_{15} x_{17} x_{18} -  x_{1} x_{8} x_{10} x_{16} x_{19}, x_{8} x_{9} x_{15} x_{17} x_{18} -  x_{1} x_{2} x_{10} x_{16} x_{19}, x_{1} x_{10} x_{15} x_{17} x_{18} -  x_{2} x_{8} x_{9} x_{16} x_{19}, x_{2} x_{10} x_{15} x_{17} x_{18} -  x_{1} x_{8} x_{9} x_{16} x_{19}, x_{8} x_{10} x_{15} x_{17} x_{18} -  x_{1} x_{2} x_{9} x_{16} x_{19}, x_{9} x_{10} x_{15} x_{17} x_{18} -  x_{1} x_{2} x_{8} x_{16} x_{19}, x_{1} x_{11} x_{15} x_{17} x_{18} -  x_{12} x_{13} x_{14} x_{16} x_{19}, x_{1} x_{12} x_{15} x_{17} x_{18} -  x_{11} x_{13} x_{14} x_{16} x_{19}, x_{11} x_{12} x_{15} x_{17} x_{18} -  x_{1} x_{13} x_{14} x_{16} x_{19}, x_{1} x_{13} x_{15} x_{17} x_{18} -  x_{11} x_{12} x_{14} x_{16} x_{19}, x_{11} x_{13} x_{15} x_{17} x_{18} -  x_{1} x_{12} x_{14} x_{16} x_{19}, x_{12} x_{13} x_{15} x_{17} x_{18} -x_
{1} x_{11} x_{14} x_{16} x_{19}, x_{1} x_{14} x_{15} x_{17} x_{18} -  x_{11} x_{12} x_{13} x_{16} x_{19}, x_{11} x_{14} x_{15} x_{17} x_{18} -  x_{1} x_{12} x_{13} x_{16} x_{19}, x_{12} x_{14} x_{15} x_{17} x_{18} -  x_{1} x_{11} x_{13} x_{16} x_{19}, x_{13} x_{14} x_{15} x_{17} x_{18} -  x_{1} x_{11} x_{12} x_{16} x_{19}, x_{1} x_{2} x_{16} x_{17} x_{18} -  x_{8} x_{9} x_{10} x_{15} x_{19}, x_{1} x_{8} x_{16} x_{17} x_{18} -  x_{2} x_{9} x_{10} x_{15} x_{19}, x_{2} x_{8} x_{16} x_{17} x_{18} -  x_{1} x_{9} x_{10} x_{15} x_{19}, x_{1} x_{9} x_{16} x_{17} x_{18} -  x_{2} x_{8} x_{10} x_{15} x_{19}, x_{2} x_{9} x_{16} x_{17} x_{18} -  x_{1} x_{8} x_{10} x_{15} x_{19}, x_{8} x_{9} x_{16} x_{17} x_{18} -  x_{1} x_{2} x_{10} x_{15} x_{19}, x_{1} x_{10} x_{16} x_{17} x_{18} -  x_{2} x_{8} x_{9} x_{15} x_{19}, x_{2} x_{10} x_{16} x_{17} x_{18} -  x_{1} x_{8} x_{9} x_{15} x_{19}, x_{8} x_{10} x_{16} x_{17} x_{18} -  x_{1} x_{2} x_{9} x_{15} x_{19}, x_{9} x_{10} x_{16} x_{17} x_{18} -  x_{1} x_{2} x_{8} x_{15} x_{19},
 x_{1} x_{11} x_{16} x_{17} x_{18} -  x_{12} x_{13} x_{14} x_{15} x_{19}, x_{1} x_{12} x_{16} x_{17} x_{18} -  x_{11} x_{13} x_{14} x_{15} x_{19}, x_{11} x_{12} x_{16} x_{17} x_{18} -  x_{1} x_{13} x_{14} x_{15} x_{19}, x_{1} x_{13} x_{16} x_{17} x_{18} -  x_{11} x_{12} x_{14} x_{15} x_{19}, x_{11} x_{13} x_{16} x_{17} x_{18} -  x_{1} x_{12} x_{14} x_{15} x_{19}, x_{12} x_{13} x_{16} x_{17} x_{18} -  x_{1} x_{11} x_{14} x_{15} x_{19}, x_{1} x_{14} x_{16} x_{17} x_{18} -  x_{11} x_{12} x_{13} x_{15} x_{19}, x_{11} x_{14} x_{16} x_{17} x_{18} -  x_{1} x_{12} x_{13} x_{15} x_{19}, x_{12} x_{14} x_{16} x_{17} x_{18} -  x_{1} x_{11} x_{13} x_{15} x_{19}, x_{13} x_{14} x_{16} x_{17} x_{18} -  x_{1} x_{11} x_{12} x_{15} x_{19}, x_{1} x_{15} x_{16} x_{17} x_{18} -  x_{11} x_{12} x_{13} x_{14} x_{19}, x_{2} x_{15} x_{16} x_{17} x_{18} -  x_{1} x_{8} x_{9} x_{10} x_{19}, x_{8} x_{15} x_{16} x_{17} x_{18} -  x_{1} x_{2} x_{9} x_{10} x_{19}, x_{9} x_{15} x_{16} x_{17} x_{18} -  x_{1} x_{2} x_{8} x_{10} x_{19}, x_{10} x_{
15} x_{16} x_{17} x_{18} -  x_{1} x_{2} x_{8} x_{9} x_{19}, x_{11} x_{15} x_{16} x_{17} x_{18} -  x_{1} x_{12} x_{13} x_{14} x_{19}, x_{12} x_{15} x_{16} x_{17} x_{18} -  x_{1} x_{11} x_{13} x_{14} x_{19}, x_{13} x_{15} x_{16} x_{17} x_{18} -  x_{1} x_{11} x_{12} x_{14} x_{19}, x_{14} x_{15} x_{16} x_{17} x_{18} -  x_{1} x_{11} x_{12} x_{13} x_{19}, x_{2} x_{3} x_{4} x_{5} -  x_{14} x_{17} x_{18} x_{19}, x_{2} x_{3} x_{7} x_{8} -  x_{13} x_{14} x_{15} x_{19}, x_{4} x_{5} x_{7} x_{8} -  x_{13} x_{15} x_{17} x_{18}, x_{2} x_{4} x_{6} x_{9} -  x_{11} x_{14} x_{16} x_{18}, x_{3} x_{5} x_{6} x_{9} -  x_{11} x_{16} x_{17} x_{19}, x_{5} x_{6} x_{7} x_{10} -  x_{12} x_{15} x_{16} x_{17}, x_{4} x_{6} x_{8} x_{10} -  x_{12} x_{13} x_{16} x_{18}, x_{3} x_{7} x_{9} x_{10} -  x_{11} x_{12} x_{15} x_{19}, x_{2} x_{8} x_{9} x_{10} -  x_{11} x_{12} x_{13} x_{14}, x_{2} x_{4} x_{6} x_{11} -  x_{9} x_{14} x_{16} x_{18}, x_{3} x_{5} x_{6} x_{11} -  x_{9} x_{16} x_{17} x_{19}, x_{2} x_{4} x_{9} x_{11} -  x_{6} x_{14} x_{16}x_{
18}, x_{3} x_{5} x_{9} x_{11} -  x_{6} x_{16} x_{17} x_{19}, x_{2} x_{6} x_{9} x_{11} -  x_{4} x_{14} x_{16} x_{18}, x_{3} x_{6} x_{9} x_{11} -  x_{5} x_{16} x_{17} x_{19}, x_{4} x_{6} x_{9} x_{11} -  x_{2} x_{14} x_{16} x_{18}, x_{5} x_{6} x_{9} x_{11} -  x_{3} x_{16} x_{17} x_{19}, x_{3} x_{7} x_{9} x_{11} -  x_{10} x_{12} x_{15} x_{19}, x_{2} x_{8} x_{9} x_{11} -  x_{10} x_{12} x_{13} x_{14}, x_{3} x_{7} x_{10} x_{11} -  x_{9} x_{12} x_{15} x_{19}, x_{2} x_{8} x_{10} x_{11} -  x_{9} x_{12} x_{13} x_{14}, x_{2} x_{9} x_{10} x_{11} -  x_{8} x_{12} x_{13} x_{14}, x_{3} x_{9} x_{10} x_{11} -  x_{7} x_{12} x_{15} x_{19}, x_{7} x_{9} x_{10} x_{11} -  x_{3} x_{12} x_{15} x_{19}, x_{8} x_{9} x_{10} x_{11} -  x_{2} x_{12} x_{13} x_{14}, x_{5} x_{6} x_{7} x_{12} -  x_{10} x_{15} x_{16} x_{17}, x_{4} x_{6} x_{8} x_{12} -  x_{10} x_{13} x_{16} x_{18}, x_{3} x_{7} x_{9} x_{12} -  x_{10} x_{11} x_{15} x_{19}, x_{2} x_{8} x_{9} x_{12} -  x_{10} x_{11} x_{13} x_{14}, x_{4} x_{6} x_{10} x_{12} -  x_{8} x_{13} x_{16} x_{18}
, x_{5} x_{6} x_{10} x_{12} -  x_{7} x_{15} x_{16} x_{17}, x_{3} x_{7} x_{10} x_{12} -  x_{9} x_{11} x_{15} x_{19}, x_{5} x_{7} x_{10} x_{12} -  x_{6} x_{15} x_{16} x_{17}, x_{6} x_{7} x_{10} x_{12} -  x_{5} x_{15} x_{16} x_{17}, x_{2} x_{8} x_{10} x_{12} -  x_{9} x_{11} x_{13} x_{14}, x_{4} x_{8} x_{10} x_{12} -  x_{6} x_{13} x_{16} x_{18}, x_{6} x_{8} x_{10} x_{12} -  x_{4} x_{13} x_{16} x_{18}, x_{2} x_{9} x_{10} x_{12} -  x_{8} x_{11} x_{13} x_{14}, x_{3} x_{9} x_{10} x_{12} -  x_{7} x_{11} x_{15} x_{19}, x_{7} x_{9} x_{10} x_{12} -  x_{3} x_{11} x_{15} x_{19}, x_{8} x_{9} x_{10} x_{12} -  x_{2} x_{11} x_{13} x_{14}, x_{3} x_{7} x_{11} x_{12} -  x_{9} x_{10} x_{15} x_{19}, x_{2} x_{8} x_{11} x_{12} -  x_{9} x_{10} x_{13} x_{14}, x_{2} x_{9} x_{11} x_{12} -  x_{8} x_{10} x_{13} x_{14}, x_{3} x_{9} x_{11} x_{12} -  x_{7} x_{10} x_{15} x_{19}, x_{7} x_{9} x_{11} x_{12} -  x_{3} x_{10} x_{15} x_{19}, x_{8} x_{9} x_{11} x_{12} -  x_{2} x_{10} x_{13} x_{14}, x_{2} x_{10} x_{11} x_{12} -  x_{8} x_{9} x_{13}x_{
14}, x_{3} x_{10} x_{11} x_{12} -  x_{7} x_{9} x_{15} x_{19}, x_{7} x_{10} x_{11} x_{12} -  x_{3} x_{9} x_{15} x_{19}, x_{8} x_{10} x_{11} x_{12} -  x_{2} x_{9} x_{13} x_{14}, x_{9} x_{10} x_{11} x_{12} -  x_{3} x_{7} x_{15} x_{19}, x_{2} x_{3} x_{7} x_{13} -  x_{8} x_{14} x_{15} x_{19}, x_{4} x_{5} x_{7} x_{13} -  x_{8} x_{15} x_{17} x_{18}, x_{2} x_{3} x_{8} x_{13} -  x_{7} x_{14} x_{15} x_{19}, x_{4} x_{5} x_{8} x_{13} -  x_{7} x_{15} x_{17} x_{18}, x_{4} x_{6} x_{8} x_{13} -  x_{10} x_{12} x_{16} x_{18}, x_{2} x_{7} x_{8} x_{13} -  x_{3} x_{14} x_{15} x_{19}, x_{3} x_{7} x_{8} x_{13} -  x_{2} x_{14} x_{15} x_{19}, x_{4} x_{7} x_{8} x_{13} -  x_{5} x_{15} x_{17} x_{18}, x_{5} x_{7} x_{8} x_{13} -  x_{4} x_{15} x_{17} x_{18}, x_{2} x_{8} x_{9} x_{13} -  x_{10} x_{11} x_{12} x_{14}, x_{4} x_{6} x_{10} x_{13} -  x_{8} x_{12} x_{16} x_{18}, x_{2} x_{8} x_{10} x_{13} -  x_{9} x_{11} x_{12} x_{14}, x_{4} x_{8} x_{10} x_{13} -  x_{6} x_{12} x_{16} x_{18}, x_{6} x_{8} x_{10} x_{13} -  x_{4} x_{12} x_{16} x_{18}, 
x_{2} x_{9} x_{10} x_{13} -  x_{8} x_{11} x_{12} x_{14}, x_{8} x_{9} x_{10} x_{13} -  x_{2} x_{11} x_{12} x_{14}, x_{2} x_{8} x_{11} x_{13} -  x_{9} x_{10} x_{12} x_{14}, x_{2} x_{9} x_{11} x_{13} -  x_{8} x_{10} x_{12} x_{14}, x_{8} x_{9} x_{11} x_{13} -  x_{2} x_{10} x_{12} x_{14}, x_{2} x_{10} x_{11} x_{13} -  x_{8} x_{9} x_{12} x_{14}, x_{8} x_{10} x_{11} x_{13} -  x_{2} x_{9} x_{12} x_{14}, x_{9} x_{10} x_{11} x_{13} -  x_{2} x_{8} x_{12} x_{14}, x_{4} x_{6} x_{12} x_{13} -  x_{8} x_{10} x_{16} x_{18}, x_{2} x_{8} x_{12} x_{13} -  x_{9} x_{10} x_{11} x_{14}, x_{4} x_{8} x_{12} x_{13} -  x_{6} x_{10} x_{16} x_{18}, x_{6} x_{8} x_{12} x_{13} -  x_{4} x_{10} x_{16} x_{18}, x_{2} x_{9} x_{12} x_{13} -  x_{8} x_{10} x_{11} x_{14}, x_{8} x_{9} x_{12} x_{13} -  x_{2} x_{10} x_{11} x_{14}, x_{2} x_{10} x_{12} x_{13} -  x_{8} x_{9} x_{11} x_{14}, x_{4} x_{10} x_{12} x_{13} -  x_{6} x_{8} x_{16} x_{18}, x_{6} x_{10} x_{12} x_{13} -  x_{4} x_{8} x_{16} x_{18}, x_{8} x_{10} x_{12} x_{13} -  x_{4} x_{6} x_{16} x_{18}
, x_{9} x_{10} x_{12} x_{13} -  x_{2} x_{8} x_{11} x_{14}, x_{2} x_{11} x_{12} x_{13} -  x_{8} x_{9} x_{10} x_{14}, x_{8} x_{11} x_{12} x_{13} -  x_{2} x_{9} x_{10} x_{14}, x_{9} x_{11} x_{12} x_{13} -  x_{2} x_{8} x_{10} x_{14}, x_{10} x_{11} x_{12} x_{13} -  x_{2} x_{8} x_{9} x_{14}, x_{2} x_{3} x_{4} x_{14} -  x_{5} x_{17} x_{18} x_{19}, x_{2} x_{3} x_{5} x_{14} -  x_{4} x_{17} x_{18} x_{19}, x_{2} x_{4} x_{5} x_{14} -  x_{3} x_{17} x_{18} x_{19}, x_{3} x_{4} x_{5} x_{14} -  x_{2} x_{17} x_{18} x_{19}, x_{2} x_{4} x_{6} x_{14} -  x_{9} x_{11} x_{16} x_{18}, x_{2} x_{3} x_{7} x_{14} -  x_{8} x_{13} x_{15} x_{19}, x_{2} x_{3} x_{8} x_{14} -  x_{7} x_{13} x_{15} x_{19}, x_{2} x_{7} x_{8} x_{14} -  x_{3} x_{13} x_{15} x_{19}, x_{3} x_{7} x_{8} x_{14} -  x_{2} x_{13} x_{15} x_{19}, x_{2} x_{4} x_{9} x_{14} -  x_{6} x_{11} x_{16} x_{18}, x_{2} x_{6} x_{9} x_{14} -  x_{4} x_{11} x_{16} x_{18}, x_{4} x_{6} x_{9} x_{14} -  x_{2} x_{11} x_{16} x_{18}, x_{2} x_{4} x_{11} x_{14} -  x_{6} x_{9} x_{16} x_{18}, x_{2}x_{
6} x_{11} x_{14} -  x_{4} x_{9} x_{16} x_{18}, x_{4} x_{6} x_{11} x_{14} -  x_{2} x_{9} x_{16} x_{18}, x_{2} x_{9} x_{11} x_{14} -  x_{4} x_{6} x_{16} x_{18}, x_{4} x_{9} x_{11} x_{14} -  x_{2} x_{6} x_{16} x_{18}, x_{6} x_{9} x_{11} x_{14} -  x_{2} x_{4} x_{16} x_{18}, x_{2} x_{3} x_{13} x_{14} -  x_{7} x_{8} x_{15} x_{19}, x_{2} x_{7} x_{13} x_{14} -  x_{3} x_{8} x_{15} x_{19}, x_{3} x_{7} x_{13} x_{14} -  x_{2} x_{8} x_{15} x_{19}, x_{2} x_{8} x_{13} x_{14} -  x_{3} x_{7} x_{15} x_{19}, x_{3} x_{8} x_{13} x_{14} -  x_{2} x_{7} x_{15} x_{19}, x_{7} x_{8} x_{13} x_{14} -  x_{2} x_{3} x_{15} x_{19}, x_{2} x_{3} x_{7} x_{15} -  x_{8} x_{13} x_{14} x_{19}, x_{4} x_{5} x_{7} x_{15} -  x_{8} x_{13} x_{17} x_{18}, x_{5} x_{6} x_{7} x_{15} -  x_{10} x_{12} x_{16} x_{17}, x_{2} x_{3} x_{8} x_{15} -  x_{7} x_{13} x_{14} x_{19}, x_{4} x_{5} x_{8} x_{15} -  x_{7} x_{13} x_{17} x_{18}, x_{2} x_{7} x_{8} x_{15} -  x_{3} x_{13} x_{14} x_{19}, x_{3} x_{7} x_{8} x_{15} -  x_{2} x_{13} x_{14} x_{19}, x_{4} x_{7} x_{8} x_{15}
 -  x_{5} x_{13} x_{17} x_{18}, x_{5} x_{7} x_{8} x_{15} -  x_{4} x_{13} x_{17} x_{18}, x_{3} x_{7} x_{9} x_{15} -  x_{10} x_{11} x_{12} x_{19}, x_{5} x_{6} x_{10} x_{15} -  x_{7} x_{12} x_{16} x_{17}, x_{3} x_{7} x_{10} x_{15} -  x_{9} x_{11} x_{12} x_{19}, x_{5} x_{7} x_{10} x_{15} -  x_{6} x_{12} x_{16} x_{17}, x_{6} x_{7} x_{10} x_{15} -  x_{5} x_{12} x_{16} x_{17}, x_{3} x_{9} x_{10} x_{15} -  x_{7} x_{11} x_{12} x_{19}, x_{7} x_{9} x_{10} x_{15} -  x_{3} x_{11} x_{12} x_{19}, x_{3} x_{7} x_{11} x_{15} -  x_{9} x_{10} x_{12} x_{19}, x_{3} x_{9} x_{11} x_{15} -  x_{7} x_{10} x_{12} x_{19}, x_{7} x_{9} x_{11} x_{15} -  x_{3} x_{10} x_{12} x_{19}, x_{3} x_{10} x_{11} x_{15} -  x_{7} x_{9} x_{12} x_{19}, x_{7} x_{10} x_{11} x_{15} -  x_{3} x_{9} x_{12} x_{19}, x_{9} x_{10} x_{11} x_{15} -  x_{3} x_{7} x_{12} x_{19}, x_{5} x_{6} x_{12} x_{15} -  x_{7} x_{10} x_{16} x_{17}, x_{3} x_{7} x_{12} x_{15} -  x_{9} x_{10} x_{11} x_{19}, x_{5} x_{7} x_{12} x_{15} -  x_{6} x_{10} x_{16} x_{17}, x_{6} x_{7} x_{12} x_{
15} -  x_{5} x_{10} x_{16} x_{17}, x_{3} x_{9} x_{12} x_{15} -  x_{7} x_{10} x_{11} x_{19}, x_{7} x_{9} x_{12} x_{15} -  x_{3} x_{10} x_{11} x_{19}, x_{3} x_{10} x_{12} x_{15} -  x_{7} x_{9} x_{11} x_{19}, x_{5} x_{10} x_{12} x_{15} -  x_{6} x_{7} x_{16} x_{17}, x_{6} x_{10} x_{12} x_{15} -  x_{5} x_{7} x_{16} x_{17}, x_{7} x_{10} x_{12} x_{15} -  x_{3} x_{9} x_{11} x_{19}, x_{9} x_{10} x_{12} x_{15} -  x_{3} x_{7} x_{11} x_{19}, x_{3} x_{11} x_{12} x_{15} -  x_{7} x_{9} x_{10} x_{19}, x_{7} x_{11} x_{12} x_{15} -  x_{3} x_{9} x_{10} x_{19}, x_{9} x_{11} x_{12} x_{15} -  x_{3} x_{7} x_{10} x_{19}, x_{10} x_{11} x_{12} x_{15} -  x_{3} x_{7} x_{9} x_{19}, x_{2} x_{3} x_{13} x_{15} -  x_{7} x_{8} x_{14} x_{19}, x_{4} x_{5} x_{13} x_{15} -  x_{7} x_{8} x_{17} x_{18}, x_{2} x_{7} x_{13} x_{15} -  x_{3} x_{8} x_{14} x_{19}, x_{3} x_{7} x_{13} x_{15} -  x_{2} x_{8} x_{14} x_{19}, x_{4} x_{7} x_{13} x_{15} -  x_{5} x_{8} x_{17} x_{18}, x_{5} x_{7} x_{13} x_{15} -  x_{4} x_{8} x_{17} x_{18}, x_{2} x_{8} x_{13} x_{15} 
-  x_{3} x_{7} x_{14} x_{19}, x_{3} x_{8} x_{13} x_{15} -  x_{2} x_{7} x_{14} x_{19}, x_{4} x_{8} x_{13} x_{15} -  x_{5} x_{7} x_{17} x_{18}, x_{5} x_{8} x_{13} x_{15} -  x_{4} x_{7} x_{17} x_{18}, x_{7} x_{8} x_{13} x_{15} -  x_{2} x_{3} x_{14} x_{19}, x_{2} x_{3} x_{14} x_{15} -  x_{7} x_{8} x_{13} x_{19}, x_{2} x_{7} x_{14} x_{15} -  x_{3} x_{8} x_{13} x_{19}, x_{3} x_{7} x_{14} x_{15} -  x_{2} x_{8} x_{13} x_{19}, x_{2} x_{8} x_{14} x_{15} -  x_{3} x_{7} x_{13} x_{19}, x_{3} x_{8} x_{14} x_{15} -  x_{2} x_{7} x_{13} x_{19}, x_{7} x_{8} x_{14} x_{15} -  x_{2} x_{3} x_{13} x_{19}, x_{2} x_{13} x_{14} x_{15} -  x_{3} x_{7} x_{8} x_{19}, x_{3} x_{13} x_{14} x_{15} -  x_{2} x_{7} x_{8} x_{19}, x_{7} x_{13} x_{14} x_{15} -  x_{2} x_{3} x_{8} x_{19}, x_{8} x_{13} x_{14} x_{15} -  x_{2} x_{3} x_{7} x_{19}, x_{2} x_{4} x_{6} x_{16} -  x_{9} x_{11} x_{14} x_{18}, x_{3} x_{5} x_{6} x_{16} -  x_{9} x_{11} x_{17} x_{19}, x_{5} x_{6} x_{7} x_{16} -  x_{10} x_{12} x_{15} x_{17}, x_{4} x_{6} x_{8} x_{16} -  x_{10} x_{12}
 x_{13} x_{18}, x_{2} x_{4} x_{9} x_{16} -  x_{6} x_{11} x_{14} x_{18}, x_{3} x_{5} x_{9} x_{16} -  x_{6} x_{11} x_{17} x_{19}, x_{2} x_{6} x_{9} x_{16} -  x_{4} x_{11} x_{14} x_{18}, x_{3} x_{6} x_{9} x_{16} -  x_{5} x_{11} x_{17} x_{19}, x_{4} x_{6} x_{9} x_{16} -  x_{2} x_{11} x_{14} x_{18}, x_{5} x_{6} x_{9} x_{16} -  x_{3} x_{11} x_{17} x_{19}, x_{4} x_{6} x_{10} x_{16} -  x_{8} x_{12} x_{13} x_{18}, x_{5} x_{6} x_{10} x_{16} -  x_{7} x_{12} x_{15} x_{17}, x_{5} x_{7} x_{10} x_{16} -  x_{6} x_{12} x_{15} x_{17}, x_{6} x_{7} x_{10} x_{16} -  x_{5} x_{12} x_{15} x_{17}, x_{4} x_{8} x_{10} x_{16} -  x_{6} x_{12} x_{13} x_{18}, x_{6} x_{8} x_{10} x_{16} -  x_{4} x_{12} x_{13} x_{18}, x_{2} x_{4} x_{11} x_{16} -  x_{6} x_{9} x_{14} x_{18}, x_{3} x_{5} x_{11} x_{16} -  x_{6} x_{9} x_{17} x_{19}, x_{2} x_{6} x_{11} x_{16} -  x_{4} x_{9} x_{14} x_{18}, x_{3} x_{6} x_{11} x_{16} -  x_{5} x_{9} x_{17} x_{19}, x_{4} x_{6} x_{11} x_{16} -  x_{2} x_{9} x_{14} x_{18}, x_{5} x_{6} x_{11} x_{16} -  x_{3} x_{9} x_{17} x_
{19}, x_{2} x_{9} x_{11} x_{16} -  x_{4} x_{6} x_{14} x_{18}, x_{3} x_{9} x_{11} x_{16} -  x_{5} x_{6} x_{17} x_{19}, x_{4} x_{9} x_{11} x_{16} -  x_{2} x_{6} x_{14} x_{18}, x_{5} x_{9} x_{11} x_{16} -  x_{3} x_{6} x_{17} x_{19}, x_{6} x_{9} x_{11} x_{16} -  x_{3} x_{5} x_{17} x_{19}, x_{4} x_{6} x_{12} x_{16} -  x_{8} x_{10} x_{13} x_{18}, x_{5} x_{6} x_{12} x_{16} -  x_{7} x_{10} x_{15} x_{17}, x_{5} x_{7} x_{12} x_{16} -  x_{6} x_{10} x_{15} x_{17}, x_{6} x_{7} x_{12} x_{16} -  x_{5} x_{10} x_{15} x_{17}, x_{4} x_{8} x_{12} x_{16} -  x_{6} x_{10} x_{13} x_{18}, x_{6} x_{8} x_{12} x_{16} -  x_{4} x_{10} x_{13} x_{18}, x_{4} x_{10} x_{12} x_{16} -  x_{6} x_{8} x_{13} x_{18}, x_{5} x_{10} x_{12} x_{16} -  x_{6} x_{7} x_{15} x_{17}, x_{6} x_{10} x_{12} x_{16} -  x_{4} x_{8} x_{13} x_{18}, x_{7} x_{10} x_{12} x_{16} -  x_{5} x_{6} x_{15} x_{17}, x_{8} x_{10} x_{12} x_{16} -  x_{4} x_{6} x_{13} x_{18}, x_{4} x_{6} x_{13} x_{16} -  x_{8} x_{10} x_{12} x_{18}, x_{4} x_{8} x_{13} x_{16} -  x_{6} x_{10} x_{12} x_{
18}, x_{6} x_{8} x_{13} x_{16} -  x_{4} x_{10} x_{12} x_{18}, x_{4} x_{10} x_{13} x_{16} -  x_{6} x_{8} x_{12} x_{18}, x_{6} x_{10} x_{13} x_{16} -  x_{4} x_{8} x_{12} x_{18}, x_{8} x_{10} x_{13} x_{16} -  x_{4} x_{6} x_{12} x_{18}, x_{4} x_{12} x_{13} x_{16} -  x_{6} x_{8} x_{10} x_{18}, x_{6} x_{12} x_{13} x_{16} -  x_{4} x_{8} x_{10} x_{18}, x_{8} x_{12} x_{13} x_{16} -  x_{4} x_{6} x_{10} x_{18}, x_{10} x_{12} x_{13} x_{16} -  x_{4} x_{6} x_{8} x_{18}, x_{2} x_{4} x_{14} x_{16} -  x_{6} x_{9} x_{11} x_{18}, x_{2} x_{6} x_{14} x_{16} -  x_{4} x_{9} x_{11} x_{18}, x_{4} x_{6} x_{14} x_{16} -  x_{2} x_{9} x_{11} x_{18}, x_{2} x_{9} x_{14} x_{16} -  x_{4} x_{6} x_{11} x_{18}, x_{4} x_{9} x_{14} x_{16} -  x_{2} x_{6} x_{11} x_{18}, x_{6} x_{9} x_{14} x_{16} -  x_{2} x_{4} x_{11} x_{18}, x_{2} x_{11} x_{14} x_{16} -  x_{4} x_{6} x_{9} x_{18}, x_{4} x_{11} x_{14} x_{16} -  x_{2} x_{6} x_{9} x_{18}, x_{6} x_{11} x_{14} x_{16} -  x_{2} x_{4} x_{9} x_{18}, x_{9} x_{11} x_{14} x_{16} -  x_{2} x_{4} x_{6} x_{18}, x_{
5} x_{6} x_{15} x_{16} -  x_{7} x_{10} x_{12} x_{17}, x_{5} x_{7} x_{15} x_{16} -  x_{6} x_{10} x_{12} x_{17}, x_{6} x_{7} x_{15} x_{16} -  x_{5} x_{10} x_{12} x_{17}, x_{5} x_{10} x_{15} x_{16} -  x_{6} x_{7} x_{12} x_{17}, x_{6} x_{10} x_{15} x_{16} -  x_{5} x_{7} x_{12} x_{17}, x_{7} x_{10} x_{15} x_{16} -  x_{5} x_{6} x_{12} x_{17}, x_{5} x_{12} x_{15} x_{16} -  x_{6} x_{7} x_{10} x_{17}, x_{6} x_{12} x_{15} x_{16} -  x_{5} x_{7} x_{10} x_{17}, x_{7} x_{12} x_{15} x_{16} -  x_{5} x_{6} x_{10} x_{17}, x_{10} x_{12} x_{15} x_{16} -  x_{5} x_{6} x_{7} x_{17}, x_{2} x_{3} x_{4} x_{17} -  x_{5} x_{14} x_{18} x_{19}, x_{2} x_{3} x_{5} x_{17} -  x_{4} x_{14} x_{18} x_{19}, x_{2} x_{4} x_{5} x_{17} -  x_{3} x_{14} x_{18} x_{19}, x_{3} x_{4} x_{5} x_{17} -  x_{2} x_{14} x_{18} x_{19}, x_{3} x_{5} x_{6} x_{17} -  x_{9} x_{11} x_{16} x_{19}, x_{4} x_{5} x_{7} x_{17} -  x_{8} x_{13} x_{15} x_{18}, x_{4} x_{5} x_{8} x_{17} -  x_{7} x_{13} x_{15} x_{18}, x_{4} x_{7} x_{8} x_{17} -  x_{5} x_{13} x_{15} x_{18}, x_{5} x_{
7} x_{8} x_{17} -  x_{4} x_{13} x_{15} x_{18}, x_{3} x_{5} x_{9} x_{17} -  x_{6} x_{11} x_{16} x_{19}, x_{3} x_{6} x_{9} x_{17} -  x_{5} x_{11} x_{16} x_{19}, x_{5} x_{6} x_{9} x_{17} -  x_{3} x_{11} x_{16} x_{19}, x_{3} x_{5} x_{11} x_{17} -  x_{6} x_{9} x_{16} x_{19}, x_{3} x_{6} x_{11} x_{17} -  x_{5} x_{9} x_{16} x_{19}, x_{5} x_{6} x_{11} x_{17} -  x_{3} x_{9} x_{16} x_{19}, x_{3} x_{9} x_{11} x_{17} -  x_{5} x_{6} x_{16} x_{19}, x_{5} x_{9} x_{11} x_{17} -  x_{3} x_{6} x_{16} x_{19}, x_{6} x_{9} x_{11} x_{17} -  x_{3} x_{5} x_{16} x_{19}, x_{4} x_{5} x_{13} x_{17} -  x_{7} x_{8} x_{15} x_{18}, x_{4} x_{7} x_{13} x_{17} -  x_{5} x_{8} x_{15} x_{18}, x_{5} x_{7} x_{13} x_{17} -  x_{4} x_{8} x_{15} x_{18}, x_{4} x_{8} x_{13} x_{17} -  x_{5} x_{7} x_{15} x_{18}, x_{5} x_{8} x_{13} x_{17} -  x_{4} x_{7} x_{15} x_{18}, x_{7} x_{8} x_{13} x_{17} -  x_{4} x_{5} x_{15} x_{18}, x_{2} x_{3} x_{14} x_{17} -  x_{4} x_{5} x_{18} x_{19}, x_{2} x_{4} x_{14} x_{17} -  x_{3} x_{5} x_{18} x_{19}, x_{3} x_{4} x_{14} x_{17}
 -  x_{2} x_{5} x_{18} x_{19}, x_{2} x_{5} x_{14} x_{17} -  x_{3} x_{4} x_{18} x_{19}, x_{3} x_{5} x_{14} x_{17} -  x_{2} x_{4} x_{18} x_{19}, x_{4} x_{5} x_{14} x_{17} -  x_{2} x_{3} x_{18} x_{19}, x_{4} x_{5} x_{15} x_{17} -  x_{7} x_{8} x_{13} x_{18}, x_{4} x_{7} x_{15} x_{17} -  x_{5} x_{8} x_{13} x_{18}, x_{5} x_{7} x_{15} x_{17} -  x_{4} x_{8} x_{13} x_{18}, x_{4} x_{8} x_{15} x_{17} -  x_{5} x_{7} x_{13} x_{18}, x_{5} x_{8} x_{15} x_{17} -  x_{4} x_{7} x_{13} x_{18}, x_{7} x_{8} x_{15} x_{17} -  x_{4} x_{5} x_{13} x_{18}, x_{4} x_{13} x_{15} x_{17} -  x_{5} x_{7} x_{8} x_{18}, x_{5} x_{13} x_{15} x_{17} -  x_{4} x_{7} x_{8} x_{18}, x_{7} x_{13} x_{15} x_{17} -  x_{4} x_{5} x_{8} x_{18}, x_{8} x_{13} x_{15} x_{17} -  x_{4} x_{5} x_{7} x_{18}, x_{3} x_{5} x_{16} x_{17} -  x_{6} x_{9} x_{11} x_{19}, x_{3} x_{6} x_{16} x_{17} -  x_{5} x_{9} x_{11} x_{19}, x_{5} x_{6} x_{16} x_{17} -  x_{3} x_{9} x_{11} x_{19}, x_{3} x_{9} x_{16} x_{17} -  x_{5} x_{6} x_{11} x_{19}, x_{5} x_{9} x_{16} x_{17} -  x_{3} x_{6} 
x_{11} x_{19}, x_{6} x_{9} x_{16} x_{17} -  x_{3} x_{5} x_{11} x_{19}, x_{3} x_{11} x_{16} x_{17} -  x_{5} x_{6} x_{9} x_{19}, x_{5} x_{11} x_{16} x_{17} -  x_{3} x_{6} x_{9} x_{19}, x_{6} x_{11} x_{16} x_{17} -  x_{3} x_{5} x_{9} x_{19}, x_{9} x_{11} x_{16} x_{17} -  x_{3} x_{5} x_{6} x_{19}, x_{2} x_{3} x_{4} x_{18} -  x_{5} x_{14} x_{17} x_{19}, x_{2} x_{3} x_{5} x_{18} -  x_{4} x_{14} x_{17} x_{19}, x_{2} x_{4} x_{5} x_{18} -  x_{3} x_{14} x_{17} x_{19}, x_{3} x_{4} x_{5} x_{18} -  x_{2} x_{14} x_{17} x_{19}, x_{2} x_{3} x_{14} x_{18} -  x_{4} x_{5} x_{17} x_{19}, x_{2} x_{4} x_{14} x_{18} -  x_{3} x_{5} x_{17} x_{19}, x_{3} x_{4} x_{14} x_{18} -  x_{2} x_{5} x_{17} x_{19}, x_{2} x_{5} x_{14} x_{18} -  x_{3} x_{4} x_{17} x_{19}, x_{3} x_{5} x_{14} x_{18} -  x_{2} x_{4} x_{17} x_{19}, x_{4} x_{5} x_{14} x_{18} -  x_{2} x_{3} x_{17} x_{19}, x_{2} x_{3} x_{17} x_{18} -  x_{4} x_{5} x_{14} x_{19}, x_{2} x_{4} x_{17} x_{18} -  x_{3} x_{5} x_{14} x_{19}, x_{3} x_{4} x_{17} x_{18} -  x_{2} x_{5} x_{14} x_{19}, 
x_{2} x_{5} x_{17} x_{18} -  x_{3} x_{4} x_{14} x_{19}, x_{3} x_{5} x_{17} x_{18} -  x_{2} x_{4} x_{14} x_{19}, x_{4} x_{5} x_{17} x_{18} -  x_{2} x_{3} x_{14} x_{19}, x_{2} x_{14} x_{17} x_{18} -  x_{3} x_{4} x_{5} x_{19}, x_{3} x_{14} x_{17} x_{18} -  x_{2} x_{4} x_{5} x_{19}, x_{4} x_{14} x_{17} x_{18} -  x_{2} x_{3} x_{5} x_{19}, x_{5} x_{14} x_{17} x_{18} -  x_{2} x_{3} x_{4} x_{19}, x_{1}^{2} - 1, x_{2}^{2} - 1, x_{3}^{2} - 1, x_{4}^{2} - 1, x_{5}^{2} - 1, x_{6}^{2} - 1, x_{7}^{2} - 1, x_{8}^{2} - 1, x_{9}^{2} - 1, x_{10}^{2} - 1, x_{11}^{2} - 1, x_{12}^{2} - 1, x_{13}^{2} - 1, x_{14}^{2} - 1, x_{15}^{2} - 1, x_{16}^{2} - 1, x_{17}^{2} - 1, x_{18}^{2} - 1, x_{19}^{2} - 1 \rangle$

\subsection{Error Correcting capability of the Schubert Code $\bf C_{(2,4)}(2,5)(\FF_2)$:}
The minimum of the total degree of each of the above binomial except binomials of the form $x_i^2-1$,for  $ i=1,2,\ldots,19$ is $4$ and hence by, theorem $4$ above, we get $t=3$, which is in accordance with our known results(\cite{Edgar, MS}), because for this Schubert code, we know that $d=8$, and hence $t=3$.\\

\subsection{Decoding with $[19,5,8]$-linear Schubert Code $\bf C_{(2,4)}(2,5)(\FF_2)$:}
By using theorem $3$ above from \cite{Edgar}, we can tackle the decoding problem as follows:
\begin{center}
\begin{tabular}{|c|c|c|}
\hline
Received word & Canonical form of &Received word is decoded to \\ & this word w.r.t. $\mathcal{G}_{(2,4)}(2,5)$ & \\
\hline
$x_1x_6x_7x_9x_{11}x_{13}x_{14}x_{18}x_{19}$	&	$x_5x_9x_{10}$		&		$x_1x_5x_6x_7x_{10}x_{11}x_{13}x_{14}x_{18}x_{19}$ \\
\hline
$x_1x_5x_7x_9x_{11}x_{13}x_{14}x_{18}x_{19}$	&	$x_6x_9x_{10}$		&		$x_1x_5x_6x_7x_{10}x_{11}x_{13}x_{14}x_{18}x_{19}$ \\
\hline
$x_1x_5x_8x_{10}x_{11}x_{13}x_{14}x_{18}x_{19}$	&	$x_6x_7x_8$		&		$x_1x_5x_6x_7x_{10}x_{11}x_{13}x_{14}x_{18}x_{19}$ \\
\hline
$x_1x_5x_7x_8x_{11}x_{13}x_{14}x_{18}x_{19}$	&	$x_6x_8x_{10}$		&		$x_1x_5x_6x_7x_{10}x_{11}x_{13}x_{14}x_{18}x_{19}$ \\
\hline
\end{tabular}
\end{center}
In order to decode the codes different from those listed in above table, we can use Maximum likelihood decoding rule\cite{Xing} or Minimum distance decoding rule\cite{Xing}. In this paper, we have given decoding of Schubert codes in some special cases. It will be intersting to see Gr\"obner bases for Grassmann codes associated with higher dimensional varieties like Grassmann varieties and their decoding.\\\\\\


\end{document}